\documentclass{mn2e}
\usepackage[pdftex]{graphicx}

\title[HerMES: Clusters of Dusty Galaxies]
{HerMES: Clusters of Dusty Galaxies Uncovered by {\em Herschel}\thanks{Based in part on observations with {\em Herschel}, an ESA space observatory with science instruments provided by European-led Principal Investigator consortia and with important participation from NASA} and {\em Planck}\thanks{Based in part on observations obtained with {\em Planck} (http://www.esa.int/{\em Planck}), an ESA science mission with instruments and contributions directly funded by ESA Member States, NASA \& Canada}}
\author[D.L.~Clements et al.]
{\parbox{\textwidth}{\raggedright D.L.~Clements$^{1}$\thanks{E-mail: \texttt{d.clements@imperial.ac.uk}},
F.G.~Braglia$^{1}$,
A.~Hyde$^{1}$,
I.~P{\'e}rez-Fournon$^{2,3}$,
J.~Bock$^{4,5}$,
A.~Cava$^{6}$,
S.~Chapman$^{20}$,
A.~Conley$^{7}$,
A.~Cooray$^{4,17}$,
D.~Farrah$^{9}$,
E.A.~Gonz\'alez~Solares$^{10}$,
L.~Marchetti$^{11,12}$,
G.~Marsden$^{13}$,
S.J.~Oliver$^{8}$,
I.G.~Roseboom$^{8,14}$,
B.~Schulz$^{4,15}$,
A.J.~Smith$^{8}$,
M.~Vaccari$^{12,16}$,
J.~Vieira$^{4}$,
M.~Viero$^{4}$,
L.~Wang$^{8}$,
J.~Wardlow$^{17}$,
M.~Zemcov$^{4,5}$,
G.~de Zotti$^{18,19}$}\vspace{0.4cm}\\
\parbox{\textwidth}{\raggedright
{\tiny $^{1}$Astrophysics Group, Imperial College London, Blackett Laboratory, Prince Consort Road, London SW7 2AZ, UK\\
$^{2}$Instituto de Astrof{\'\i}sica de Canarias (IAC), E-38200 La Laguna, Tenerife, Spain\\
$^{3}$Departamento de Astrof{\'\i}sica, Universidad de La Laguna (ULL), E-38205 La Laguna, Tenerife, Spain\\
$^{4}$California Institute of Technology, 1200 E. California Blvd., Pasadena, CA 91125, USA\\
$^{5}$Jet Propulsion Laboratory, 4800 Oak Grove Drive, Pasadena, CA 91109, USA\\
$^{6}$Departamento de Astrof\'isica, Facultad de CC. F\'isicas, Universidad Complutense de Madrid, E-28040 Madrid, Spain\\
$^{7}$Center for Astrophysics and Space Astronomy 389-UCB, University of Colorado, Boulder, CO 80309, USA\\
$^{8}$Astronomy Centre, Dept. of Physics \& Astronomy, University of Sussex, Brighton BN1 9QH, UK\\
$^{9}$Department of Physics, Virginia Tech, Blacksburg, VA 24061, USA\\
$^{10}$Institute of Astronomy, University of Cambridge, Madingley Road, Cambridge CB3 0HA, UK\\
$^{11}$Department of Physical Sciences, The Open University, Milton Keynes MK7 6AA, UK\\
$^{12}$Dipartimento di Fisica e Astronomia, Universit\`{a} di Padova, vicolo Osservatorio, 3, 35122 Padova, Italy\\
$^{13}$Department of Physics \& Astronomy, University of British Columbia, 6224 Agricultural Road, Vancouver, BC V6T~1Z1, Canada\\
$^{14}$Institute for Astronomy, University of Edinburgh, Royal Observatory, Blackford Hill, Edinburgh EH9 3HJ, UK\\
$^{15}$Infrared Processing and Analysis Center, MS 100-22, California Institute of Technology, JPL, Pasadena, CA 91125, USA\\
$^{16}$Astrophysics Group, Physics Department, University of the Western Cape, Private Bag X17, 7535, Bellville, Cape Town, South Africa\\
$^{17}$Dept. of Physics \& Astronomy, University of California, Irvine, CA 92697, USA\\
$^{18}$INAF-OAPd, Vicolo dell'Osservatorio 5, 35122 Padova, Italy\\
$^{19}$SISSA, via Bonomea 265, I 34136 Trieste, Italy\\
$^{20}$Department of Physics and Atmospheric Science, Dalhousie University, Coburg Road Halifax, B3H 4R2, Canada\\}}}
\date{}
\pagerange{\pageref{firstpage}--\pageref{lastpage}}
\pubyear{}

\begin{document}

\maketitle

\label{firstpage}

\begin{abstract}
The potential for {\em Planck} to detect clusters of dusty, star-forming galaxies at z$>$1 is tested by examining the {\em Herschel}-SPIRE images of {\em Planck} Early Release Compact Source Catalog (ERCSC) sources lying in fields observed by the HerMES survey. Of the 16 {\em Planck} sources that lie in the $\sim$90 sq. deg. examined, we find that twelve are associated with single bright {\em Herschel} sources. The remaining four are associated with overdensities of 
{\em Herschel} sources, making them candidate clusters of dusty, star-forming galaxies. We use complementary optical/NIR data for these `clumps' to test this idea, and find evidence for the presence of galaxy clusters in all four cases. We use photometric redshifts and red sequence galaxies to estimate the redshifts of these clusters, finding that they range from 0.8 to 2.3. These redshifts imply that the {\em Herschel} sources in these clusters, which contribute to the detected {\em Planck} flux, are forming stars very rapidly, with typical total cluster star formation rates $> 1000 M_{\odot}/yr$. The high redshift clusters discovered in these observations are used to constrain the epoch of cluster galaxy formation, finding that the galaxies in our clusters are 1-1.5 Gy old at $z\sim1-2$. Prospects for the discovery of further clusters of dusty galaxies are discussed, using not only all sky {\em Planck} surveys, but also deeper, smaller area, {\em Herschel} surveys. 
\end{abstract}


\section{Introduction}

The discovery of distant far-IR luminous galaxies by submm imagers (eg. Hughes et al., 1998; Smail et al., 1997)
and the discovery of the Cosmic Infrared Background (Puget et al., 1996; Fixsen et al., 1998) have demonstrated the importance of the far-IR/submm bands in determining a complete picture of the history of galaxy formation and evolution. The high redshift (z$\sim$2--3) sources detected in these submillimetre (submm) surveys are expected to be the progenitors of the massive elliptical galaxies that we see today (Lilly et al., 1999; Farrah et al., 2006; Swinbank et al., 2008). Within hierarchical clustering models of large scale structure and galaxy formation we would expect that the most massive elliptical galaxies will form in the cores of what will become today's most massive galaxy clusters. Granato et al. (2004) and others predict that many galaxies in such regions will undergo simultaneous starbursts, detectable as dust obscured submm galaxies (SMGs), and producing integrated {\it clumps} of proto-spheroidal galaxies.

Hints of such objects may already have been found by clustering studies with Spitzer (eg. Magliocchetti et al., 2007), as an overdensity of submm sources (Chapman et al. 2009), or through using z=1.7---2.8 quasars as signposts for possible protoclusters (Stevens et al., 2010). The latter study finds far more submm-bright companions to quasars than would be expected if they were in typical field environments, implying the presence of a cluster of dusty galaxies. Studies of the regions around high redshift radio galaxies, whether in the near-IR (Kodama et al., 2007), or at {\em Herschel} wavelengths (Voltchanov et al., 2013; Rigby et al., 2013) find similar results. Meanwhile, the highest redshift protocluster currently known, at z$\sim$5.3, includes at least one submm luminous object (Capak et al., 2011). While this object is extreme, such sources may need to be quite common if the recent discovery of a mature galaxy cluster at z=2.07 (Gobat et al., 2011), with a fully formed red-sequence of galaxies with ages $\ge 1.3$Gyr, is representative of a significant population. 

The dusty star-forming phase of a protocluster, as discussed by Granato et al. (2004), is likely to be quite short, so the objects will be rare on the sky. Fortunately, the new generation of far-IR/submm satellites, {\em Herschel} and {\em Planck}, are up to this challenge. The {\em Herschel} Space Observatory (Pilbratt et al., 2010) allows large areas of the sky to be covered to sensitive flux levels at wavelengths corresponding to the peak of the dust spectral energy distribution (SED) of high redshift starbursts at $\sim$100$\mu$m in the emitted frame (eg. Negrello et al., 2010; Wardlow et al., 2013). In particular, the HerMES survey (Oliver et al., 2012), the largest {\em Herschel} key programme in terms of observing time, has covered $\sim$380 sq. deg. of the sky at a variety of depths and is thus well suited to the search for rare dusty high redshift sources. The availability of a wide range of multiwavelength complementary data in the HerMES fields is particularly useful as it allows any unusual sources found to be `followed-up' with extant data at a wide range of wavelengths. Meanwhile, the H-ATLAS survey (Eales et al., 2010), the largest area {\em Herschel} survey at $\sim$550 sq. deg., is also suited to the search for rare objects (eg. Heranz et al. (2012), Fu et al., (2012), Clements et al. in prep). {\em Planck} provides all sky coverage at wavelengths matching the longest {\em Herschel} bands  (eg. {\em Planck} Collaboration, 2011a). While less sensitive than {\em Herschel}, and with a much larger beam, {\em Planck}'s area coverage is unmatched, and its longer wavelength channels can be very useful in selecting high redshift objects. In principle, {\em Planck} can find cold compact structures and {\em Herschel} can then confirm that these are associated with clumps of galaxies, potentially at high redshift.

Negrello et al. (2005) examined the effect of clustering on the extragalactic sources that would be detected by instruments with large beams, such as the {\em Planck} High Frequency Instrument (HFI) which has a $\sim$5 arcmin beam full width half maximum (FWHM). They concluded that if the beam is approximately matched to the clustering scale, then the large beam instrument will detect the clustered sources as unresolved, or marginally resolved, discrete sources, and that these clusters will make a significant contribution to the source counts measured in such surveys. Now that the first data products from the {\em Planck} survey have been released, we can test these ideas by examining {\em Herschel} images of sources listed in the {\em Planck} compact source catalog. Clusters of dusty star forming galaxies along the lines of those proposed by Negrello et al. (2005) will appear as groups of discreet objects in the {\em Herschel} maps, while contaminating sources, such as foreground bright early-type galaxies or cirrus dust, will appear very different, respectively as either single bright sources or extended diffuse emission. This paper presents the first results of our efforts to cross match {\em Planck} and {\em Herschel} sources in the HerMES survey. Other studies using data from H-ATLAS (Herranz et al., 2012; Clements et al., in prep) and targeted observations of sources selected by their colours from the {\em Planck} maps, rather than the {\em Planck} catalogs used here (Montier et al. in prep), are also underway.

The rest of this paper is organised as follows. In Section 2 we describe the {\em Planck} and {\em Herschel} data used to find candidate clusters of dusty galaxies. In Section 3 we describe the process used to match {\em Planck} and {\em Herschel} sources and then describe the {\em Planck} and {\em Herschel} properties of the resulting sample. In Section 4 we present mutiwavelength complementary data and observations for the four candidate dusty galaxy clusters we identify, while in Section 5 we discuss the results of these observations, which confirm that the objects are indeed galaxy clusters. In Section 6 we discuss these results and their implications for the properties of these clusters and for galaxy and cluster evolution more generally. We draw our conclusions in Section 7. Throughout this paper we use the standard concordance cosmology with $H_0 = 70\, \mathrm{km}\, \mathrm{s}^{-1} \, \mathrm{Mpc}^{-1}, \Omega_M = 0.3$ and $\Omega_{\Lambda}=0.7$.

\section{The HerMES and {\em Planck} Surveys}

In order to search for clusters of dusty galaxies, sensitive large area surveys in the far-IR or submillimetre bands are required. The {\em Herschel} (Pilbrat et al., 2010) and {\em Planck} ({\em Planck} Collaboration, 2011a) spacecraft are ideally suited to this purpose. {\em Planck} is an all sky survey mission operating at nine different frequencies from 30 to 857 GHz (1cm to 350$\mu$m in wavelength). Its main goal is to study the cosmic microwave background (CMB), but as a byproduct it is producing all sky surveys in all its observational bands that can be used to look for foreground non-CMB sources, both compact and extended. {\em Herschel}, conversely, is an observatory mission, with instruments capable of diffraction limited imaging from 70 to 500$\mu$m in wavelength (4.3 THz to 600GHz in terms of frequency). With its larger primary mirror and larger detector arrays, {\em Herschel} can reach higher angular resolutions and higher sensitivities than {\em Planck}, but it is not an all sky survey instrument, with its areal coverage limited to $<$10\% of the extragalactic sky. The data from {\em Planck} and {\em Herschel} used in our search for clusters of dusty galaxies are the {\em Planck} Early Release Compact Source Catalog (ERCSC; {\em Planck} Collaboration, 2011b) and the Data Release 1 catalogs for selected fields from the HerMES\footnote{http://hermes.sussex.ac.uk} survey (Oliver et al., 2012).

\subsection{The {\em Planck} ERCSC}

The ERCSC is a catalog of compact sources found in the first all sky survey completed by {\em Planck} in its nine observational channels, covering the 30, 44 and 70 GHz bands from the Low Frequency Instrument (Mannella et al., 2011) and the 100, 143, 217, 353, 545 and 857 GHz  bands from the High Frequency Instrument (HFI Core Team, 2011). For our current purposes we are interested in looking for sources whose emission is dominated by dust. Such sources, at a wide range of redshift, will be most prominent in the highest frequency bands. The highest frequency channel on {\em Planck}, at 857GHz, corresponds to the SPIRE 350$\mu$m band. The catalogs that make up the ERCSC database ({\em Planck} Collaboration, 2011b) include monochromatic catalogs, with sources detected and fluxes extracted in each of the nine bands, as well as a bandmerged catalog including fluxes for every 857GHz detected source given for each of the highest four frequencies. We use this latter data product for the current work. It is based on sources detected in the highest frequency channel, 857GHz, corresponding to a wavelength of 350$\mu$m, and adds fluxes extracted from the {\em Planck} maps at 545, 353 and 217 GHz at the position of each 857GHz selected source. The {\em Planck} beam in each of these bands is $\sim$5 arcminutes in diameter, and all selected sources are detected to at least 5$\sigma$ sensitivity in the 857GHz channel. This sensitivity does not translate into a fixed flux limit, however, as a result of both foreground confusion, which varies from place to place on the sky, and because the {\em Planck} scanning strategy means that some parts of the sky are covered more often than others and thus reach higher sensitivities. Details of the extraction and bandmerging processes are discussed in {\em Planck} Collaboration (2011b) and in the documentation associated with the ERCSC. 

\subsection{The HerMES Survey}

The {\em Herschel} Multitiered Extragalactic Survey (HerMES) is the largest Guaranteed Time Key Project being implemented by the {\em Herschel} Space Observatory (Oliver et al., 2012). It covers a total of about 380 sq. deg. with the SPIRE instrument to provide 250, 350 and 500$\mu$m maps (Griffin et al., 2010) and, in most survey fields, uses the PACS instrument to add 100 and 160$\mu$m coverage (Poglitsch et al., 2010). The survey area is spread over most well studied extragalactic fields to ensure the availability of supporting multifrequency data. The depths reached by the HerMES survey vary from field to field to produce a classic `wedding cake' survey, with the deepest observations covering small areas and shallow observations covering large areas. For the fields of interest to the current paper, the HerMES instrumental noise ranges from 12.7-13.8mJy at 250$\mu$m, 10.5-11.3mJy at 350$\mu$m and 15.2-16.4mJy at 500$\mu$m (5$\sigma$). These values are lower than the 5$\sigma$ confusion limits in these bands of 24, 27.5 and 30.5 mJy respectively (Nguyen et al., 2010). 

We use HerMES images and catalogues produced as part of the HerMES Data Release 1 (Wang et al., in prep). The images were produced using the SPIRE-HerMES Iterative Mapmaker (SHIM). The algorithms used are discussed fully in Levenson et al. (2010) and updated in Viero et al. (submitted), while overall SPIRE calibration is discussed in Swinyard et al. (2010). Sources are extracted from these maps using the SUSSExtractor algorithm (Savage \& Oliver, 2007) in a manner similar to that described in Smith et al. (2012) for the initial catalogs released by HerMES. The source catalogs used are all monochromatic in the sense that no data at other frequencies, whether from a different SPIRE channel or other external source priors, was used in the extraction of sources at a given frequency. The completeness and reliability of the catalogs used here were checked in a similar manner to the checks applied in Smith et al. (2012), and will be discussed in detail in Wang et al. (in prep), and amount to roughly 50\% completeness and 90\% reliability at a 250$\mu$m flux of 40mJy. For our purposes here, high reliability is more important than high completeness to ensure that we minimise any contribution from unreliable, false, sources to calculations of eg. source densities.


\section{{\em Planck} Sources in HerMES}

Candidate dusty clusters can be identified by examining the SPIRE maps at the positions of {\em Planck} ERCSC sources. The most common counterparts to ERCSC sources in extragalactic fields are either nearby galaxies or high galactic latitude cirrus structures (see eg. Herranz et al., 2013; {\em Planck} collaboration, 2011c). In both cases visual inspection of the SPIRE maps is enough to determine the nature of the ERCSC source. To find candidate dusty clusters we examine the HerMES maps at the location of {\em Planck} ERCSC sources in fields where there is plentiful ancillary data to allow their further study. The HerMES maps searched include the fields listed in Table 1, details of which can be found in Oliver et al. (2012). The total area on the sky covered in this search is 91.1 sq. deg. We find 16 {\em Planck} 857GHz sources in these HerMES fields. The details of these sources are shown in Table 2, and we show 250$\mu$m maps of the regions associated with these {\em Planck} sources in Fig. 1. We also search existing source catalogs in the NASA Extragalactic Database (NED) to determine if any already known objects are associated with the {\em Planck} ERCSC sources. The results of this identification process are also shown in Table 2. 

\begin{table}
\begin{tabular}{ccc}\hline
Field Name&Field Area (sq. deg.)&ERCSC Sources\\ \hline
Bo\"{o}tes-NDWFS&10.6&7\\
XMMLSS- SWIRE&18.9&3\\
EGS&2.7&1\\
Lockman SWIRE&16.1&4\\
CDFS SWIRE&10.9&1\\
COSMOS&4.4&0\\
ELAIS S1&7.9&0\\
ELAIS N1&12.3&0\\
GOODS-N&0.6&0\\
FLS&6.7&0\\ \hline
\end{tabular}
\caption{List of HerMES fields where extensive multiwavelength complementary data exists that were examined in the search for {\em Planck} clumps, and the number of {\em Planck} ERCSC sources found. Details of the {\em Herschel} observations of these fields can be found in Oliver et al. (2012).}
\end{table}

It is clear from this analysis that there are two classes of {\em Herschel} sources associated with {\em Planck} ERCSC sources: Firstly, there are bright individual, well known objects that can be clearly distinguished in the images, most of which are also resolved by {\em Herschel}.  All of these sources were detected by IRAS. These objects are responsible for the majority of {\em Planck} ERCSC sources, accounting for 12 of 16 objects. However, there is also a small number of {\em Planck} sources (4/16) for which no bright nearby counterpart can be found. The {\em Herschel} counterparts to these sources, classified as `clumps' in Table 2, appear to correspond to groups of fainter {\em Herschel} sources in contrast to the single bright nearby sources that account for the other twelve {\em Planck} sources. This tendency can be quantified by examining the standard deviation of the 250$\mu$m flux distribution, $\sigma_{250}$, of the {\em Herschel} sources associated with each ERCSC source. For these purposes, we define association to mean that a {\em Herschel} source lies within 4.23 arcminutes of the ERCSC source position, since 4.23 arcminutes is the FWHM of the {\em Planck} beam at 857 GHz ({\em Planck} Collaboration, 2011b; see also Herranz et al., 2013). As can be seen, all `clump' sources have a low value of $\sigma_{250}$, $\leq 0.025$ Jy, while sources with bright counterparts have standard deviations twice this value since they are dominated by a single bright source. There are two exceptions to this result. NGC5641, which has a low value of $\sigma_{250}$ because it lies at the very edge of the HerMES map, and Omicron Ceti, also known as Mira. The latter is a bright extended object, the brightest of the {\em Planck} sources in all bands, and has significant structure in the {\em Herschel} images. This structure is broken into several different subcomponents by the {\em Herschel} point-source-optimised source extraction routines. The $\sigma_{250}$ value for this object will thus be unreliable and biased to a value lower than an extended source extraction method would produce. Nevertheless, the $\sigma_{250}$ values for both Omicron Ceti and NGC5641 are still $> 0.025$ Jy and thus larger than the values of any clump.

These clumps match the generic properties expected for dusty clusters as predicted by Negrello et al. (2005).  Three colour {\em Herschel} images of the sources we identify as clumps in this way are shown in Figure 2, with the 500$\mu$m image shown as red, 350$\mu$m shown as green and 250$\mu$m shown as blue, demonstrating that they appear to be local density enhancements of red {\em Herschel} sources. We next look at their more detailed properties as revealed by both {\em Planck} and {\em Herschel}.

\subsection{The {\em Planck} Properties of Clumps}

Figure 3 shows the {\em Planck} colours $F_{857}/F_{545}$ vs. $F_{545}/F_{353}$ (ie. F(350$\mu$m)/F(550$\mu$m) vs. F(550$\mu$m)/F(850$\mu$m)) for the four clumps identified above compared to the bulk of ERCSC sources, shown as small red dots, and the HerMES {\em Planck} sources that are not identified as clumps (blue dots). The colours are also compared to colour as a function of redshift for two galaxy spectral energy distribution (SED) templates from the models of Pearson et al. (2009), one for an M82 starbursting galaxy and the other for a quiescent cirrus-type galaxy. The colours of our clump sources have significant errors since they are not detected with high significance in all the relevant {\em Planck} bands. In particular the Lockman clump lacks a secure 353 GHz detection so its colours are poorly constrained. This is also true of many of the non-clump {\em Planck} sources in the HerMES fields. However, for the three remaining clumps it is clear that their {\em Planck} colours are redder than the bulk of {\em Planck} ERCSC sources, suggesting that they may lie at higher redshift than the z$<$0.1 typical of {\em Planck} ERCSC sources ({\em Planck} Consortium, 2011c). The Bo\"{o}tes clump in particular has the reddest as well as the best measured colours, which is strongly suggestive that it lies at a high redshift. In this context it is worth noting that the {\em Planck} source associated with the red, z=3.26 H-ATLAS lensed source HATLAS J114637.9-001132, discussed by Fu et al. (2012) and Herranz et al. (2013), has colours quite similar to the Bo\"{o}tes clump. This H-ATLAS source is also listed as being an extended source in the {\em Planck} ERCSC. Of the {\em Planck}-HerMES sources listed in Table 1, only two are found to be extended in the {\em Planck} ERCSC. Both of these are clump sources lacking a bright {\em Herschel} counterpart, and one of these extended sources is the Bo\"{o}tes clump.

\subsection{The {\em Herschel} Properties of Clumps}

Having investigated the {\em Planck} properties of the clump sources, the next step is to study the properties of the individual {\em Herschel} sources associated with the clumps. We extract these associated sources from the HerMES catalogs by selecting all {\em Herschel} sources that lie within 4.23 arcminutes of the position of the {\em Planck} source. We use a radius of 4.23 arcminutes for this purpose since it is the FWHM of the {\em Planck} beam at 857GHz. We find that typically sixty to seventy {\em Herschel} sources are associated with each clump in this way, though not all sources are detected in all three SPIRE bands. A comparison of the {\em Herschel} source density in the region of each clump with {\em Herschel} source densities within a degree of each clump is shown in Figure 4, where source densities are colour coded blue for 250$\mu$m sources, green for 350$\mu$m and red for 500$\mu$m. As can be seen, each of the four clumps appears to be associated with a local overdensity of {\em Herschel} sources. We calculate the statistical significance of each overdensity as follows. Firstly, we apply an adaptive kernel filter to the catalogue of {\it {\em Herschel}} sources to associate a local density value with each source (e.g. Pisani et al. 1993). The distance to the $n$-th nearest neighbour is used to calculate the kernel width at each point, where $n$ is derived from the mean density of sources in the field (we found $n = 11$ consistently for all catalogues). The density $\delta_i$ at the position of the $i$-th source is expressed as:

\[ \delta_i  = \frac{W_i}{\pi(d_{ik}\lambda_i)^2} \sum_j \exp\left[-0.5\left(\frac{d_{ij}}{d_{ik}\lambda_i}\right)^2\right] \]

where $d_{ij}$ is the distance to the $j$-th point, $d_{ik}$ is the distance to the $k$-th nearest neighbour, both from the $i$-th source, $\lambda_i$ is a smoothing factor dependent on the local density around the $i$-th source, and $W_i$ is a weighting factor (we used the flux of each source as a weight).
Then, a mean background density and r.m.s. was calculated from the density distribution using a self consistent 3$\sigma$ recursive clipping algorithm. The densities were then converted to a significance in terms of background $\sigma$. Lastly, the overdensity field thus calculated was convolved with a Gaussian kernel with a FWHM of 4.23 arcminutes (i.e. matching the {\it {\em Planck}} beam at 857 GHz). The peak value within the {\it {\em Planck}} beam was then used to represent the overall overdensity of each clump.

The statistical overdensity in each band for each clump is given in Table 3. As can be seen, all of the sources we have identified as {\em Planck} clumps, and which can be seen visually as overdensities in Figs 1, 2 and 4, are confirmed to be statistical overdensities. Genuine clusters of dusty galaxies would be expected to appear as such overdensities of {\em Herschel} sources (Negrello et al., 2005).

We can also examine the properties of the individual {\em Herschel} sources associated with these clumps. Their {\em Herschel} colours are shown in Fig. 5, together with a comparison with colour as a function of redshift for two model source SEDs, one for a star forming galaxy and the other for a more quiescent, cirrus-type galaxy. Only sources with at least a 2$\sigma$ flux measurement in all three bands are plotted. We also plot the mean colours for all the sources associated with each clump. As with the {\em Planck} colours, the {\em Herschel} colours of the clump sources are redder than would be expected for low redshift galaxies, and are broadly consistent with redshifts $\sim$1 or greater. 

The SPIRE 350$\mu$m passband is very similar to the {\em Planck} 857GHz band. The {\em Planck} fluxes for the clumps at 857GHz can thus, in principle, be directly compared to the sum of the 350$\mu$m fluxes for the associated {\em Herschel} sources. However, all the clump sources have 857GHz fluxes below $\sim$1.3 Jy. At this flux level, the {\em Planck} Collaboration (Aatrokoski et al., 2011) note that ERCSC fluxes will be `flux boosted', so the {\em Planck} fluxes are likely to be over-estimates of the actual flux associated with these sources. Herranz et al. (2013) note a similar effect in their studies of {\em Planck} sources in the H-ATLAS Phase 1 fields. When we calculate the contribution of all 350$\mu$m {\em Herschel} sources contributing to the {\em Planck} 857GHz flux of each clump, convolving with the 4.23 arcmin FWHM {\em Planck} beam profile,  it is thus not surprising that we find summed fluxes $\sim$ 2 to 3 times lower than the {\em Planck} ERCSC flux, consistent with the boosting Herranz et al. (2013) found for one of their fainter sources.

\begin{table*}
\begin{tabular}{cccccccccc} \hline
{\em Planck} Name&Field&RA&Dec&F857&F545&F353&NExt&$\sigma$(250)&ID\\ \hline
PLCKERC857 G052.87+68.36&Bo\"{o}tes&217.091&32.397&1.26$\pm$0.1&0.29$\pm$0.1&0.08$\pm$0.1&0&0.060&KUG1426+326\\
PLCKERC857 G055.79+67.45&Bo\"{o}tes&217.966&33.622&1.0$\pm$0.18&0.45$\pm$0.1&0.06$\pm$0.09&0&0.067&VV775\\
PLCKERC857 G059.17+66.46&Bo\"{o}tes&218.822&35.114&1.0$\pm$0.1&0.18$\pm$0.09&0.0$\pm$0.09&0&0.186&KUG1433+353\\
PLCKERC857 G059.93+68.74&Bo\"{o}tes&216.037&34.853&2.2$\pm$0.16&0.48$\pm$0.12&0.08$\pm$0.09&0&0.038&NGC5641$^*$\\
PLCKERC857 G060.27+67.38&Bo\"{o}tes&217.606&35.319&2.7$\pm$0.15&1.06$\pm$0.09&0.22$\pm$0.08&0&0.202&NGC5656\\
PLCKERC857 G060.37+66.55&Bo\"{o}tes&218.5786&35.559&1.24$\pm$0.15&0.81$\pm$0.1&0.36$\pm$0.09&1&0.022&Clump 1\\
PLCKERC857 G060.85+67.13&Bo\"{o}tes&217.836&35.599&1.0$\pm$0.1&0.24$\pm$0.08&0.08$\pm$0.08&0&0.097&MCG+06-32-056\\
PLCKERC857 G095.44+58.94&EGS&216.127&52.936&1.1$\pm$0.18&0.51$\pm$0.11&0.18$\pm$0.05&0&0.018&Clump 2\\
PLCKERC857 G147.66+53.84&Lockman&165.363&57.677&0.75$\pm$0.1&0.31$\pm$0.09&0.21$\pm$0.06&0&0.056&NGC3488\\
PLCKERC857 G148.24+52.44&Lockman&163.020&58.417&0.81$\pm$0.1&0.0$\pm$0.09&0.0$\pm$0.08&0&0.13&NGC3408\\
PLCKERC857 G149.59+53.66&Lockman&163.650&56.986&0.85$\pm$0.09&0.03$\pm$0.1&0.15$\pm$0.08&0&0.12&NGC3445\\
PLCKERC857 G149.81+50.11&Lockman&158.364&59.196&1.25$\pm$0.13&0.45$\pm$0.08&0.06$\pm$0.06&1&0.025&Clump 3\\
PLCKERC857 G224.76-54.44&CDFS&53.219&-28.497&0.81$\pm$0.1&0.35$\pm$0.07&0.17$\pm$0.07&0&0.019&Clump 4\\
PLCKERC857 G167.74-57.97&XMM-LSS&34.836&-2.970&4.75$\pm$0.17&1.81$\pm$0.13&0.46$\pm$0.14&0&0.033&Omicron Ceti\\
PLCKERC857 G171.77-59.54&XMM-LSS&35.402&-5.523&3.74$\pm$0.16&1.36$\pm$0.11&0.33$\pm$0.12&0&0.098&NGC895\\
PLCKERC857 G172.21-60.84&XMM-LSS&34.680&-6.625&1.83$\pm$0.12&0.84$\pm$0.15&0.06$\pm$0.11&0&0.15&NGC881\\ \hline
\end{tabular}
\caption{{\em Planck} sources in HerMES Fields. All fluxes are given in Jy. {\em Planck} bands are indicated as F857, F545, F353 for the 857, 545 and 353 GHz channels. Also shown is the standard deviation for associated {\em Herschel} 250$\mu$m sources, given as $\sigma$(250). See section 3 for details. Next is the {\em Planck} ERCSC flag for extended sources, with Next=1 indicating extension. $^*$ indicates a source at the very edge of the HerMES map. Identifications indicate the name of the foreground galaxy found to be associated with this source. 'Clump' indicates that no foreground galaxy identification was found and thus the source is a candidate dusty cluster.}
\end{table*}

\begin{table}
\begin{tabular}{cccc} \hline
Field&\multicolumn{3}{c}{Overdensity}\\
&250$\mu$m&350$\mu$m&500$\mu$m\\ \hline
Bo\"{o}tes&4.8$\sigma$&4.7$\sigma$&5.7$\sigma$\\
CDF-S&3.6$\sigma$&3.3$\sigma$&2.9$\sigma$\\
EGS&4.6$\sigma$&4.5$\sigma$&5.8$\sigma$\\
Lockman&7.2$\sigma$&4.3$\sigma$&6.7$\sigma$\\  \hline
\end{tabular}
\caption{The statistical significance of the local overdensity of {\em Herschel} sources at the position of each of the identified 'clumps' in each of the {\em Herschel} bands.}
\end{table}

\begin{figure*}
\begin{tabular}{ccc}
\includegraphics[width=4cm]{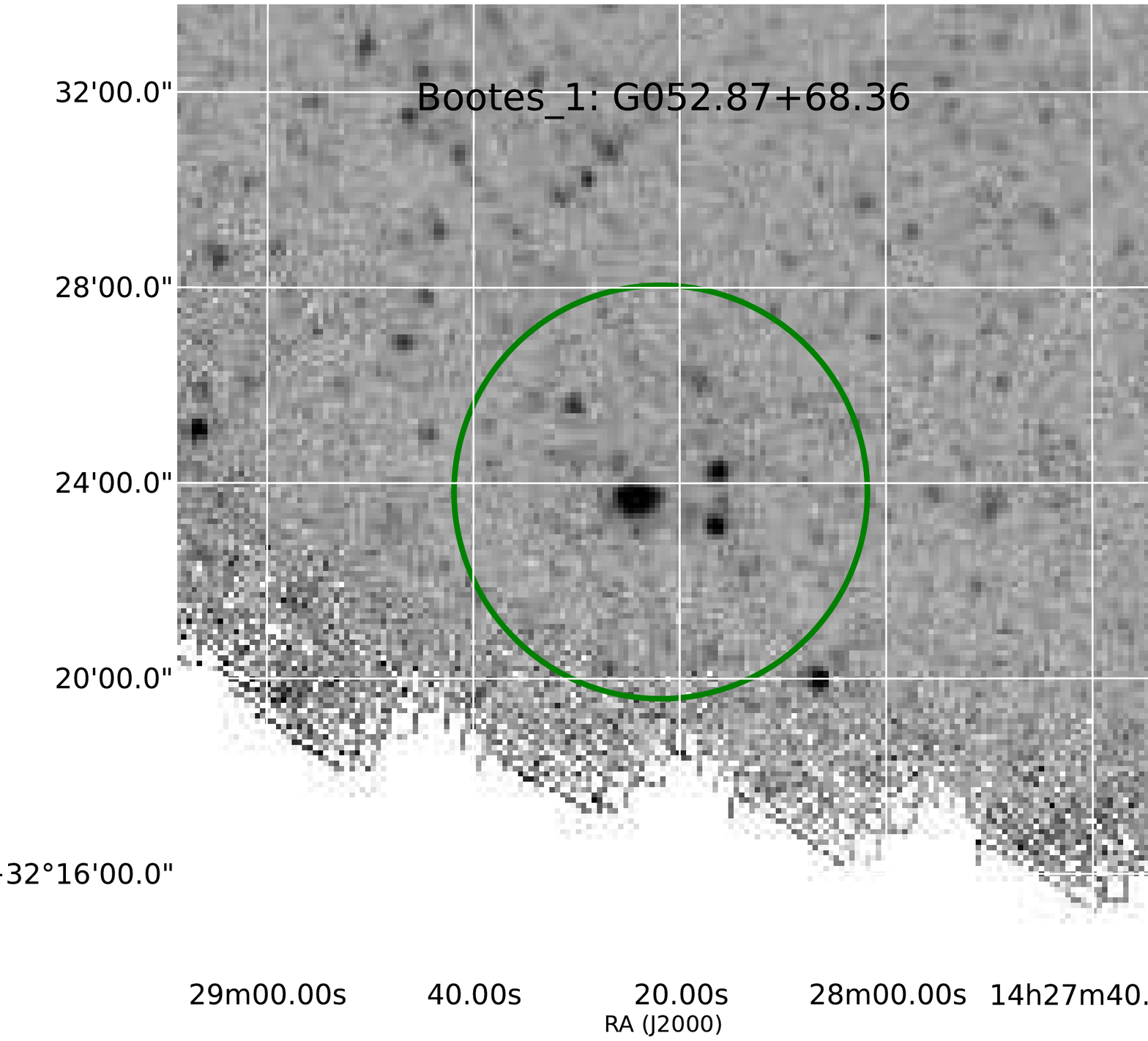}&
\includegraphics[width=4cm]{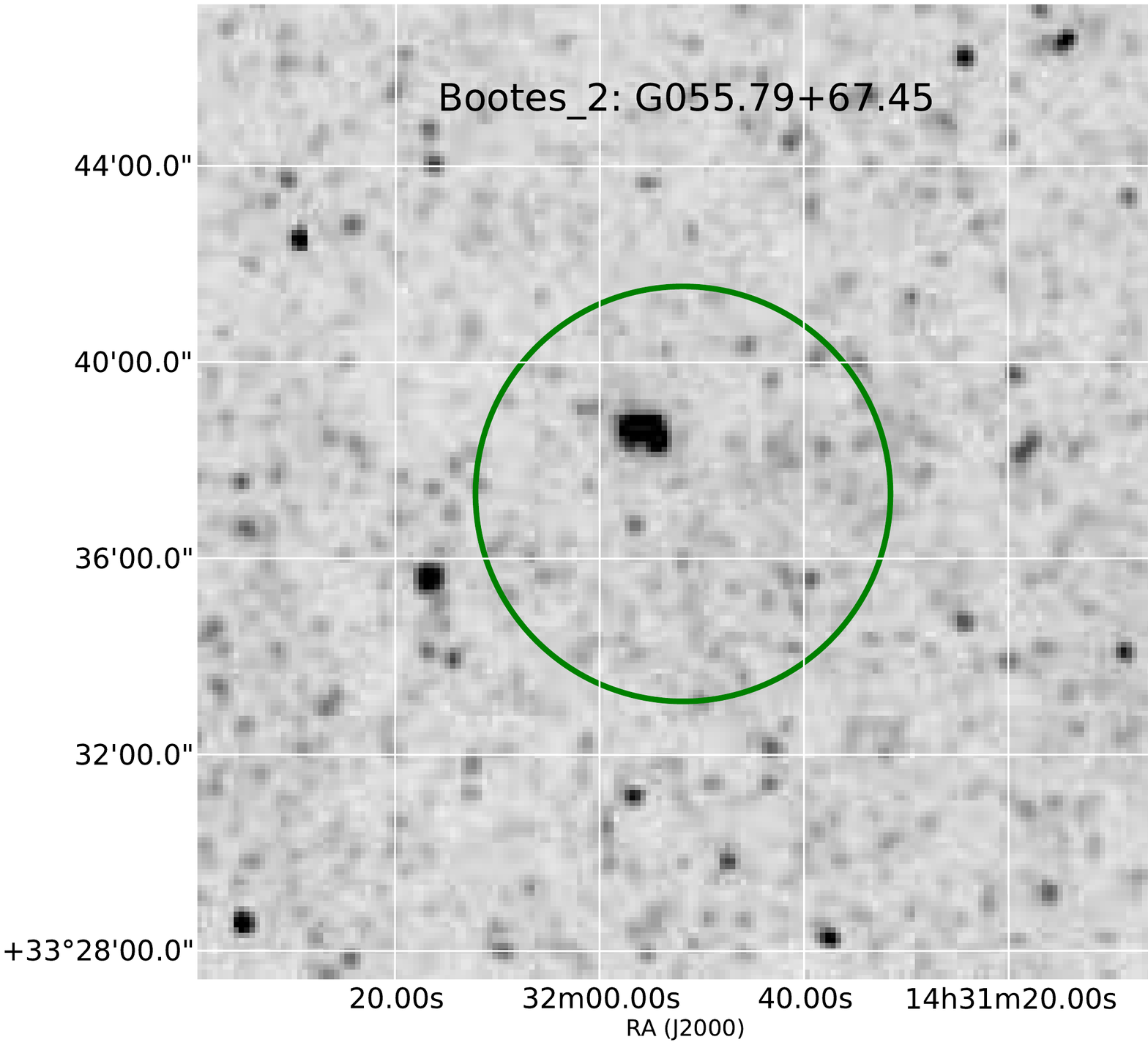} &
\includegraphics[width=4cm]{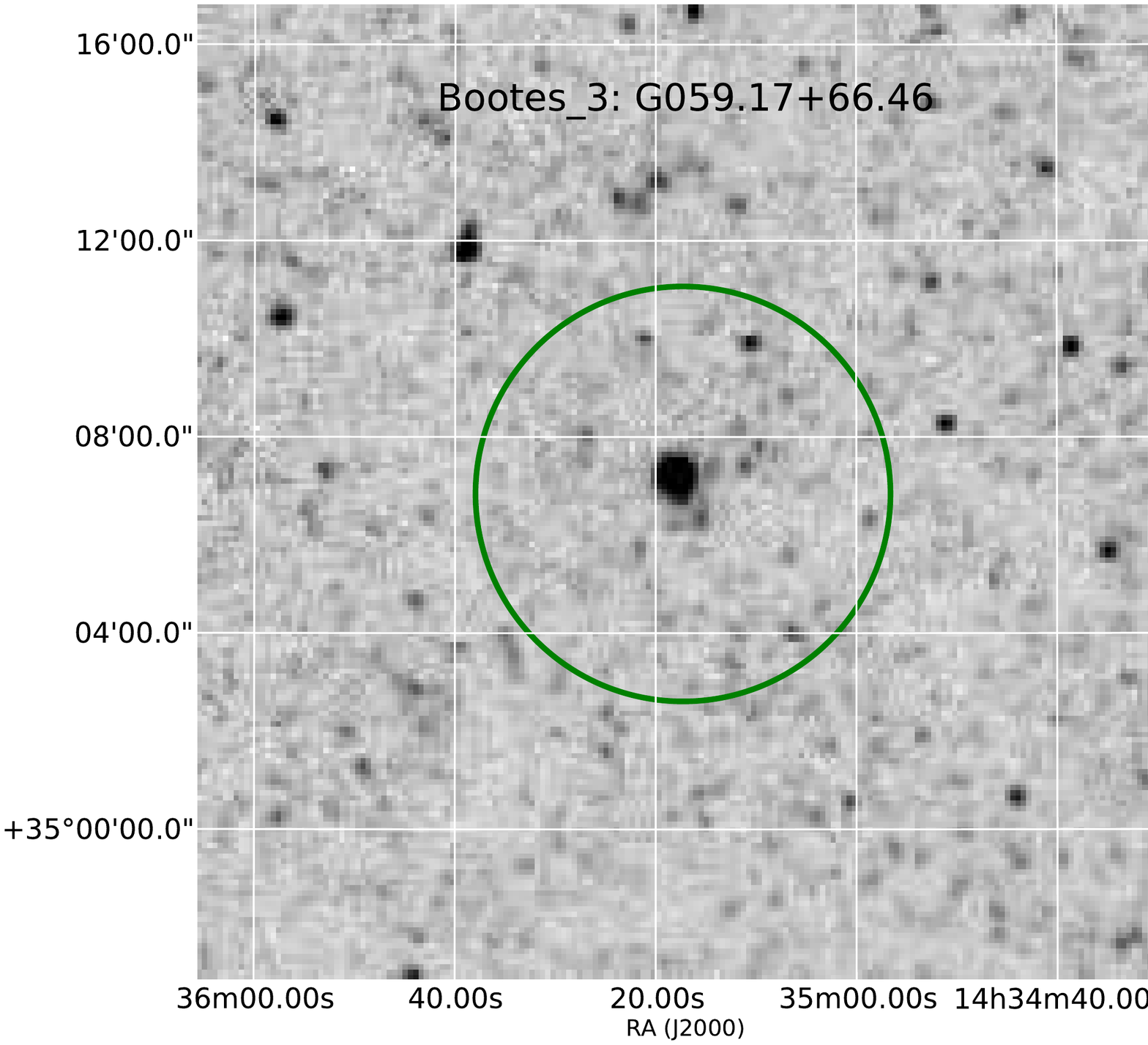}\\
\includegraphics[width=4cm]{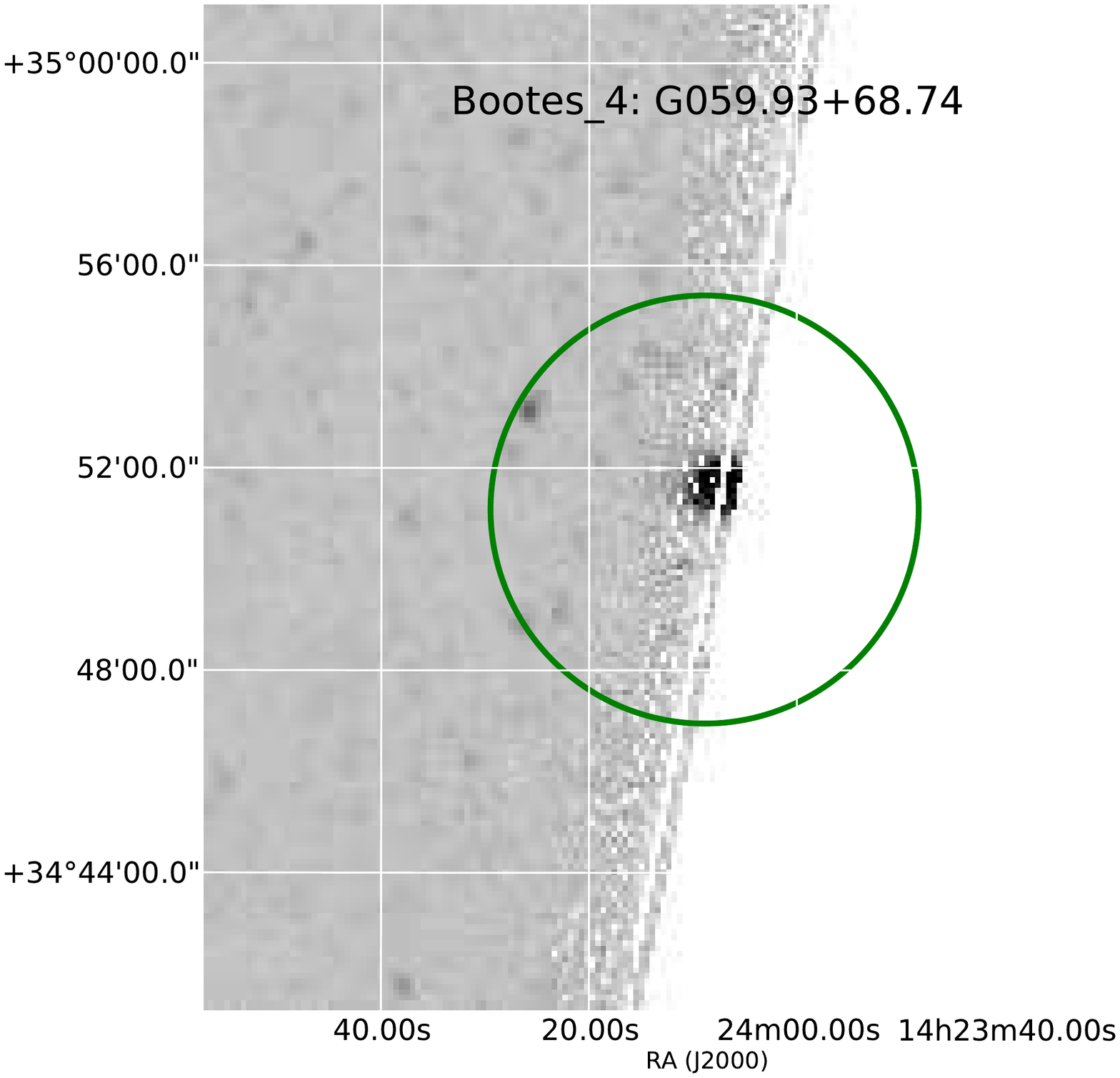}&
\includegraphics[width=4cm]{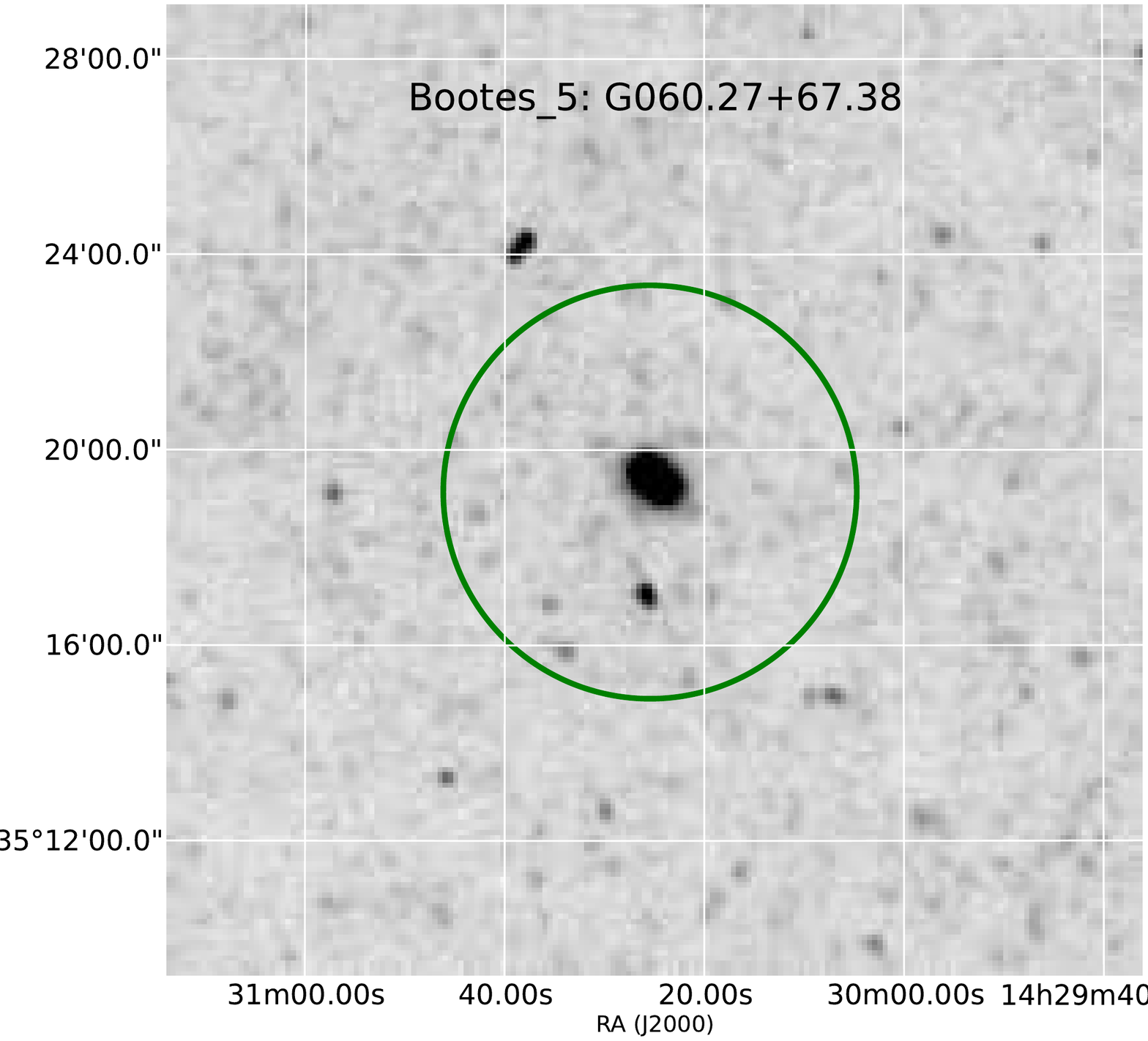}&
\includegraphics[width=4cm]{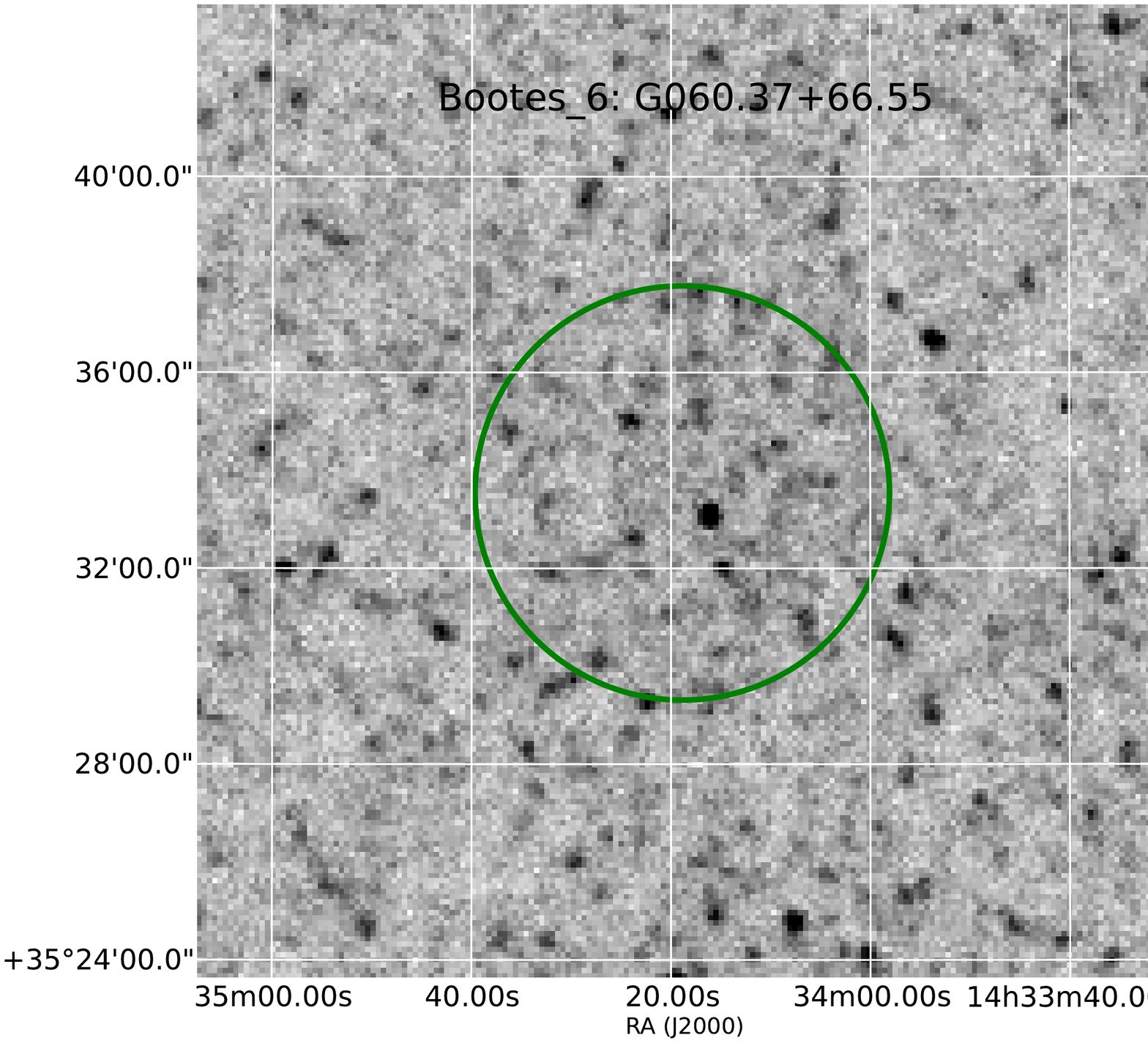}\\
\includegraphics[width=4cm]{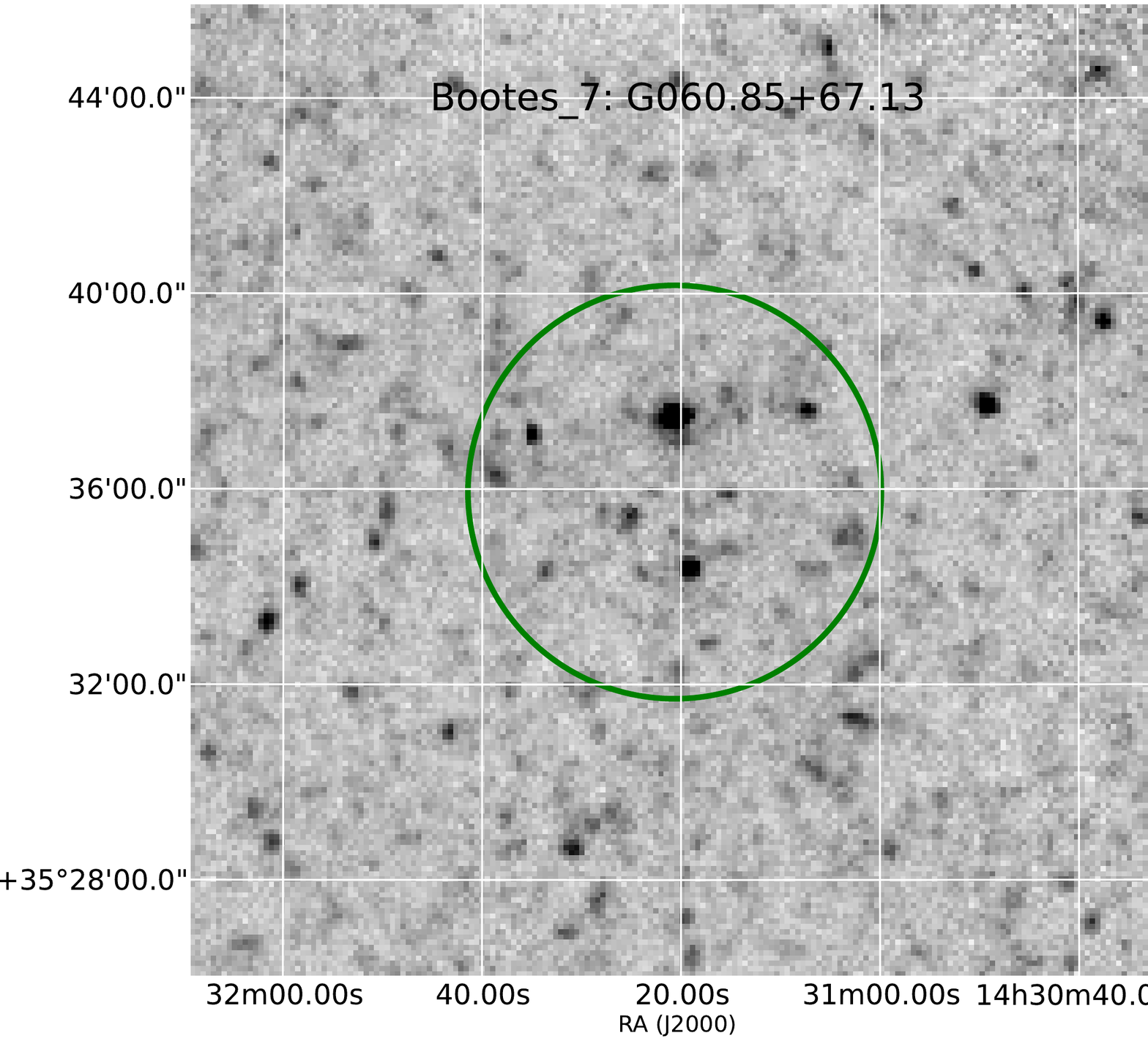}&
\includegraphics[width=4cm]{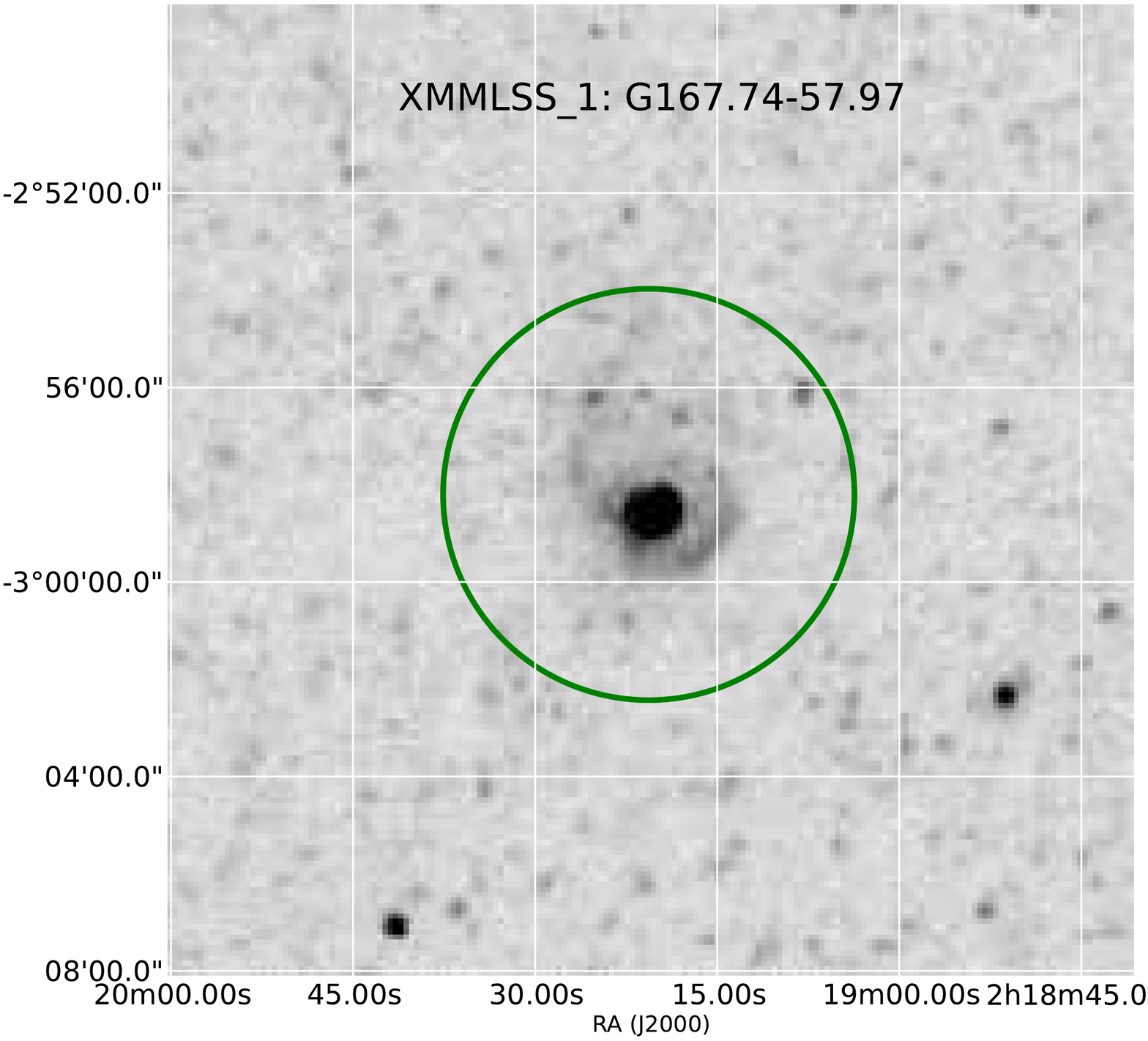}&
\includegraphics[width=4cm]{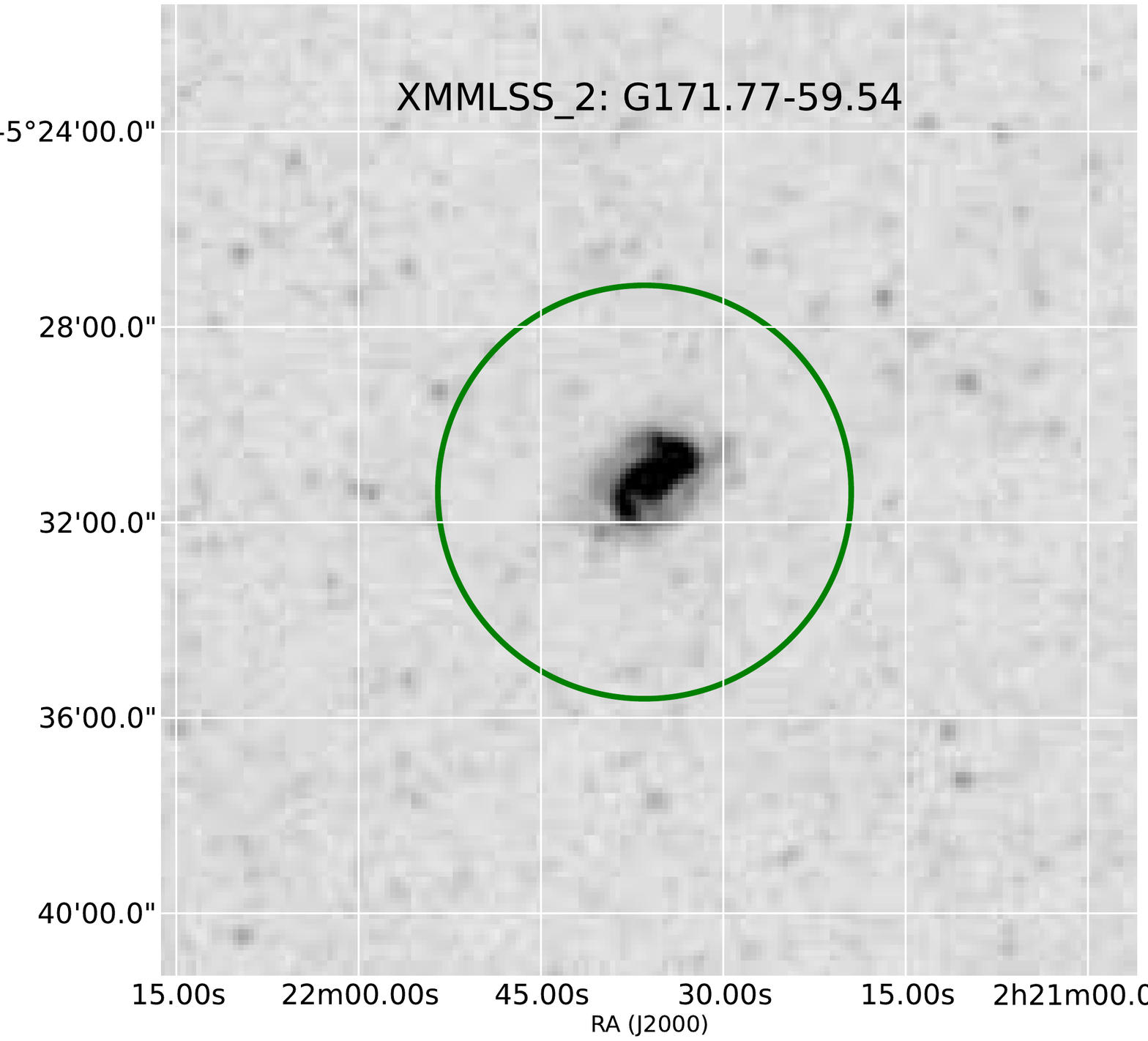}\\
\includegraphics[width=4cm]{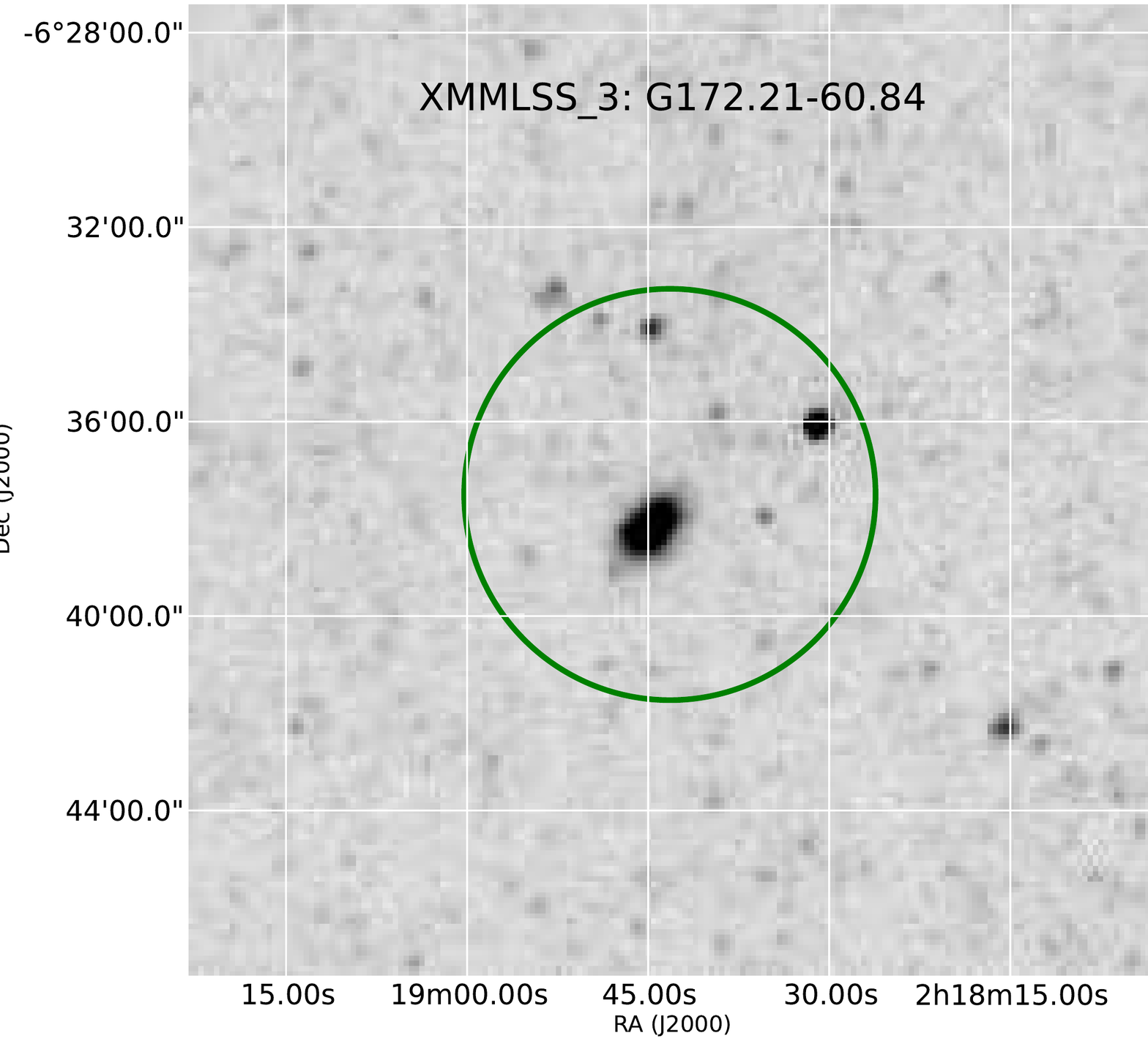}&
\includegraphics[width=4cm]{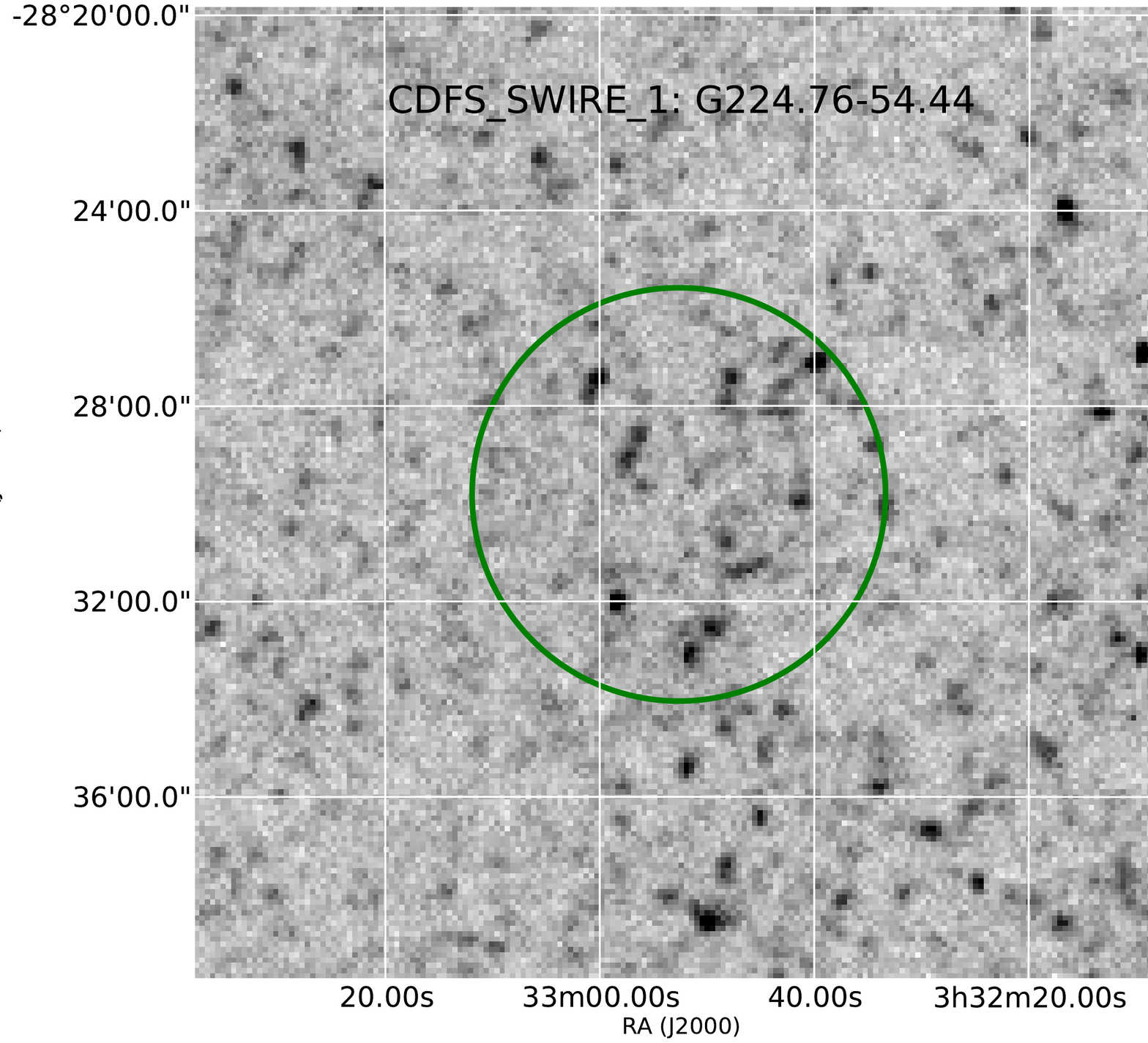}&
\includegraphics[width=4cm]{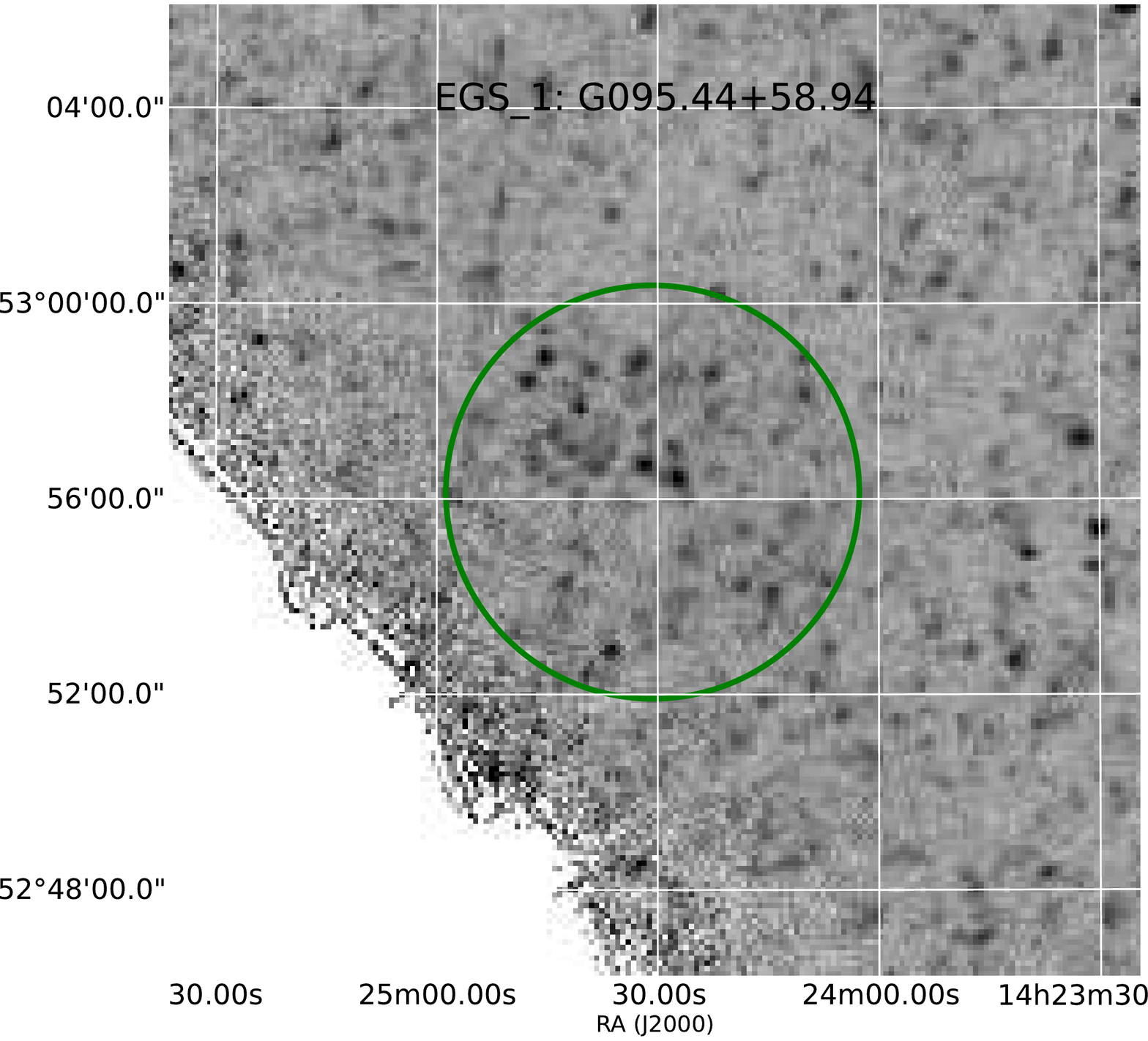}\\
\end{tabular}
\caption{{\em Herschel} SPIRE 250$\mu$m images of the {\em Planck} detected sources. Each image is 10 arc min across. The circle is centred on the position of the {\em Planck} source and has a radius of 4.23 arcmin, the FWHM of the {\em Planck} beam at 857 GHz.}
\end{figure*}

\begin{figure*}
\begin{tabular}{ccc}
\includegraphics[width=4cm]{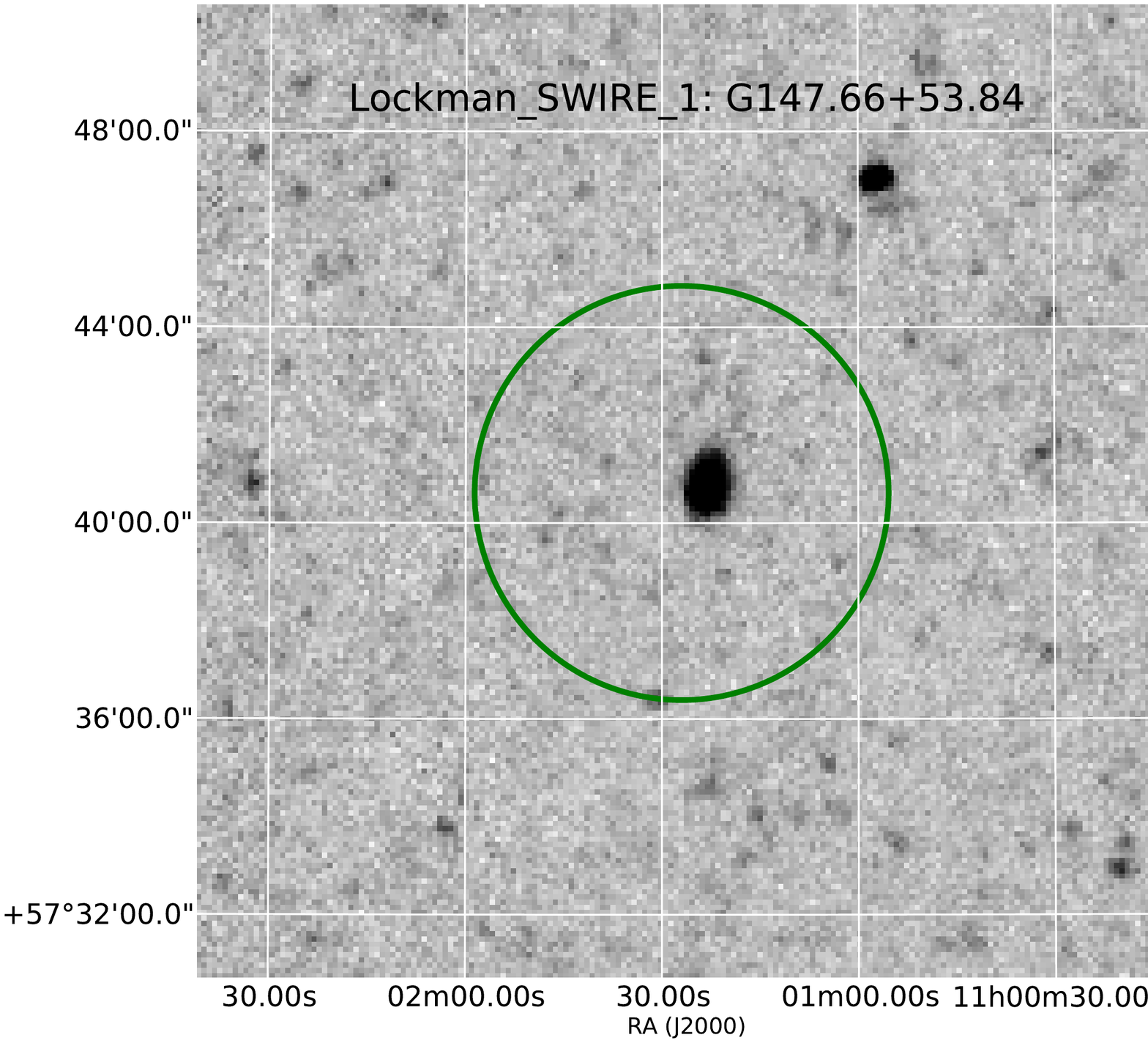}&
\includegraphics[width=4cm]{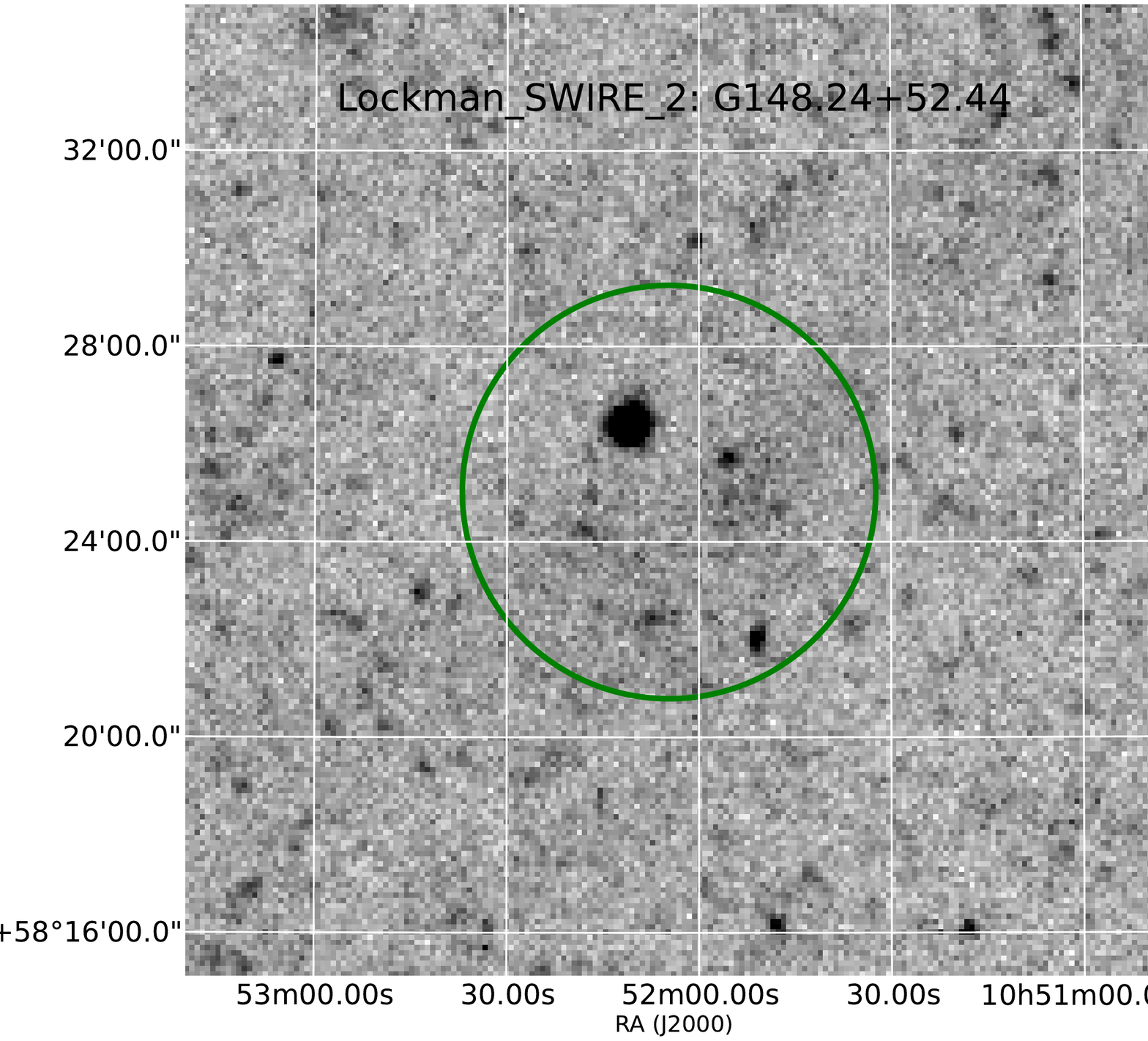}&
\includegraphics[width=4cm]{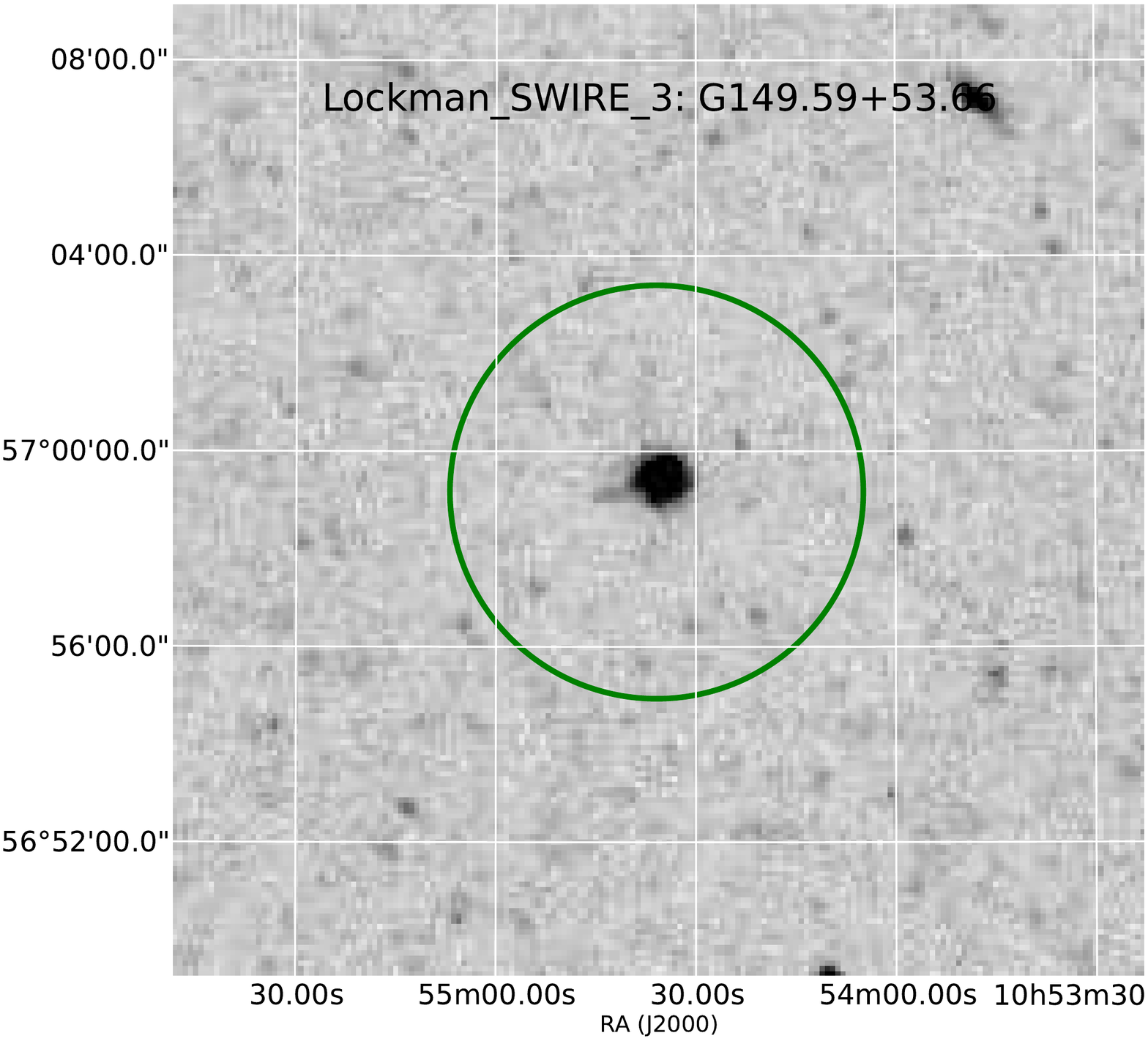}\\
\includegraphics[width=4cm]{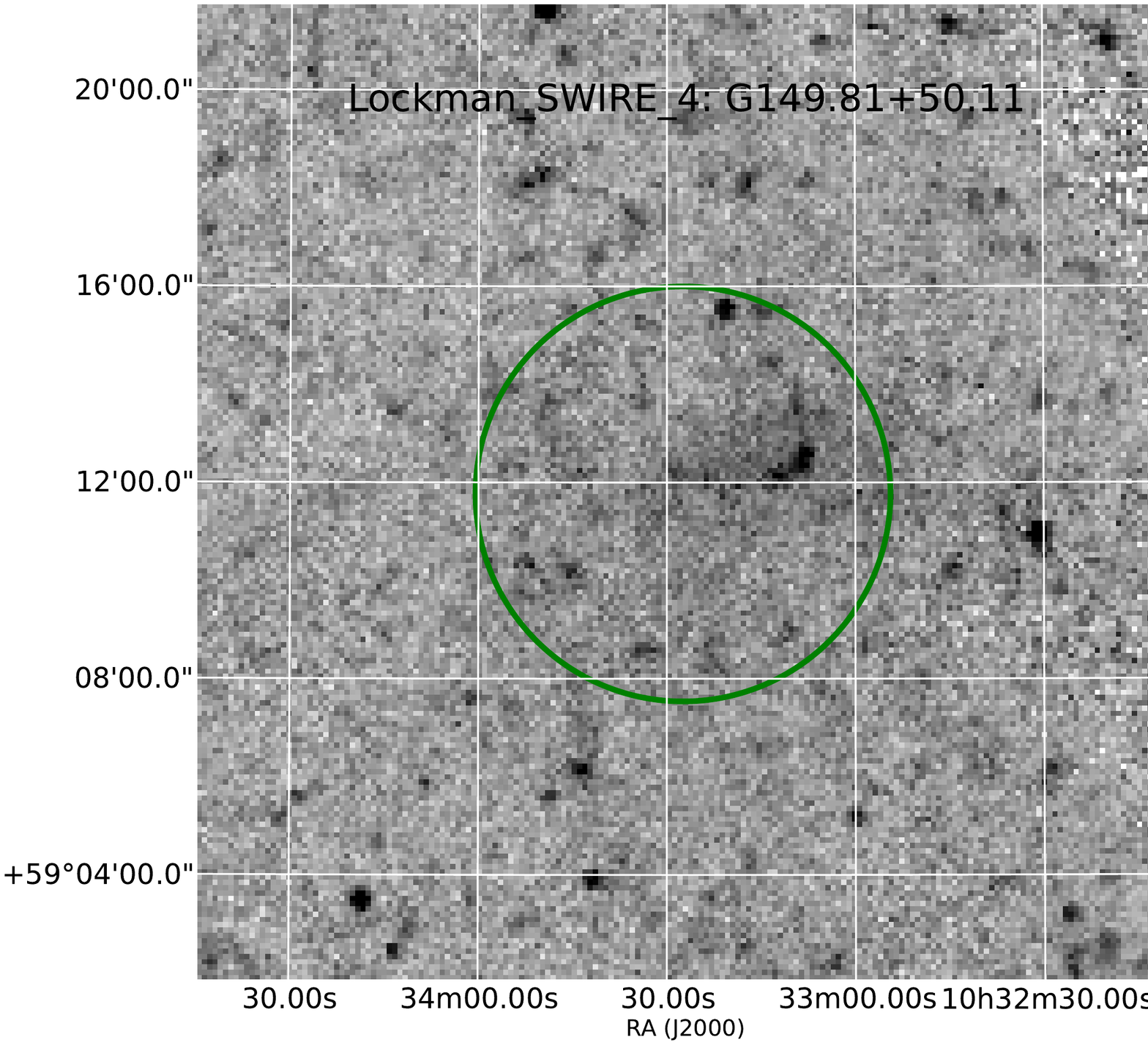}\\
\end{tabular}
\contcaption{}
\end{figure*}

%

\begin{figure*}
\begin{tabular}{cc}
\includegraphics[width=8cm]{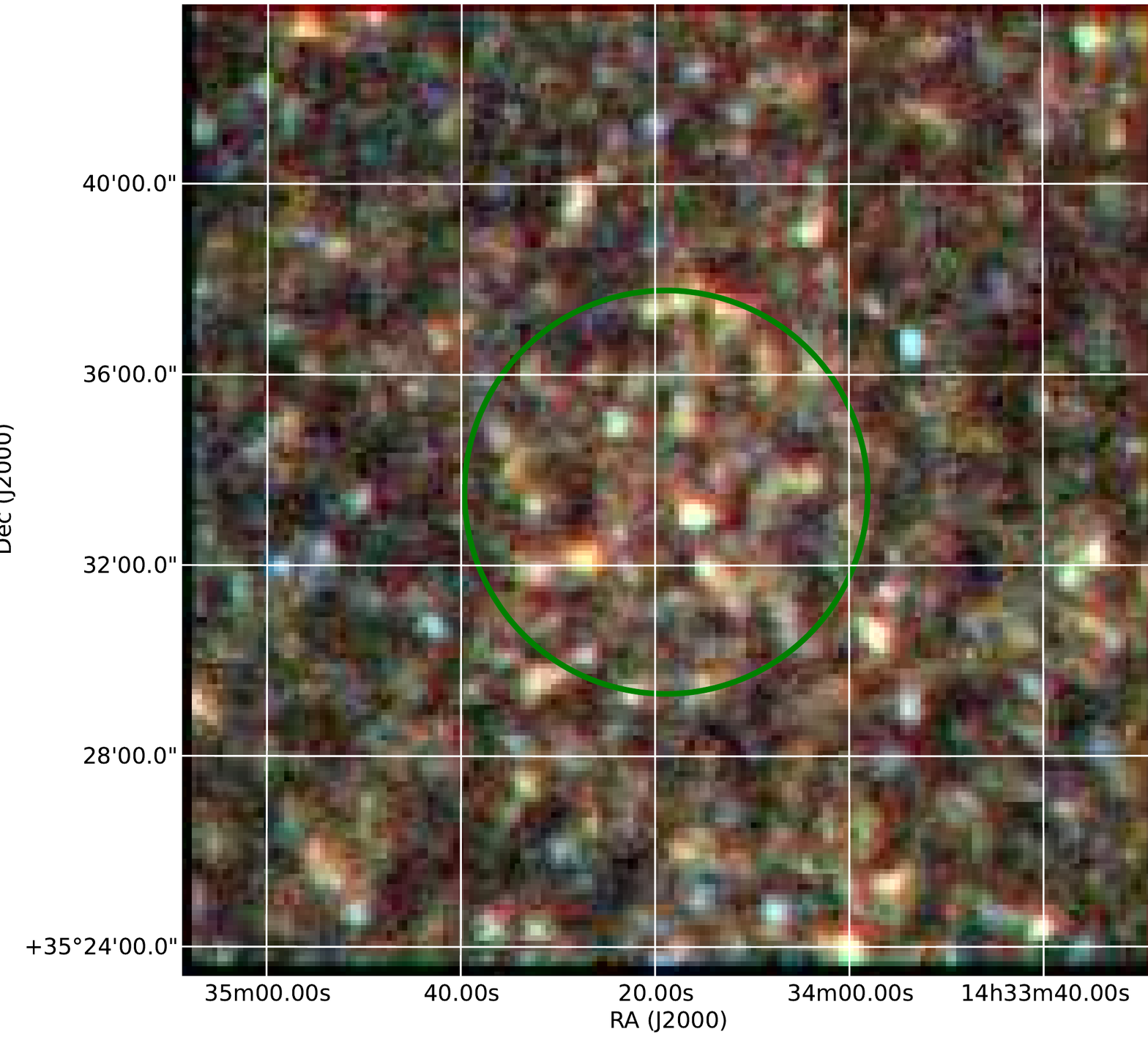}&
\includegraphics[width=8.25cm]{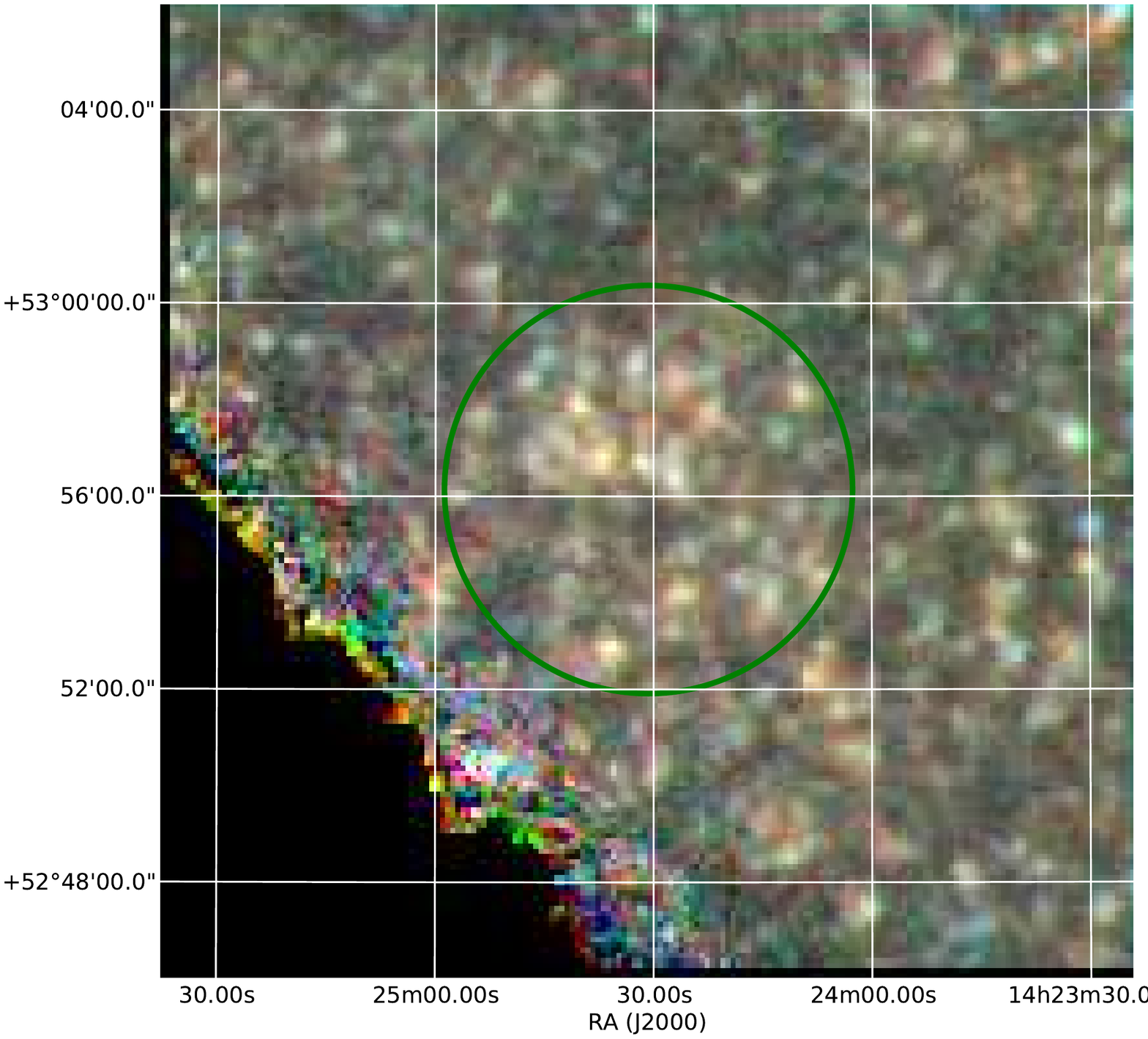}\\
\includegraphics[width=8cm]{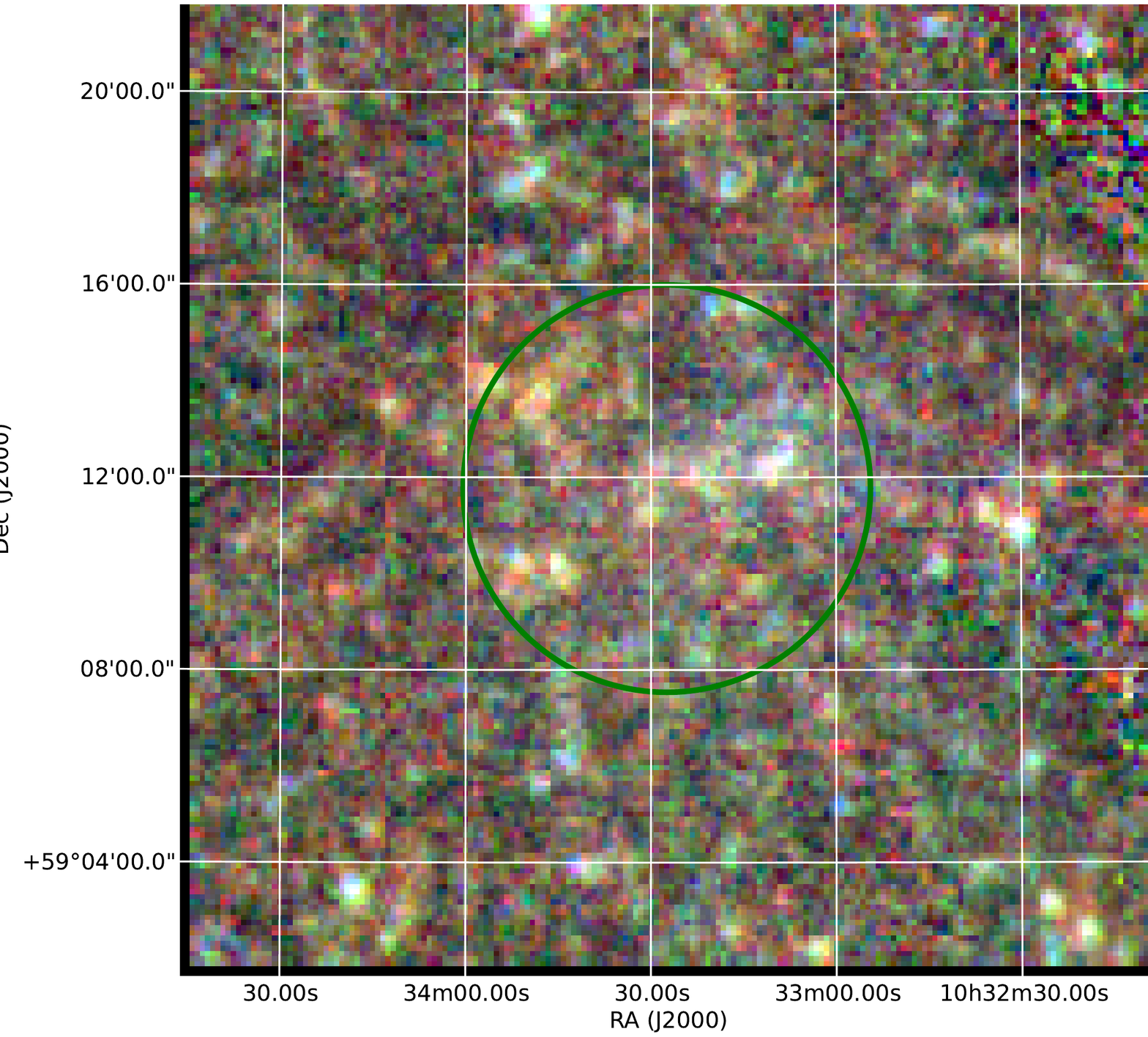}&
\includegraphics[width=8cm]{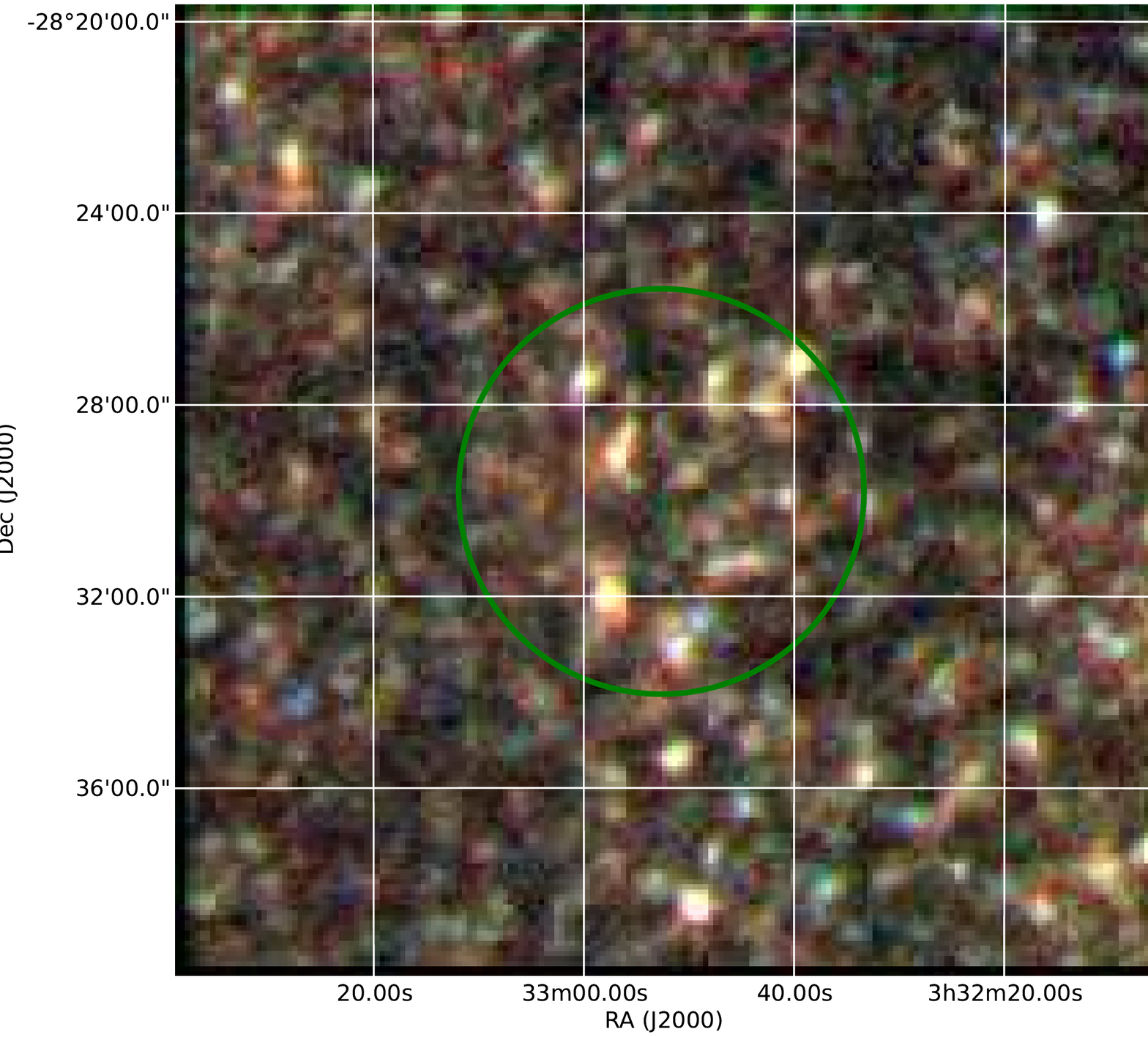}\\
\end{tabular}
\caption{Three colour {\em Herschel} images for {\em Planck} clumps. Blue = 250$\mu$m, Green=350$\mu$m, Red=500$\mu$m. Fields are Bo\"{o}tes, EGS, Lockman and CDFS clockwise from the top left. The green circle indicates the size of the {\em Planck} beam at the position of the {\em Planck} ERCSC source.}
\end{figure*}

\begin{figure}
\includegraphics[width=8cm, angle=180]{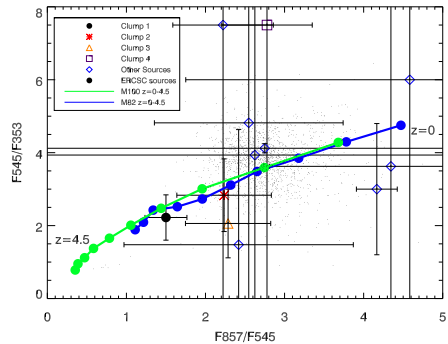}
\caption{{\em Planck} Colours for {\em Planck} Clumps. {\em Planck} colours $F_{857}/F_{545} vs. F_{545}/F_{353}$ for the {\em Planck} clumps shown for each clump, compared to the {\em Planck} colours for generic {\em Planck} ERCSC sources (shown as small black dots) and the sources in the HerMES fields not identified as clumps (shown as blue diamonds, excluding the three sources which are undetected in one or more of the {\em Planck} bands). ERCSC source colours are only plotted for sources detected at $>$5$\sigma$ in the 353 GHz band to ensure that error bars on the colours are small, and for sources with $|b| > 20$ to eliminate sources in the galactic plane. These colours are compared to the {\em Planck} colour tracks as a function of redshift for SED models of a starbursting (M82) and a quiescent galaxy (M100). Colour tracks start at z=0 and extend to z=4.5. Tick marks along the tracks are at intervals of 0.5 in redshift. This figure is discussed further in Section 3.1}
\end{figure}

\begin{figure*}
\begin{tabular}{cc}
\includegraphics[width=7cm]{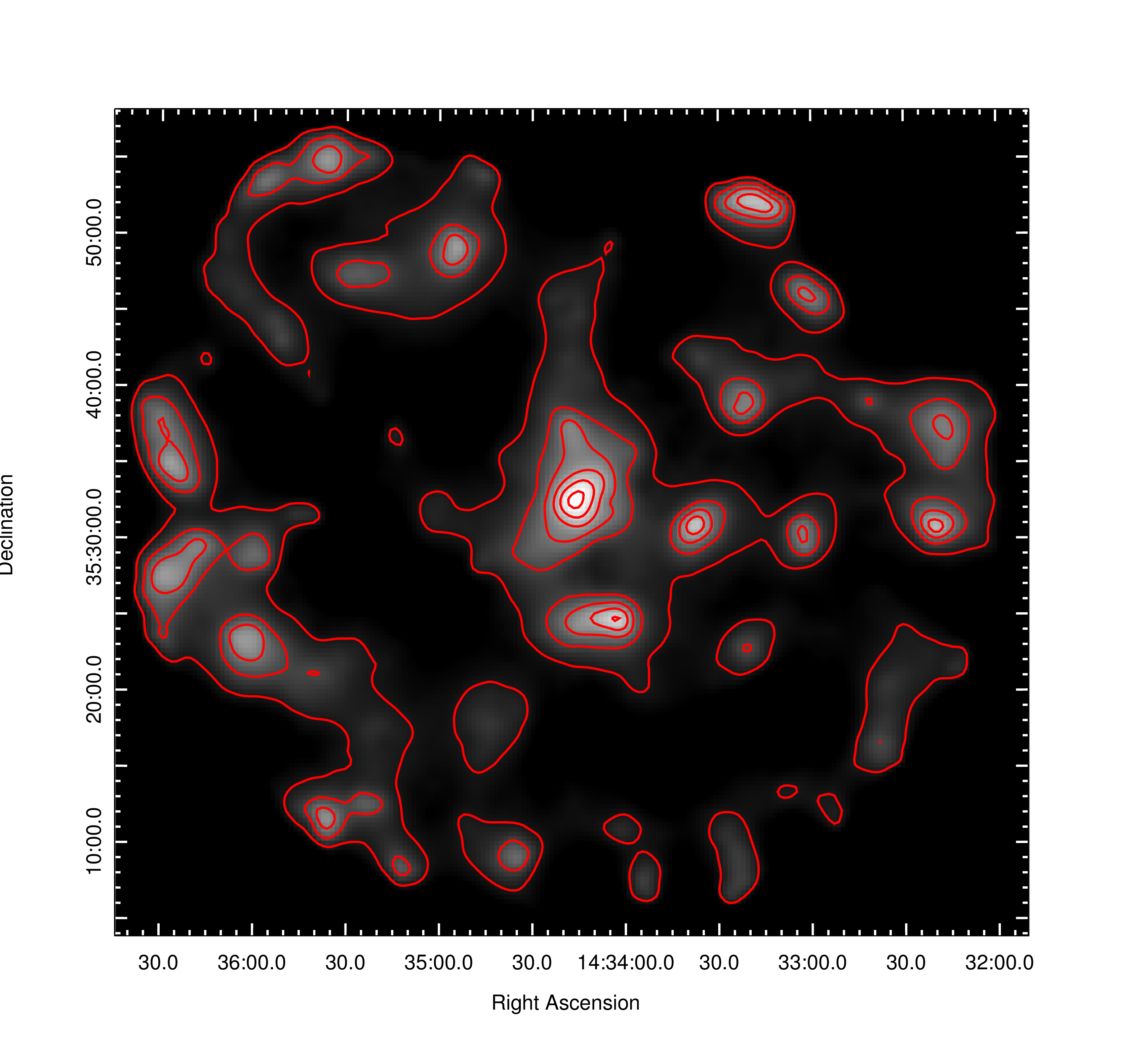}&
\includegraphics[width=7cm]{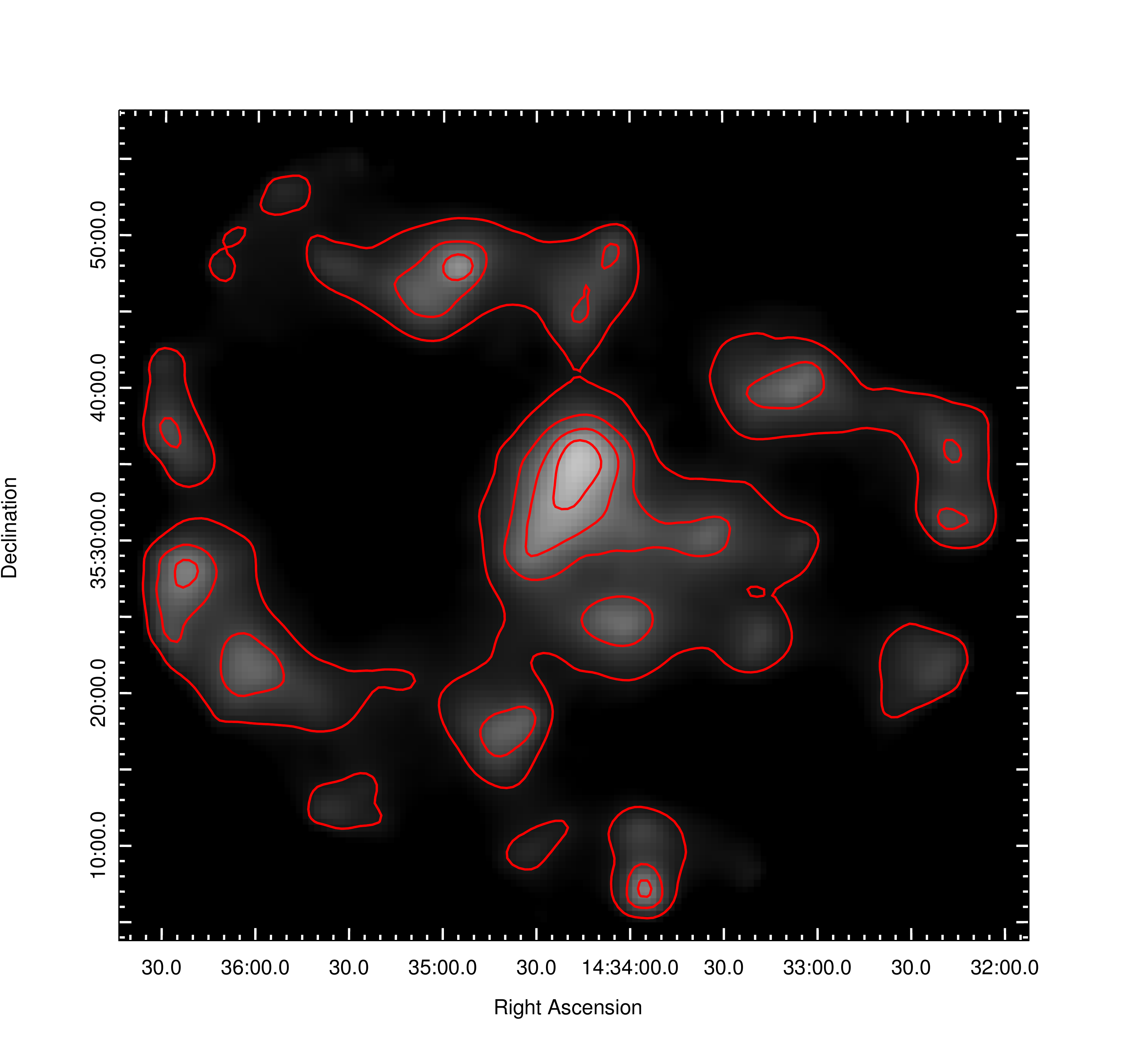}\\
\includegraphics[width=7cm]{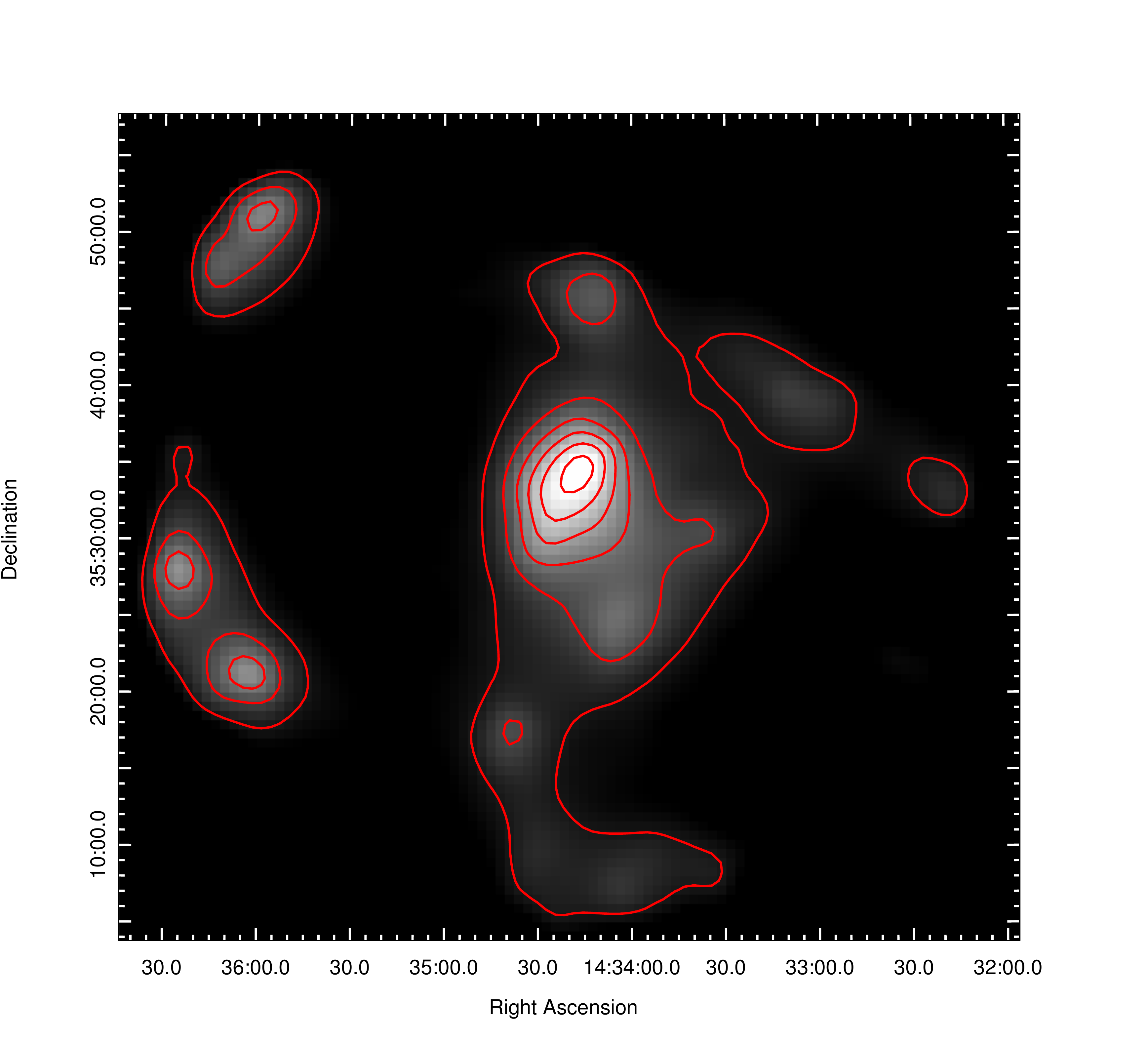}&
\includegraphics[width=7cm]{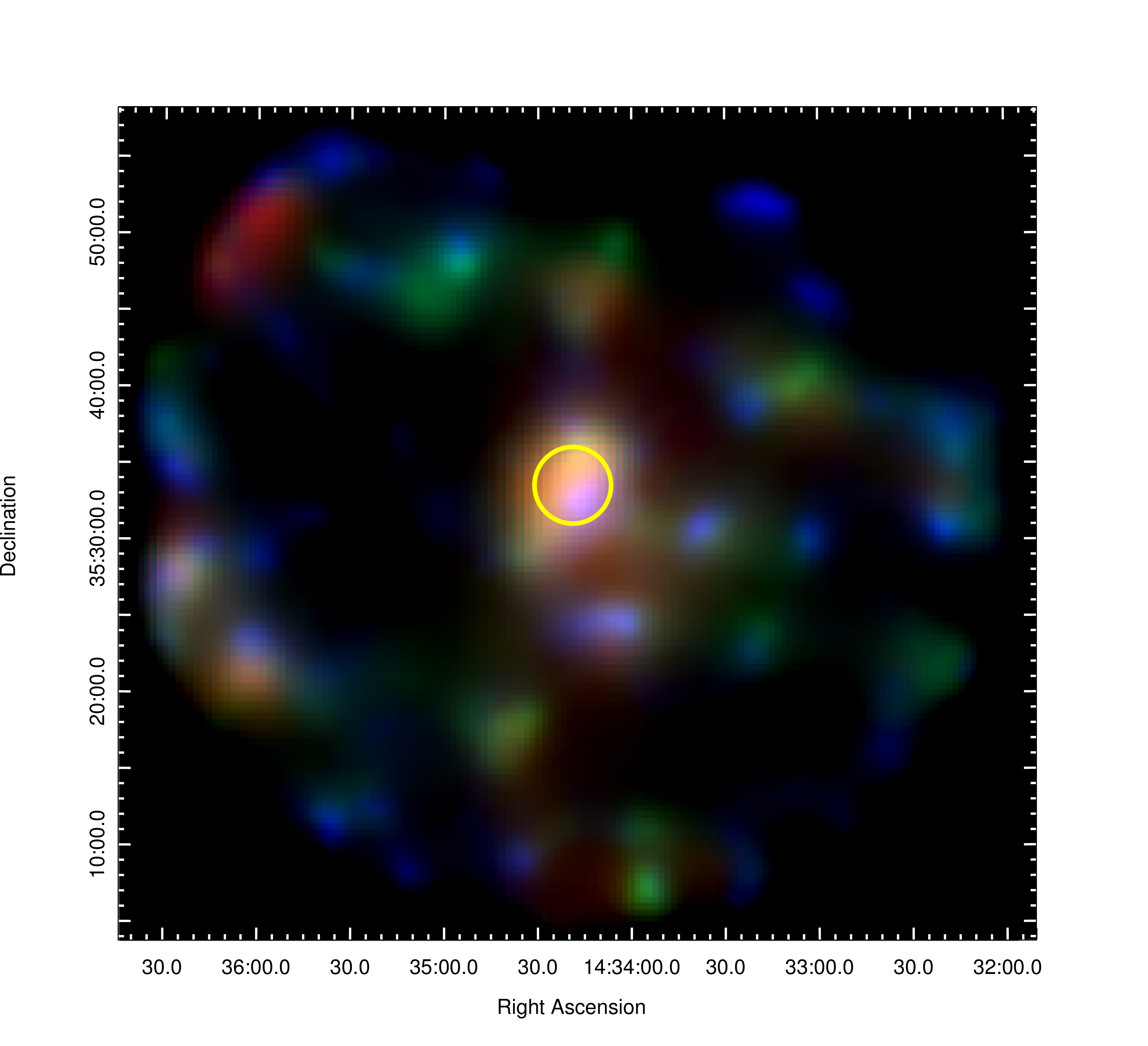}\\
\end{tabular}
\caption{Adaptively smoothed SPIRE source density map for the Bo\"{o}tes field covering the region around each {\em Planck} source at 250, 350 and 500 $\mu$m, and a colour rendition of the over density with blue representing 250, green 350 and red 500$\mu$ source overdensity. For each clump these are placed top left, top right, lower left and lower right respectively. The position of the Planck source position is indicated by a yellow circle representing the 4.23 arcmin {\em Planck} beam in the colour rendition. Contours are from 3 to 8 $\sigma$ in intervals of 1$\sigma$.}
\end{figure*}

\begin{figure*}
\begin{tabular}{cc}
\includegraphics[width=7cm]{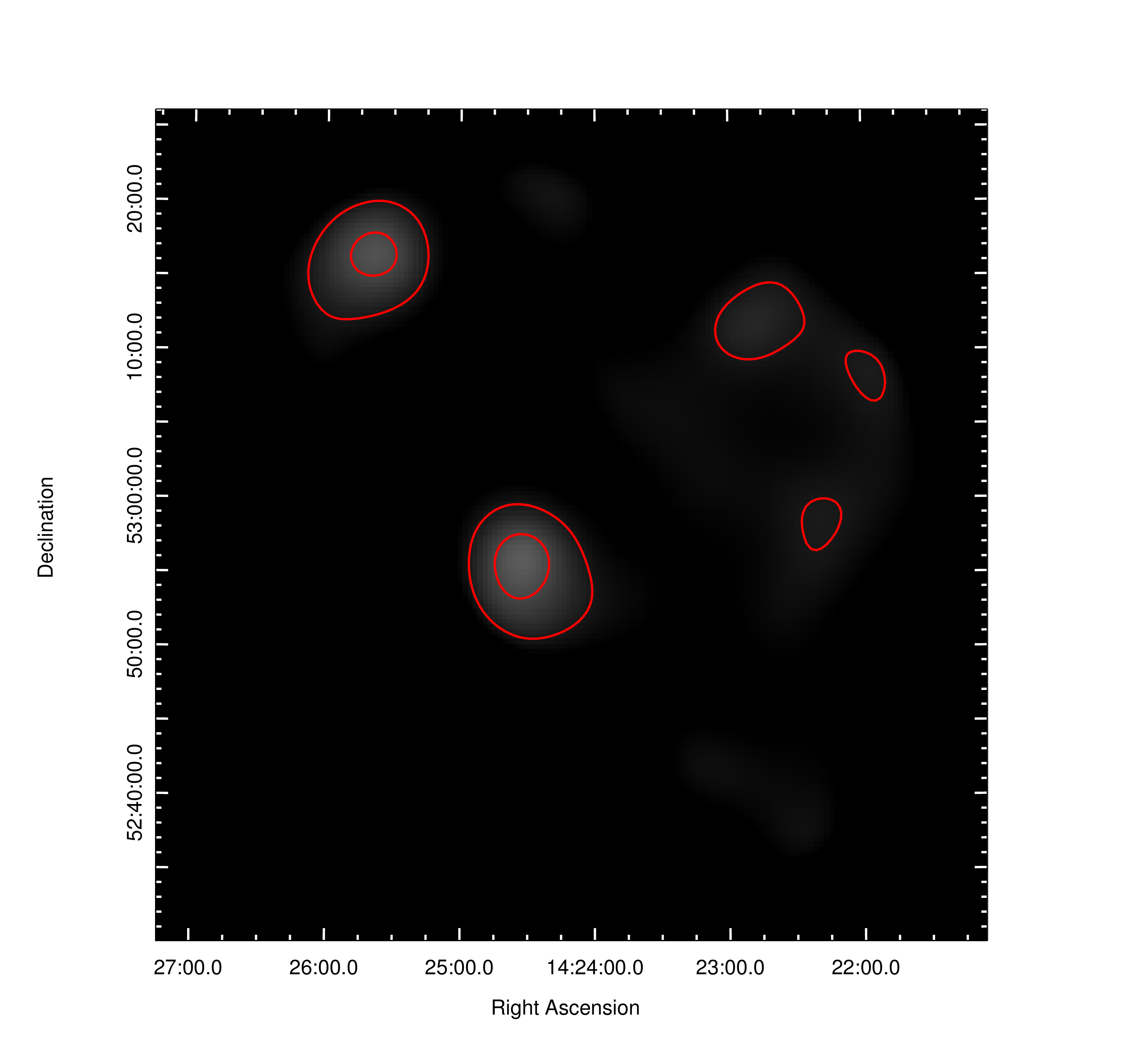}&
\includegraphics[width=7cm]{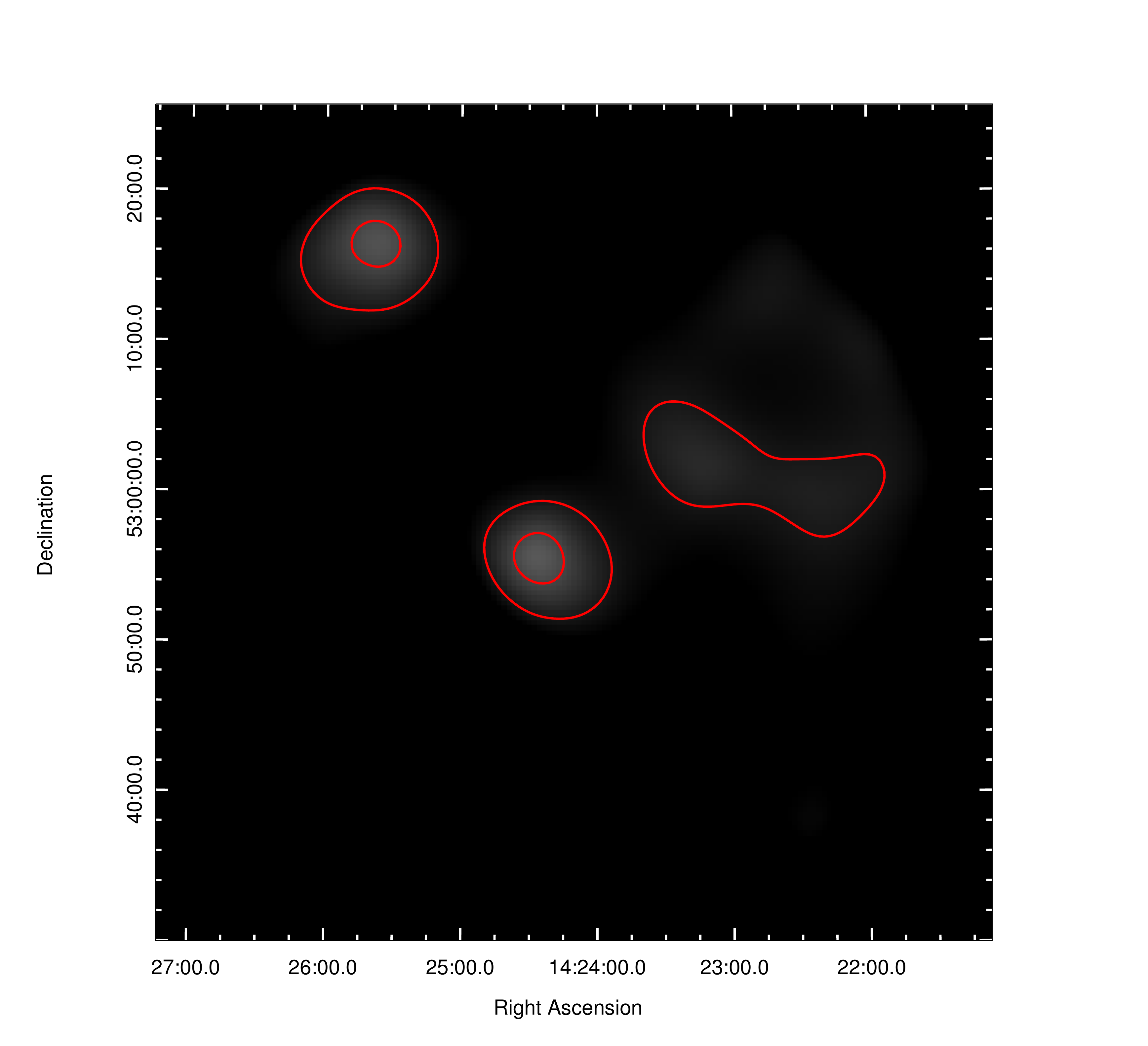}\\
\includegraphics[width=7cm]{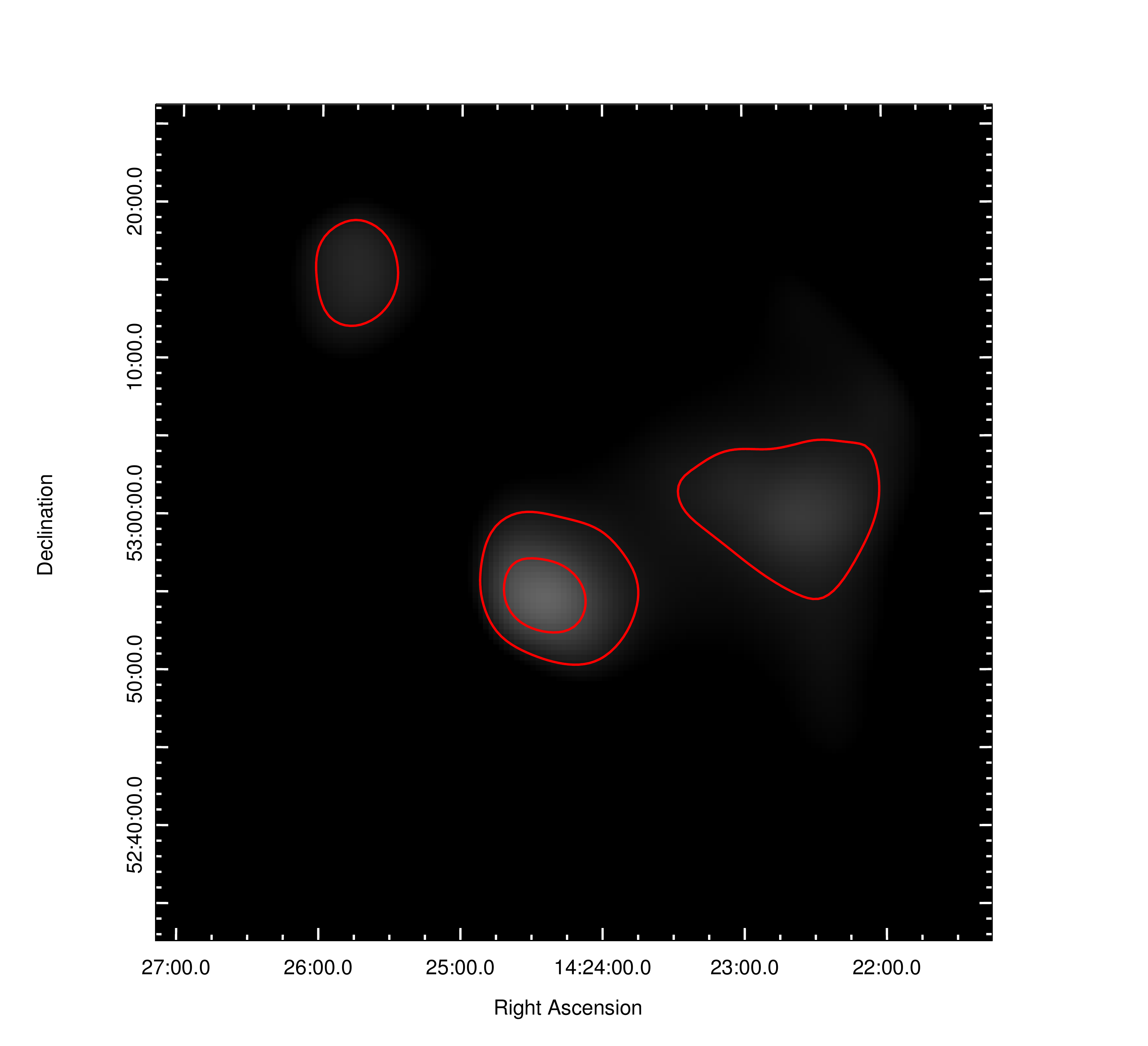}&
\includegraphics[width=7cm]{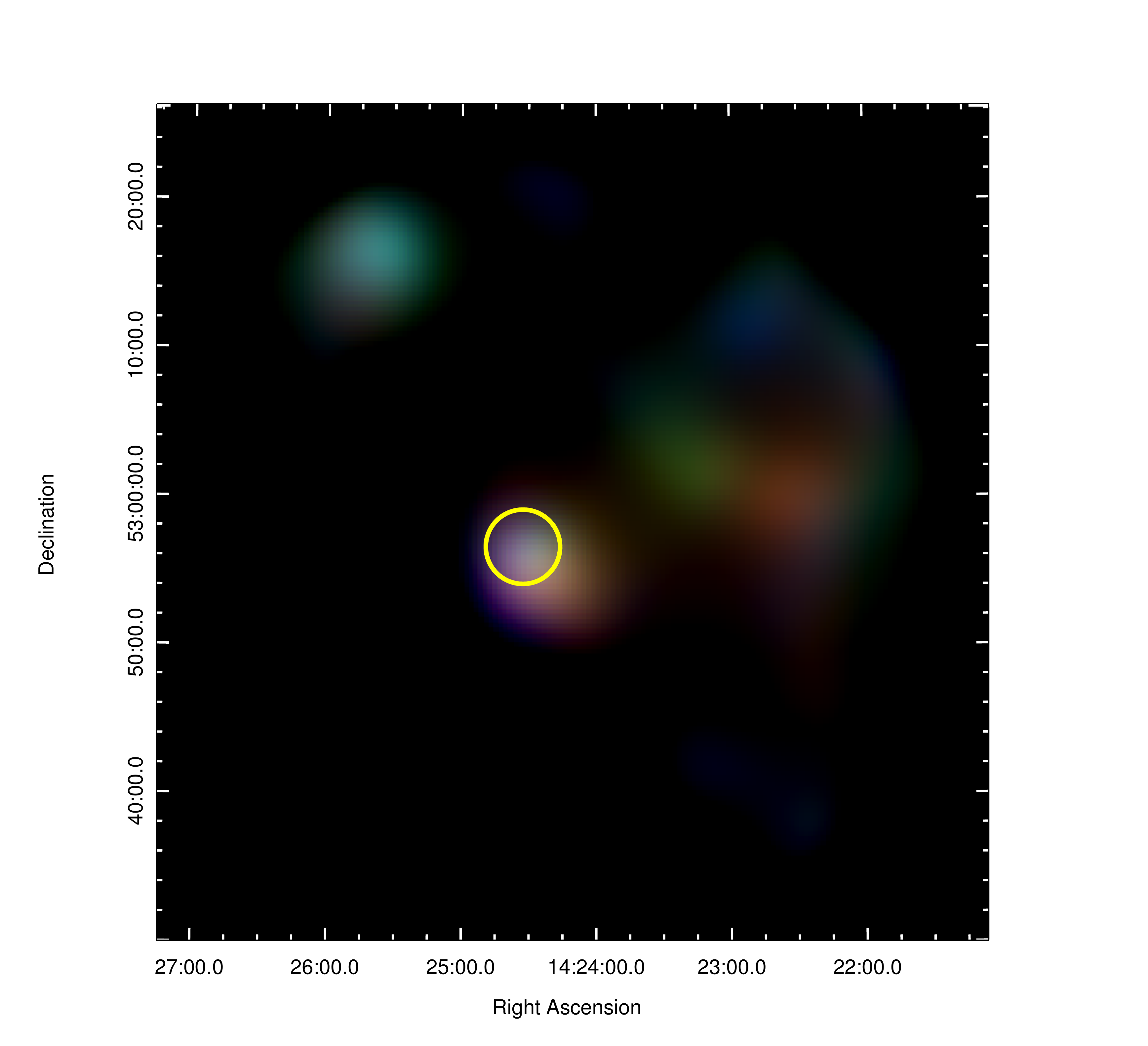}\\
\end{tabular}
\contcaption{As above for the EGS field.}
\end{figure*}

\begin{figure*}
\begin{tabular}{cc}
\includegraphics[width=7cm]{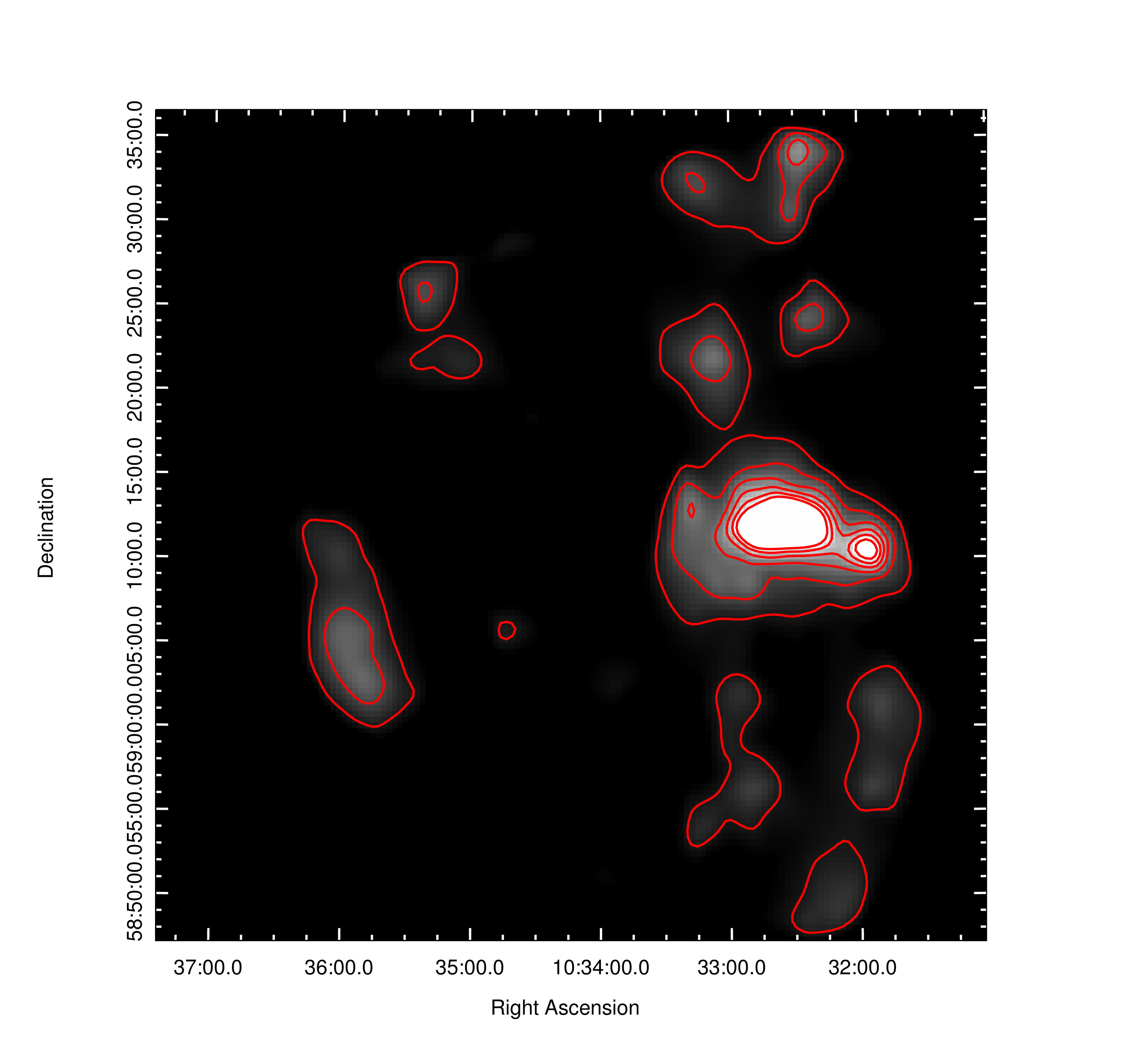}&
\includegraphics[width=7cm]{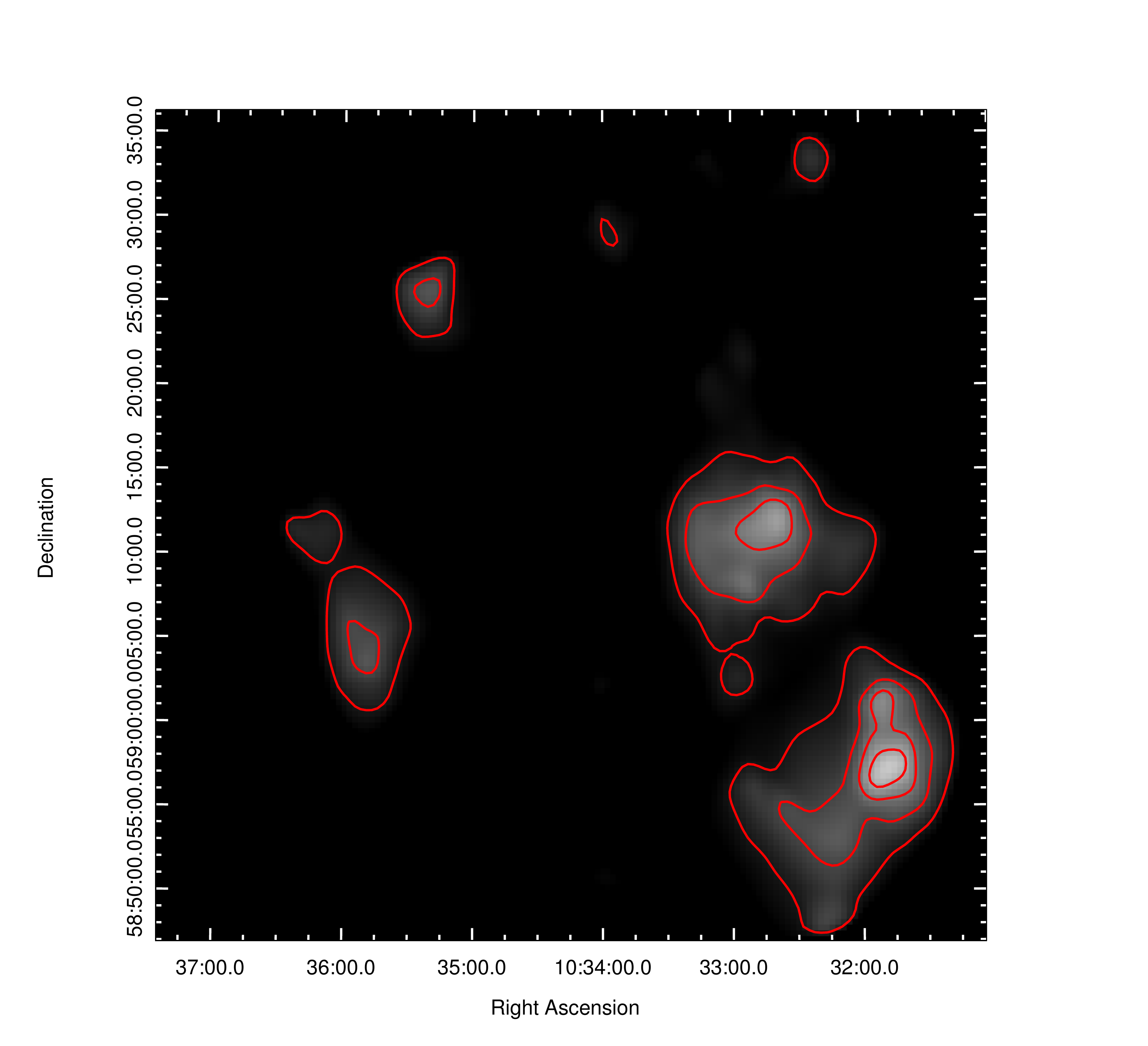}\\
\includegraphics[width=7cm]{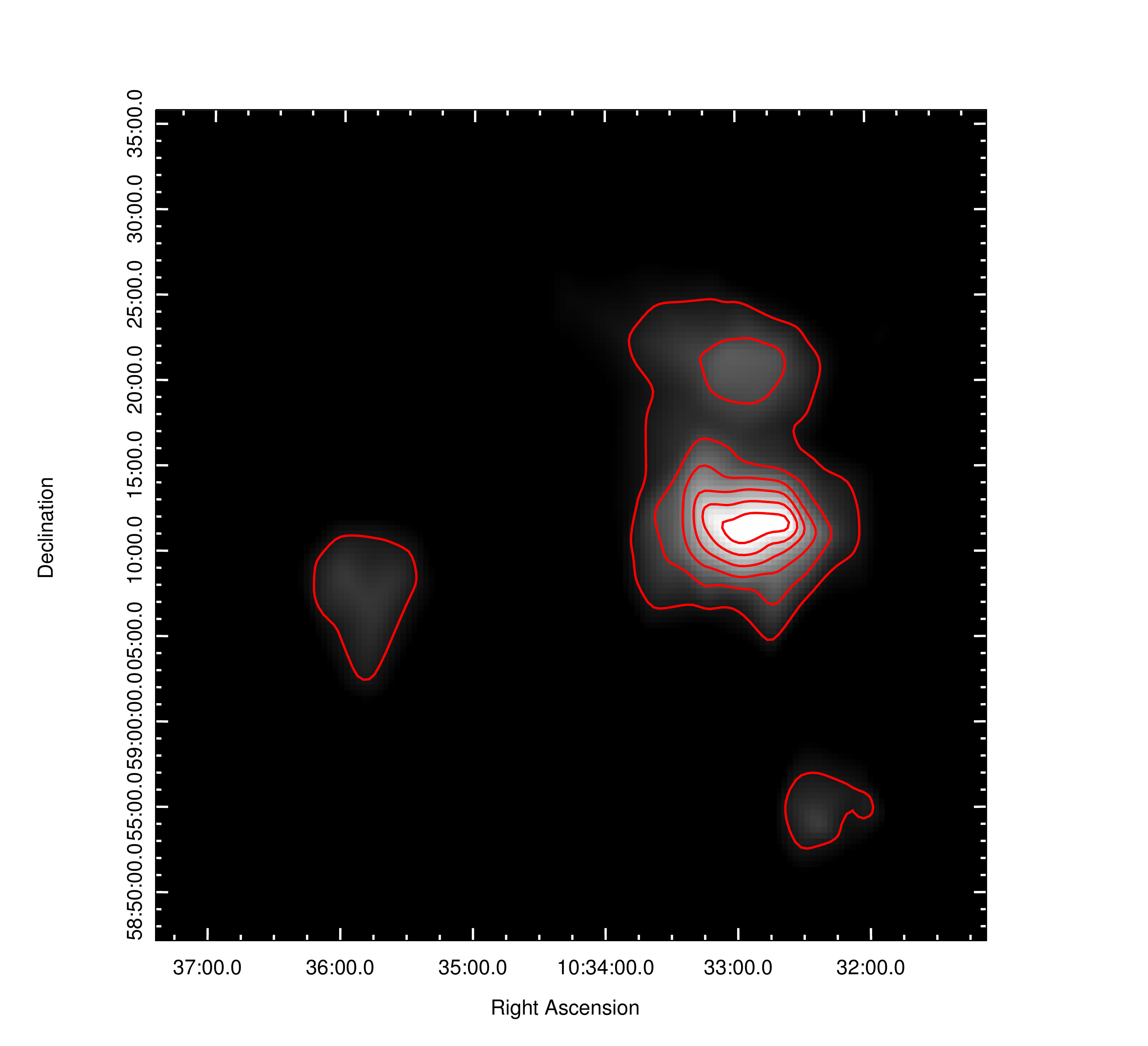}&
\includegraphics[width=7cm]{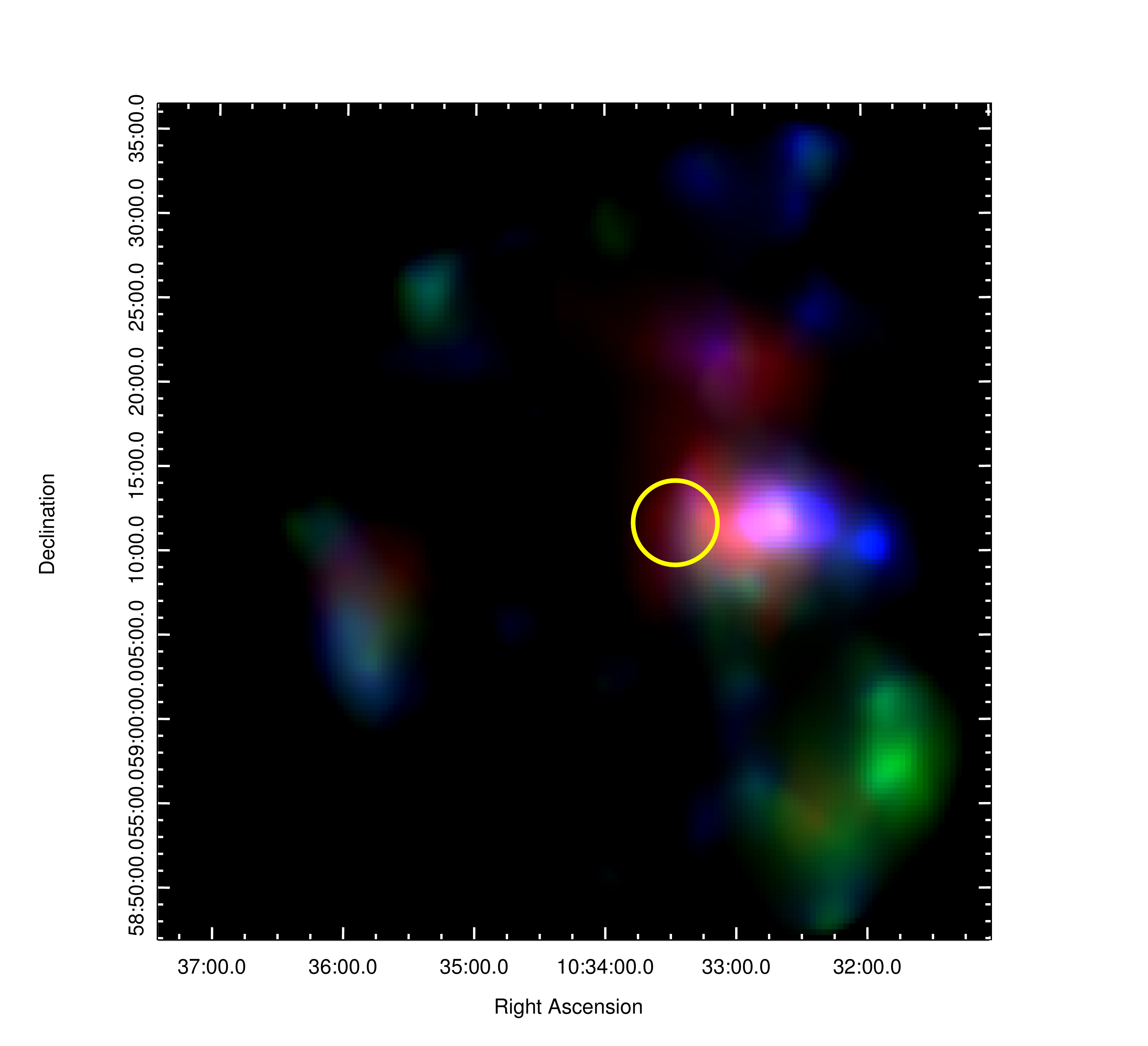}\\
\end{tabular}
\contcaption{As above for the Lockman field.}
\end{figure*}

\begin{figure*}
\begin{tabular}{cc}
\includegraphics[width=7cm]{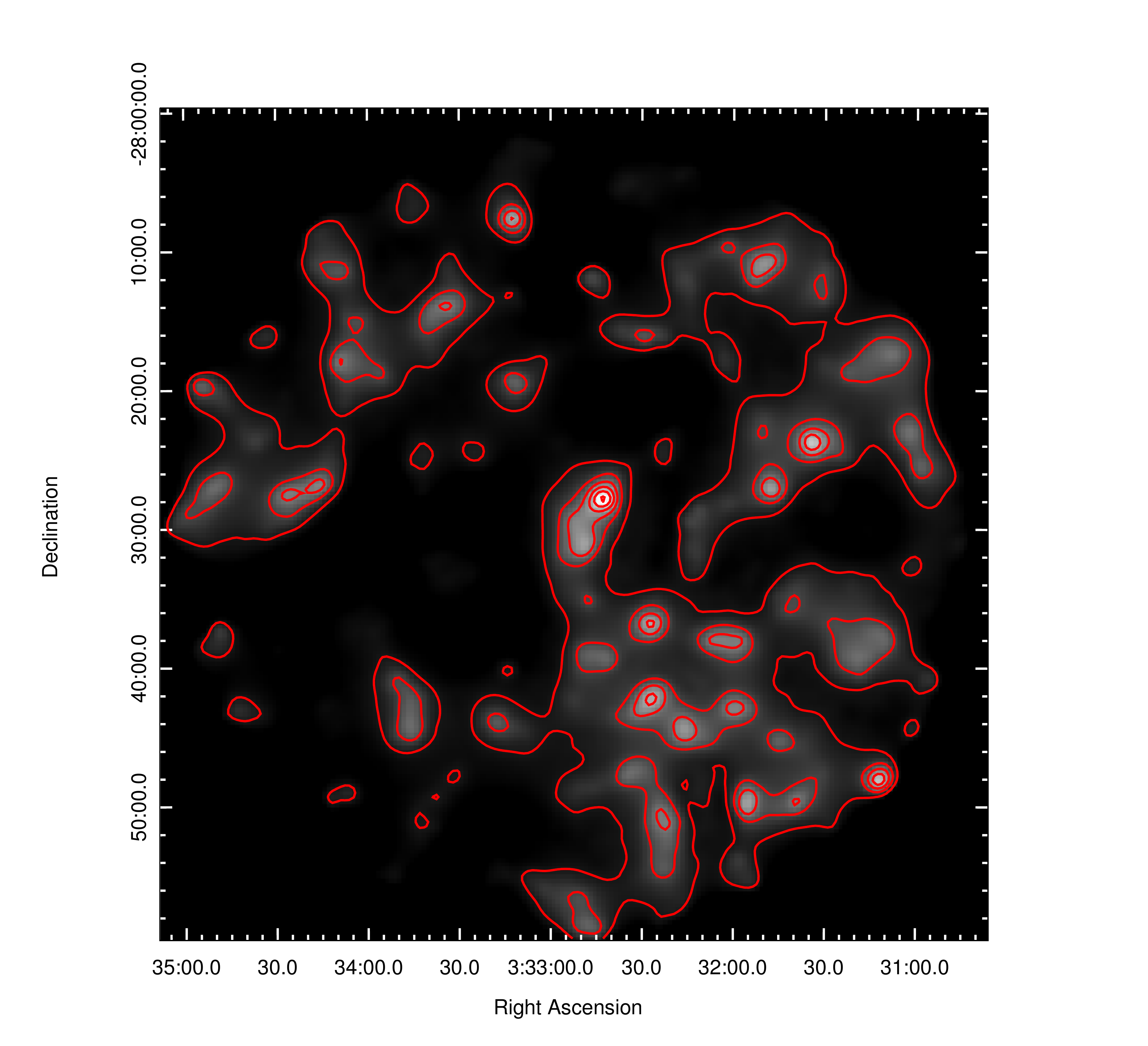}&
\includegraphics[width=7cm]{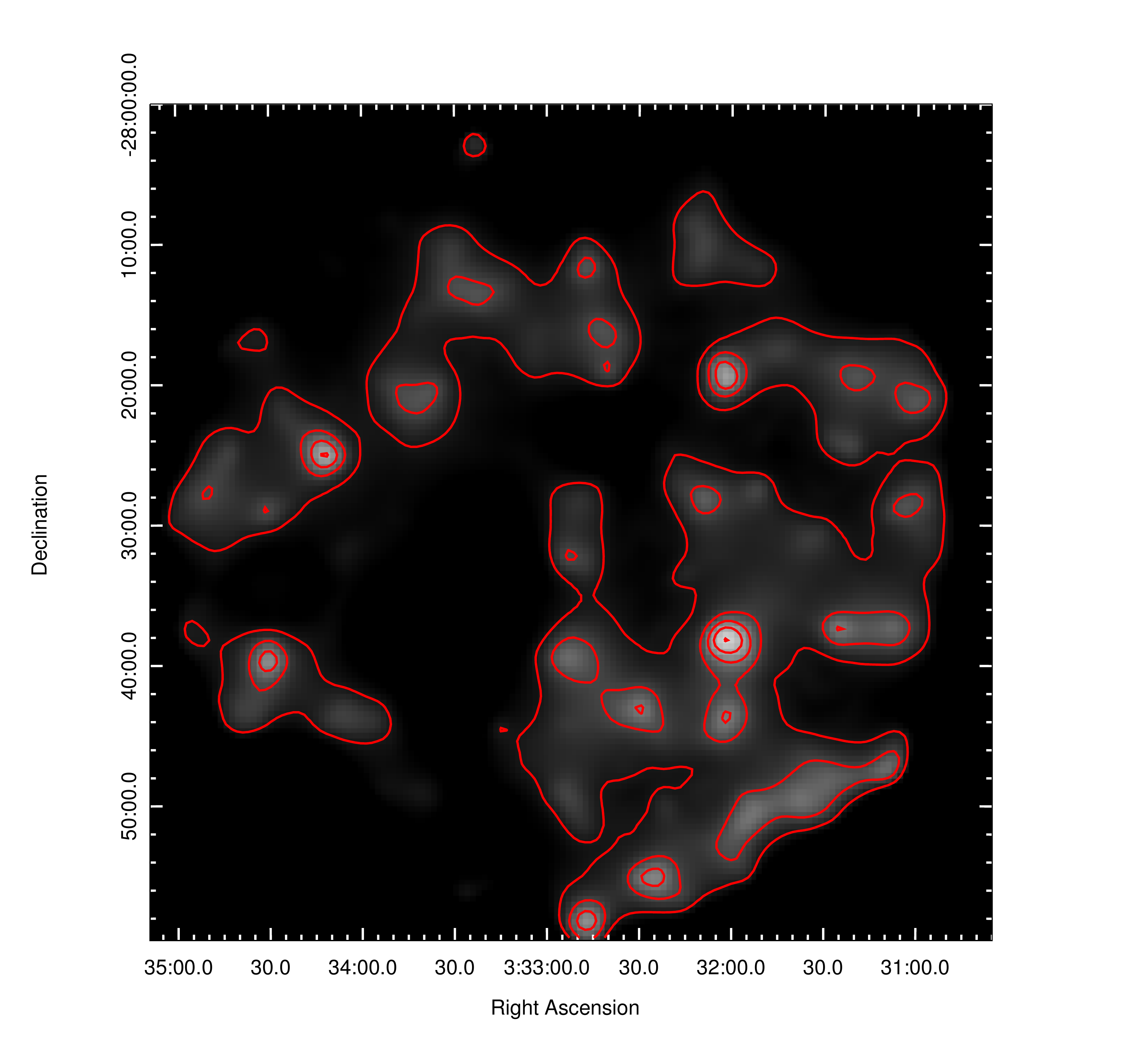}\\
\includegraphics[width=7cm]{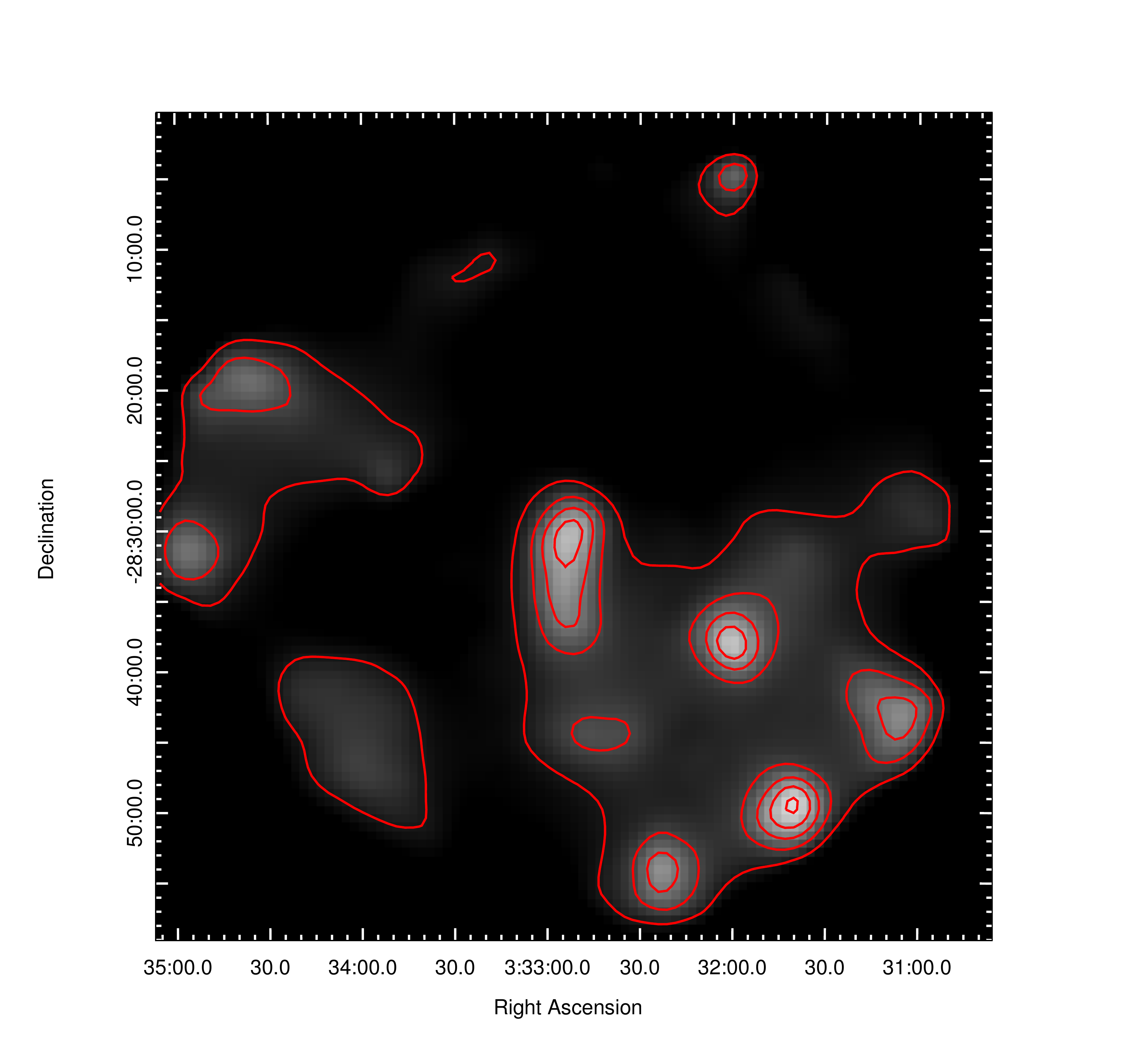}&
\includegraphics[width=7cm]{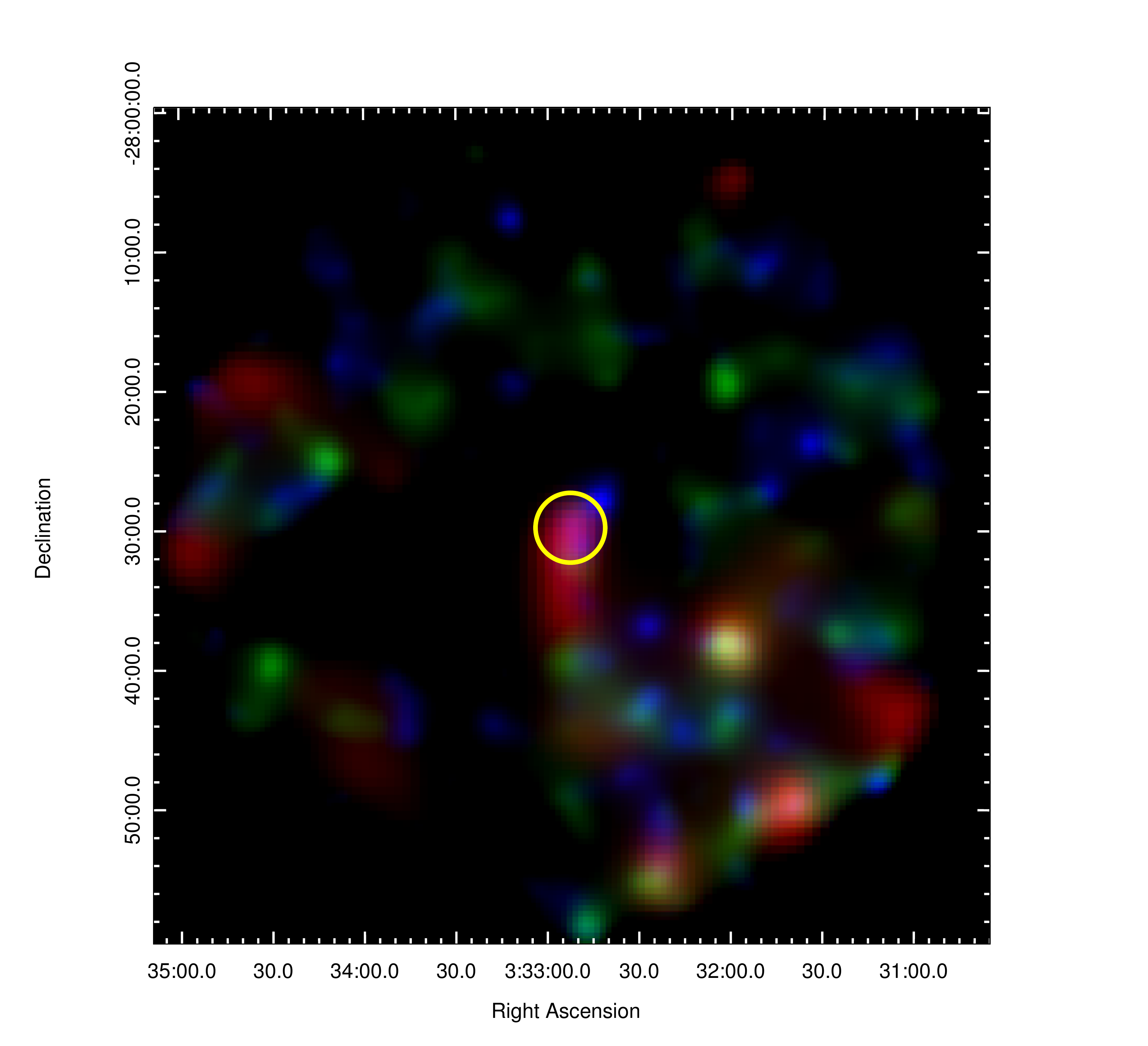}\\
\end{tabular}
\contcaption{As above for the CDFS field.}
\end{figure*}

\begin{figure}
\includegraphics[width=8cm,angle=180]{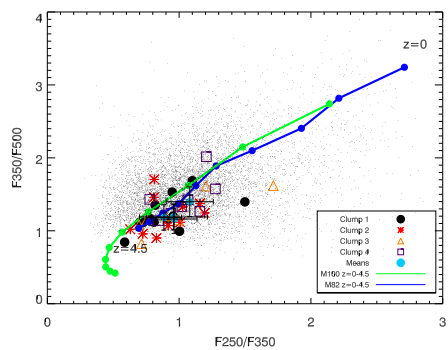}
\caption{SPIRE Colours for individual {\em Herschel} sources associated with the {\em Planck} Clumps, and for generic HerMES sources. Only sources detected with a S/N$>$2 in all three SPIRE bands are plotted. Also plotted are the mean colours for each clump. The small black dots are the SPIRE colours for all HerMES sources in the ECDFS field with S/N$>$2 in all three bands. These are compared to the SPIRE colour tracks as a function of redshift for SED models of a starbursting (M82) and a quiescent galaxy (M100) from z=0 to z=4.5 with tick marks at intervals of 0.5 in redshift, as in Fig. 3.}
\end{figure}

\section{Complementary Optical and Near-IR Data}

Our analysis of the {\em Planck} and {\em Herschel} properties of the clump sources is consistent with the idea that they are z$\geq$1 clusters of dusty galaxies, matching the general properties of the sources predicted by Negrello et al. (2005). {\em Herschel} and {\em Planck} data, however, can only go so far in revealing the nature of these objects. Confirmation that these are genuine galaxy clusters requires observational data at other wavelengths. Ideally this should include optical spectroscopy so that redshifts can be determined, but in the absence of that information we can still make significant progress in understanding these objects by examining the broadband optical and near-IR properties of sources associated with the positions of the {\em Planck} clumps. The HerMES survey was designed to take data in regions of the sky that are well studied at other wavelengths. The Bo\"{o}tes clump (Clump 1 in Table 2) lies within the NOAO NDWFS (Jannuzi \& Dey 1999; Dey et al. 2008), for which there is a wide variety of optical and near-IR data. The same is true of the CDF-S field, containing Clump 4 from Table 2, which is also complemented by Spitzer data from the SWIRE survey (Lonsdale et al., 2003). The EGS field, which contains Clump 2, has also been subject to deep multiwavelength observations by the AEGIS collaboration (Davis et al., 2007) while the Lockman field, containing Clump 3, was also observed by the SWIRE survey. However, the location of the clumps in these two latter fields is away from the areas where deep multiwavelength data is available, so new followup data for these objects has been obtained. In this section we describe the new and archival data for these fields.

\subsection{Archival data}


\subsubsection{Bo\"{o}tes}

The Bo\"{o}tes clump (Clump 1 in Table 2) lies within the area covered by the NOAO Deep-Wide Field Survey (NDWFS, DR3), which provides deep optical coverage in the {\em B$_W$RI} bands. This dataset is complemented by the $z$-Bo\"{o}tes survey of Cool (2007), which provides $z$-band coverage of the same field, and the Infrared Bo\"{o}tes Imaging Survey (IBIS; Gonzalez et al. in prep) in the JHK$_S$ near-IR bands. GALEX data are also available in both UV bands. {\it Spitzer}-IRAC 4-band photometry was derived from the {\it Spitzer} Deep-Wide Field Survey maps (SDWFS; Ashby et al. 2009), providing a catalogue of 3.6$\mu$m-detected sources (Vaccari et al., in prep.). Sources from all available catalogues were cross-matched using a search radius of 2.5$^{\prime\prime}$, providing an 11-band catalogue of 830131 sources in the whole NDWFS Bo\"{o}tes field. All photometric measurements refer to a 4$^{\prime\prime}$ aperture and were translated to the AB system. Table 4 shows the $5\sigma$ completeness magnitude and flux in each band.

\begin{table}
\caption{Magnitude and flux limits for ancillary data in the Bo\"{o}tes field. All values refer to $5\sigma$ detections in 4$^{\prime\prime}$ apertures in AB magnitudes.}
\begin{tabular}{cccc}
\hline
Band & Survey & $m_{comp}$ & $f_{comp}$ [$\mu$Jy] \\
\hline
B$_W$ & NDWFS & 24.4 & 0.6 \\
R &  NDWFS & 23.8 & 1.1 \\
I &  NDWFS & 23.2 & 1.9 \\
$z$ & $z$-Bo\"{o}tes & 22.0 & 5.7 \\
J & IBIS & 22.3 & 4.4 \\
H & IBIS & 21.7 & 7.6 \\
K$_S$ & IBIS & 21.4 & 10.0 \\
IRAC 3.6 & SDWFS & 22.4 & 4.0 \\
IRAC 4.5 & SDWFS & 22.0 & 5.8 \\
IRAC 5.8 & SDWFS & 21.1 & 13.2 \\
IRAC 8.0 & SDWFS & 21.0 & 14.5 \\
\hline
\end{tabular}
\end{table}


\subsubsection{CDF-S}

The CDF-S clump (Clump 4 in Table 2) lies within the field observed by the SWIRE survey, thus offering deep {\it Spitzer}-IRAC and MIPS photometry. Other ancillary data cover only part of the field, and include deep CTIO MOSAIC-II optical imaging in the Ugriz bands (Siana et al., private communication) and GALEX. Data were combined with a search radius of 2.5$^{\prime\prime}$, providing a 9-band catalogue of 462638 sources in the whole field. Again, all measurements refer to 4$^{\prime\prime}$ aperture photometry in the AB system. Table 5 shows the $5\sigma$ detection limit in each band.

\begin{table}
\caption{Magnitude and flux limits for ancillary data in the CDF-S field. All values refer to $5\sigma$ detections in 4$^{\prime\prime}$ apertures in AB magnitudes. all data come from the SWIRE survey and related data.}
\begin{tabular}{ccc}
\hline
Band & $m_{comp}$ & $f_{comp}$ [$\mu$Jy] \\
\hline
U & 23.0 & 2.3 \\
g & 24.2 & 0.8 \\
r & 23.6 & 1.3 \\
i & 22.5 & 3.6 \\
z & 21.6 & 8.3 \\
IRAC 3.6 & 21.8 & 6.9 \\
IRAC 4.5 & 21.1 & 13.2 \\
IRAC 5.8 & 19.8 & 43.7 \\
IRAC 8.0 & 19.6 & 52.5 \\
\hline
\end{tabular}
\end{table}

Although the central region of the CDF-S is covered by deep spectroscopy, the clump is outside this area. Nevertheless, since it is inside the SWIRE field, most sources have a SWIRE photometric redshift estimate (Rowan-Robinson et al., 2008 \& 2013).

\subsubsection{EGS and Lockman-SWIRE}

The EGS and Lockman-SWIRE clumps (clumps 2 and 3 in Table 2) both fall outside the deep multi-wavelength coverage of AEGIS and SWIRE respectively; for these two clumps, only the archival SDSS DR7 and 2MASS data were collected, providing shallow coverage in the ugriz + JHK bands. No spectroscopic data are available covering these clumps. We obtained dedicated observations of these two clumps with the NICS camera on the TNG telescope.

\subsection{Near IR Followup Observations}

The EGS and Lockman-SWIRE clumps were targeted with dedicated observations in the J and K$^{\prime}$ filters using the NICS camera on the 3.58m Telescopio Nazionale Galileoo (TNG). Observations were taken from 31 May 2011 to 1 June 2011. Atmospheric conditions for both fields were similar, with average seeing of about 1.2$^{\prime\prime}$ in J and 0.9$^{\prime\prime}$ in K$^{\prime}$. The J data were acquired with a 3-point dithering pattern, each dithered frame being integrated for 30 seconds; similarly, K$^{\prime}$ data were integrated for 10 seconds per frame with a 6-point dithering pattern.

The data were reduced with the SNAP\footnote{http://www.tng.iac.es/news/2002/09/10/snap/} pipeline, which provides dedicated software for the NICS camera. SNAP performs automatic reduction of NIR frames using a double-pass sky subtraction and coaddition process, and automatically corrects for field distortions and chip cross-talk. The standard double-pass reduction algorithm was found to provide accurate images, and changing the parameters didn't affect their quality. For the Lockman-SWIRE clump, the presence of the star 2MASS J10334014+5912390 (K$_S$ = 9.87) generates persistent ghost images due to chip cross-talk that SNAP is not able to fully remove. The resulting fake objects were masked by hand and excluded from the source extraction. Similarly, for the EGS clump the star 2MASS 14242707+5257147 (K$_S$ = 10.73) also leaves ghost images which were also masked by hand. The final coadded images cover a 4.8$^{\prime}$$\times$4.8$^{\prime}$ area, slightly larger than the {\em Planck} beam at 545 GHz.

Source detection and photometry was done with SExtractor. Photometric calibration was based on observations of the standard star Feige 67 and checked against available 2MASS photometry in the J band. We find a few stars in each field that have 2MASS measurements and are not saturated in our images: comparison with our source catalogs shows photometric offsets consistent with 0 in both fields (0.02$\pm$0.08 mag). All magnitudes were then converted to the AB system using conversion factors of 0.935 and 1.795 in J and K$^{\prime}$ respectively (calculated directly from the filter response curves). We obtain $5\sigma$ completeness magnitudes of 21.5 in J and 20.5 in K$^{\prime}$ for both fields (AB magnitudes).

Full integration times, seeing and other parameters for the observations of each field in each band are provided in Table 6.

\begin{table*}
\caption{Photometric quantities for the TNG observations of EGS and Lockman-SWIRE. Integration times are given in seconds, seeing in arcseconds, photometric zero-points in magnitudes.}
\begin{tabular}{ccccccccc}
\hline
Field & T$_{int}$(J) & T$_{int}$(K$^{\prime}$) & Seeing (J) & Seeing (K$^{\prime}$) & $Z_P$(J) & $Z_P$(K$^{\prime}$) & $N_{det}$\\
\hline
EGS & 7470 & 3600 & 1.15 & 0.88 & 26.725 & 26.073 & 339 \\
Lockman & 4050 & 3600 & 1.20 & 0.95 & 26.696 & 26.143 & 169 \\
\hline
\end{tabular}
\end{table*}

\section{The Optical/Near-IR Properties of {\em Planck} Clumps}

\subsection{Galaxy Colours and Red Sequences}

\subsubsection{EGS and Lockman-SWIRE}

For the EGS and Lockman-SWIRE clumps, we rely on our dedicated observations in the near-IR to search for a red sequence (RS; Gladders \& Yee, 2000) or overdensity of galaxies with similar colours inside the {\em Planck} beam, since the available data from SDSS proved too shallow to detect any significant overdensity of galaxies in the regions of the clumps. Moreover, near-IR colours provide a more direct comparison with redshift, as the JK' colour of passive ellipticals is expected to increase steadily up to $z \sim 4$. It is thus possible to derive a rough redshift estimate purely from the NIR colour of the RS.

The mean colour of the RS, as seen in Fig. 6, is identified by recursive 3$\sigma$ clipping in the colour-magnitude diagram (CMD) over all galaxies inside 1$^{\prime}$ from the nominal position of the clump from the {\em Planck} 545 GHz beam and brighter than the completeness limit in the K$^{\prime}$ band. After convergence, the robust mean colour and scatter are derived, and all objects inside the beam and within the RS colour range are flagged as belonging to the RS. We also independently calculate the colour of the RS by looking for overdensities of objects with a common colour inside the beam, under the assumption that the passively evolving ellipticals that form the bulk of the cluster population will be concentrated at the cluster centre. We detect in both clumps a significant overdensity of objects with the colour of the RS found in the colour-magnitude diagram.

For the EGS clump (see Fig. 7), the galaxies in the RS are mostly concentrated in a peak close (separation $d = 25^{\prime\prime}$) to the nominal location of the clump. A secondary peak about 2 arcminutes away from the main peak is also inside the {\em Planck} beam, suggesting a larger, composite structure. The mean colour of the RS is J-K$^{\prime}$=0.68$\pm$0.08. We convert this colour to a redshift estimate by assuming a passively evolving elliptical template with a single star-formation episode (generated using the Bruzual \& Charlot (2003) libraries). We investigate different formation redshifts of 3, 4, 5, 6 and 8, finding consistent results in all cases. The redshift estimate from this colour is $z = 0.81 \pm 0.10$.

The RS in the Lockman-SWIRE clump is less well defined, mostly because this field is overall less populated, suggesting a higher redshift for the clump. Nevertheless, we are able to detect a tentative red sequence associated with a significant overdensity of objects sitting at the centre of the beam (see Fig. 7), with a common colour (J-K$^{\prime}$=1.34$\pm$0.09), which corresponds to a redshift estimate of $z = 2.06 \pm 0.10$.

\begin{figure*}
\includegraphics[width=8.5cm]{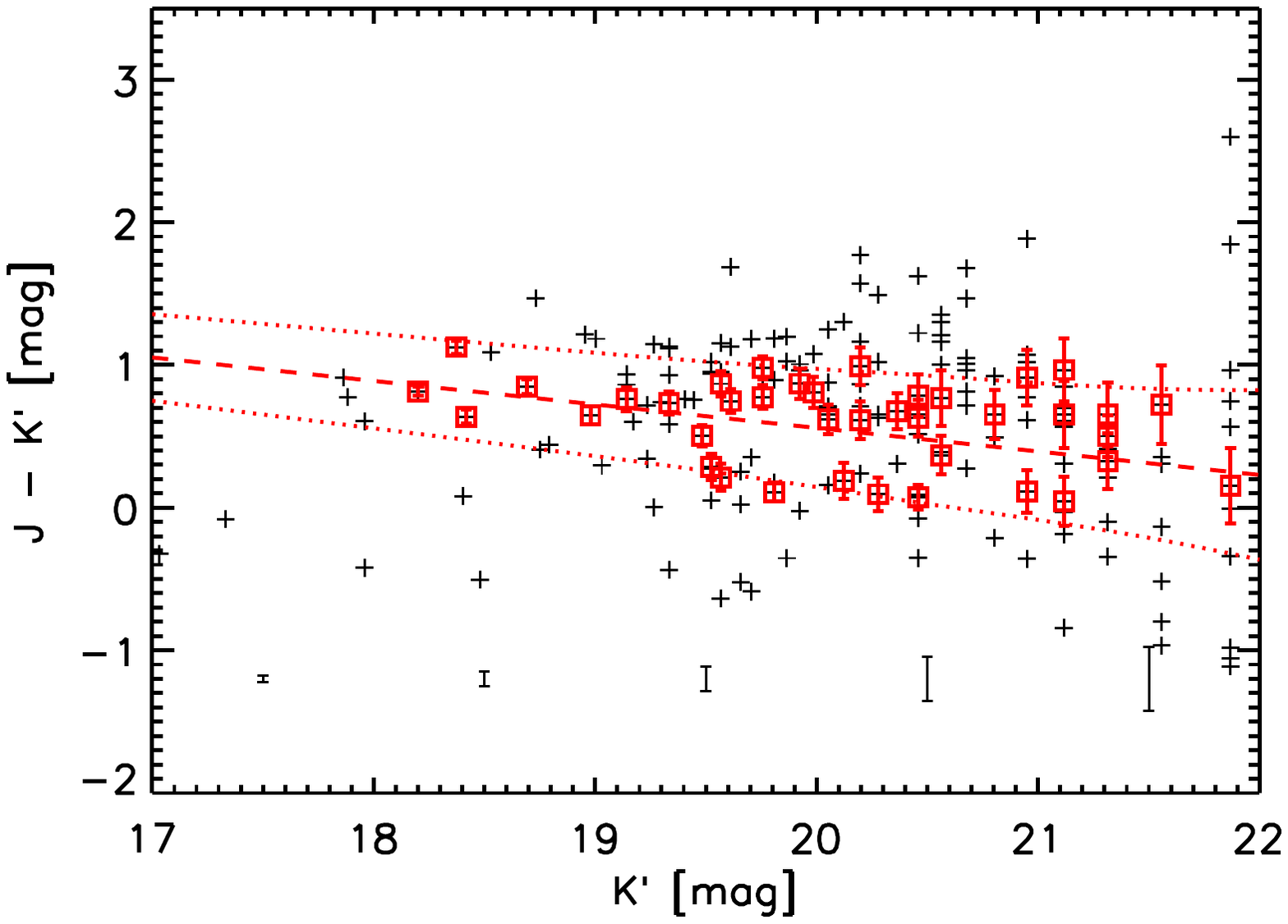}
\includegraphics[width=8.5cm]{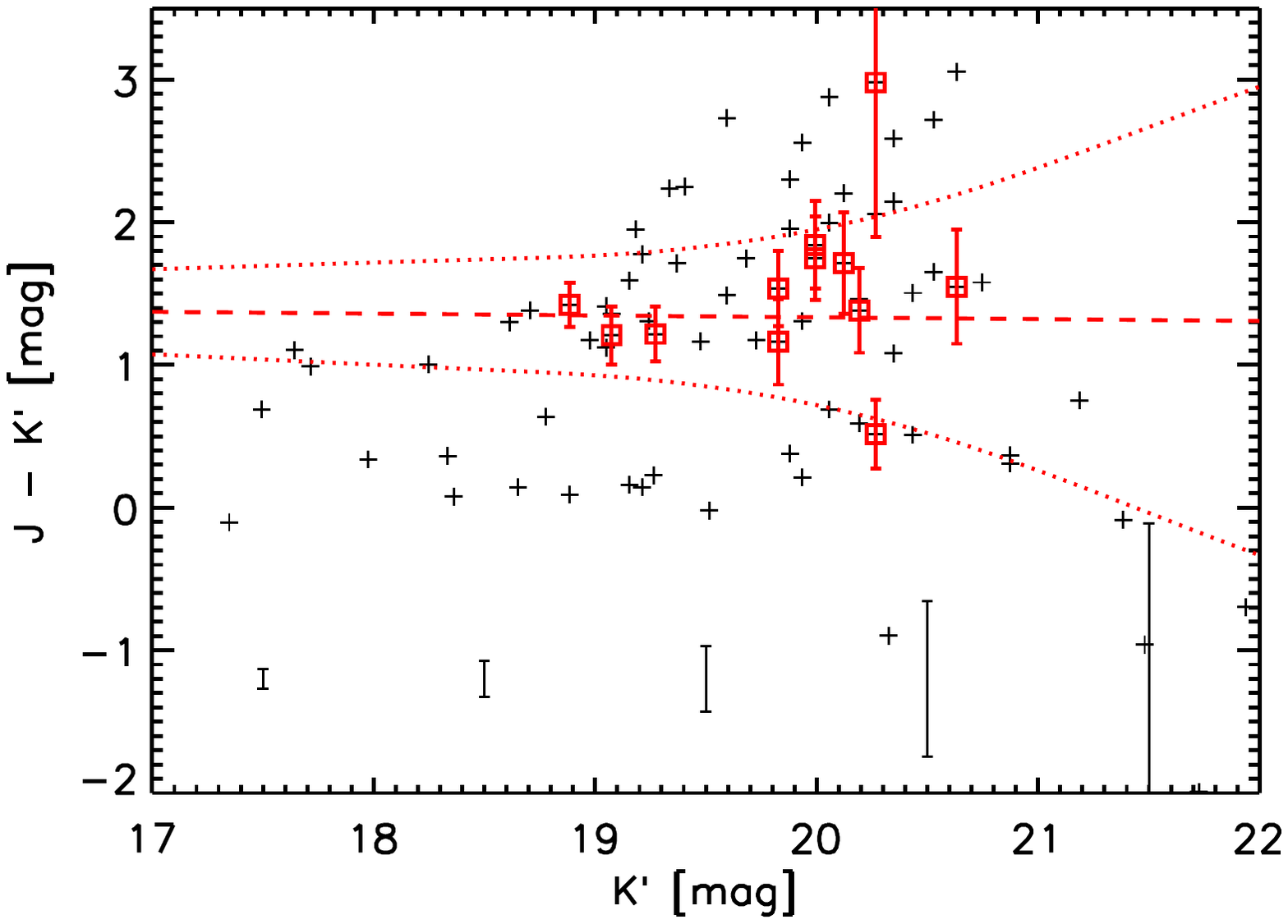}
\caption{Colour-magnitude diagram and red sequence for the EGS (left) and Lockman-SWIRE (right) clumps. Black crosses are all detected objects in the field, red boxes are objects inside the {\em Planck} beam and selected as red sequence galaxies according to the technique explained in section 5.1.1. The dashed and dotted lines show the fit to the RS and its scatter, including photometric errors. Vertical bars in the lower region of each plot show the average colour error at each magnitude interval. Magnitudes in the AB system.}
\end{figure*}

\begin{figure}
\includegraphics[width=5cm]{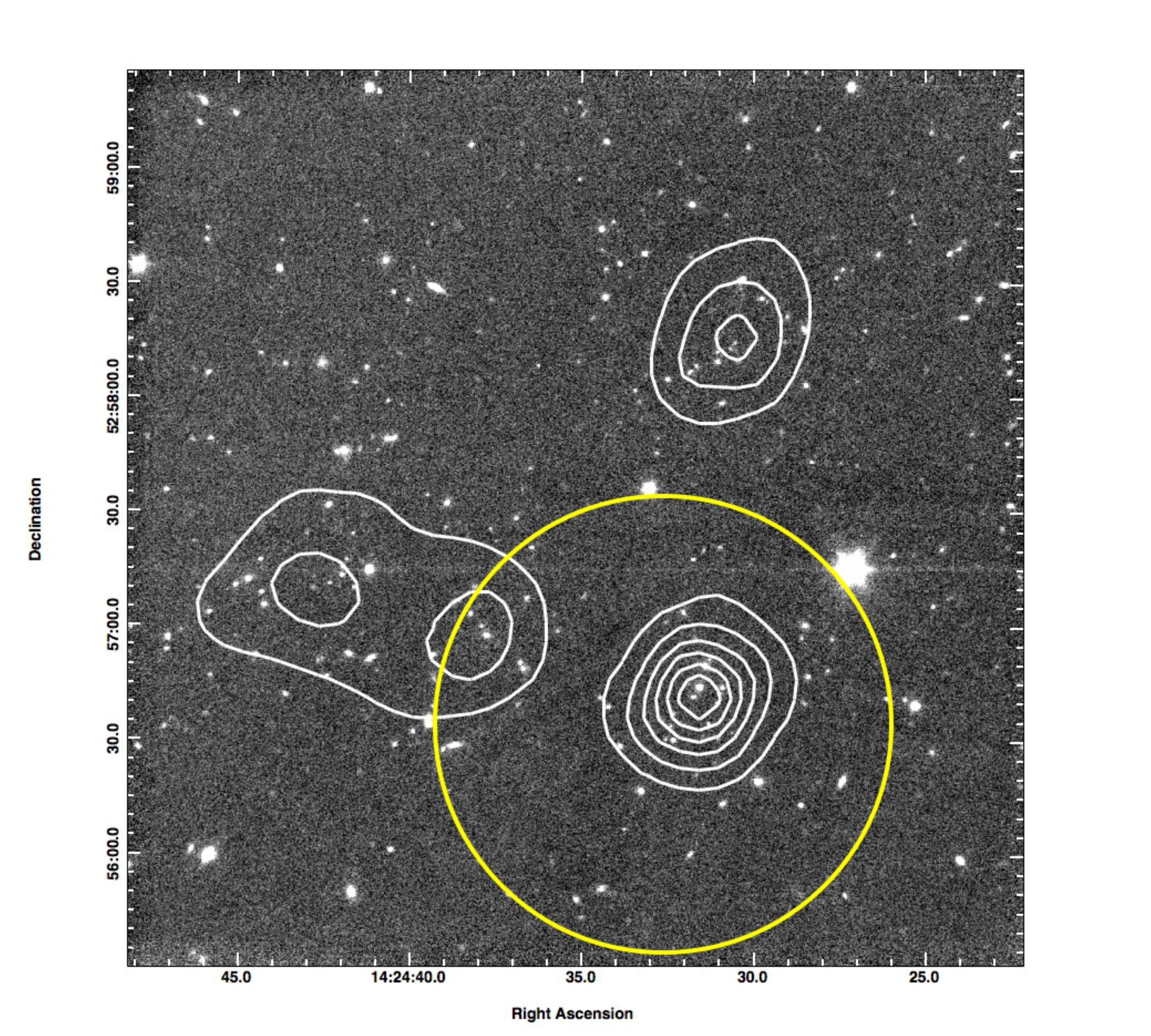}
\includegraphics[width=5cm]{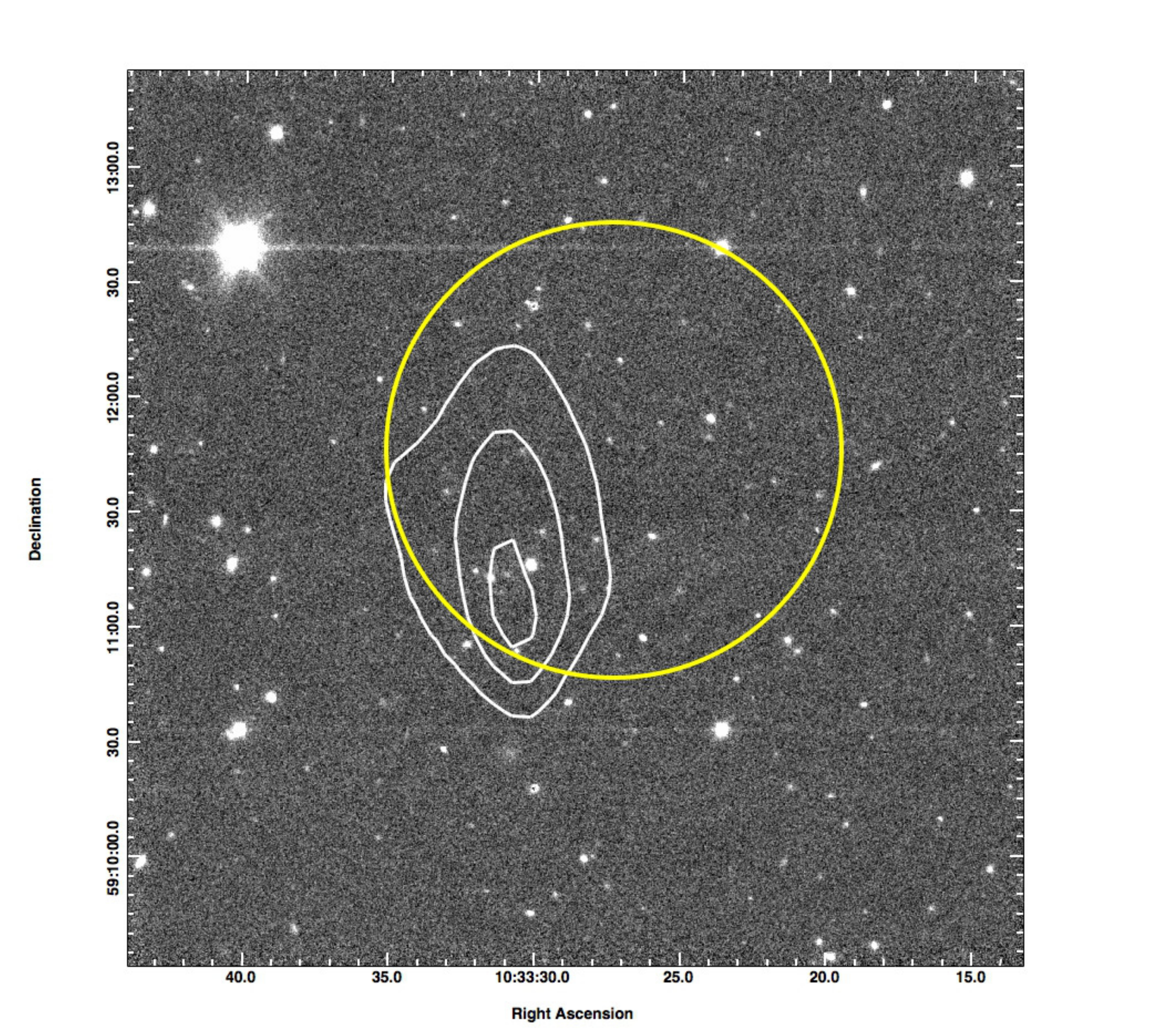}
\caption{Overdensity contours of RS galaxies in the EGS (top) and Lockman-SWIRE regions (bottom). The background image is the NICS K$^{\prime}$ unsmoothed map, the white contours mark the overdensity contours at 3--8$\sigma$. The shown field is 4.5$^{\prime}$ across, matching the extent of the {\em Planck} beam. The yellow circle has a radius of 1.1$^{\prime}$ and is centred at the position of the ERSCS source.}
\end{figure}

\subsubsection{Bo\"{o}tes}

The Bo\"{o}tes clump shows an extremely red RS, essentially detected only in the IRAC bands. We detect a RS in the $(3.6\mu\rm{m} - 4.5\mu\rm{m})$ vs. $4.5\mu\rm{m}$ plane, with a robust mean colour of 0.22$\pm$0.11. This is consistent with a very high redshift z $\geq$ 2, regardless of the evolutionary model assumed. The high-$z$ clusters (z=1.1---1.4) detected by the ISCS survey (Brodwin et al. 2006) show RSs with median colours $(i - 3.6\mu\rm{m}) = 2.7 \pm 0.3$: the RS that we detect in the IRAC bands has a mean colour of $3.07 \pm 0.42$ in the same bands, thus suggesting a possibly higher redshift than the ISCS clusters, in agreement with the IRAC colours. It must be noted here that the assumption of different evolutionary models for elliptical templates yields very different $(i - 3.6\mu\rm{m})$ colours, while the redder $(3.6\mu\rm{m} - 4.5\mu\rm{m})$ colours are scarcely affected by different evolutionary histories. This demonstrates the superior leverage of deep IRAC imaging for detecting high-$z$ galaxy. See Fig. 8.

\begin{figure}
\includegraphics[width=8cm]{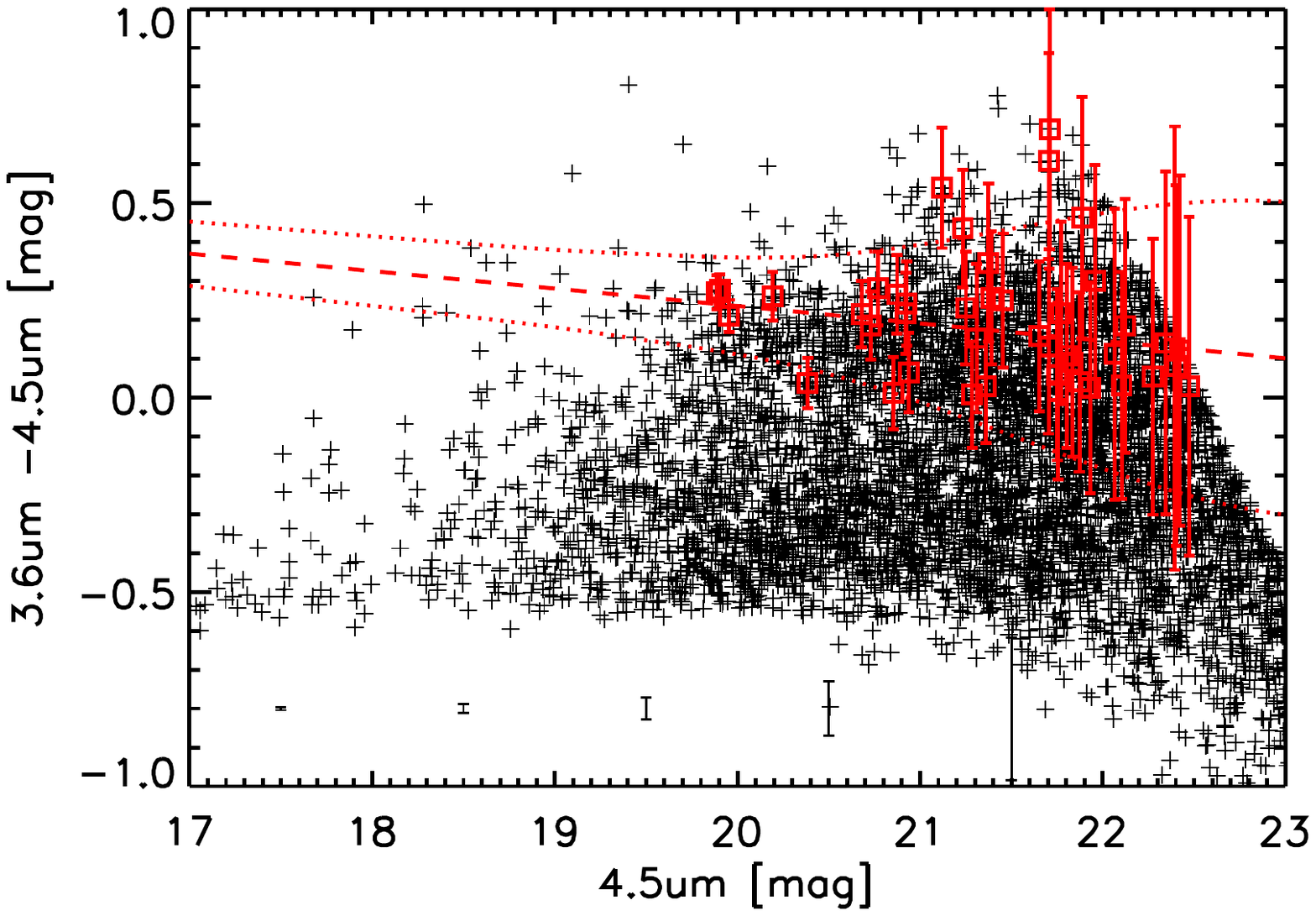}
\caption{Colour-magnitude diagram and red sequence for the Bo\"{o}tes clumps in the $(3.6\mu\rm{m} - 4.5\mu\rm{m})$ colours. Black crosses are all detected objects within 10$^{\prime}$ of the {\em Planck} clump, red boxes are objects inside the {\em Planck} beam and brighter than the completeness magnitude in the $4.5\mu\rm{m}$ band. The dashed and dotted lines show the fit to the RS and its scatter. Vertical bars in the lower region of the plot show the average colour error at each magnitude interval. Magnitudes in the AB system.}
\end{figure}

\subsubsection{CDF-S}

We look for a RS in the CDF-S clump in the $(i - 3.6\mu\rm{m})$ vs. $3.6\mu\rm{m}$ plane. We find a faint but well-defined RS, over two magnitudes, down to the completeness limit in the $3.6\mu\rm{m}$ magnitude. Its colour of 2.22 $\pm$ 0.29 is slightly bluer than the colours of the ISCS clusters, suggesting a redshift of 0.94$\pm$0.09. See Fig. 9.

\begin{figure}
\includegraphics[width=8cm]{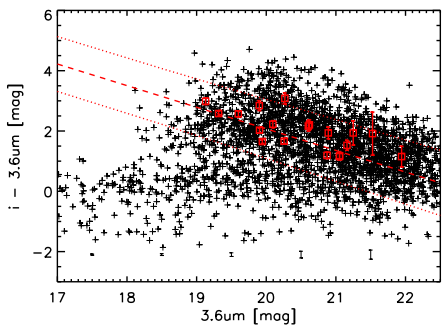}
\caption{Colour-magnitude diagram and red sequence for the CDF-S clump in the $i - 3.6\mu\rm{m}$ colours. Black crosses are all detected objects within 10$^{\prime}$ of the {\em Planck} clump, red boxes are objects inside the {\em Planck} beam and brighter than the completeness magnitude in the $3.6\mu\rm{m}$ band.  The dashed and dotted lines show the fit to the RS and its scatter. Vertical bars in the lower region of the plot show the average colour error at each magnitude interval. Magnitudes in the AB system.}
\end{figure}

\subsection{Photometric Redshift Analysis}

Photometric redshifts are calculated for all objects in the Bo\"{o}tes and CDF-S fields, using the wide photometric coverage from the optical to the IRAC bands. We use the publicly available code EAZY (Brammer, van Dokkum \& Coppi 2008), which was specifically developed to address situations where spectroscopic coverage is not available for direct comparison and calibration. We tested the code against simulated catalogues to  obtain an independent assessment of the intrinsic error on our photo-$z$. We also find that the default templates provided with EAZY provide the best photometric redshifts. Where K-band photometry is available (i.e. for the Bo\"{o}tes clump), a K-band prior is favoured over an R-band prior, as it reduces the fraction of wrong identifications. A complete treatment of the tests done and their results will be shown in Braglia et al. (in prep.).

After calculating photo-$z$'s for all objects in each field, we look for 3D overdensities (i.e. a significant peak in photometric redshift space, matched to a clustering of objects on the plane of the sky) of objects inside the regions covered by the {\em Planck} clumps. Peaks in photometric redshift space are identified by comparison of the distribution of redshifts along the line of sight of the clump and in the larger field, to track the effect of large-scale structure (walls and filaments that will appear as massive spikes of objects at the same redshift). An adaptive kernel algorithm is then used to produce an overdensity map in each redshift slice. 

\subsubsection{Bo\"{o}tes}

Photometric redshifts in the Bo\"{o}tes field are based on the multi-wavelength dataset described in Section 4.1.1. Where IRAC data are available, we rely only on the two bluer channels, 3.6 and 4.5$\mu$m, to minimise contamination of the stellar SED by any hot dust component. Based on tests against simulated catalogues and collected archival spectroscopic redshifts in this field (Vaccari et al., in prep), we find an intrinsic scatter $\Delta z/(1+z) = 0.042$ out to $z \sim 4$ in the photometric redshift measure. Comparison of the redshift distribution in the whole NDWFS Bo\"{o}tes field with that along the line of sight of the {\em Planck} clump (see Fig. 10) shows a marked spike of objects at a photometric redshift of about 2.3. This peak is also found to be associated with a strong overdensity of objects ($\sim 10\sigma$) at the position of the clump. Our analysis shows no significant overdensity of objects along the clump's line of sight at any other redshift, thus confirming this peak to be the only cluster candidate. The robust mean estimate of the redshift for this peak is $2.27 \pm 0.14$. Monte Carlo simulations, looking for photometric redshift spikes of similar significance to that found for this clump within Planck beams placed at random locations in the Bo\"{o}tes field, show that similar significance redshift spikes are found $\leq$1\% of the time. This thus represents the chance that the association of this Planck clump with a redshift spike is a false positive.

Furthermore, our overdensity is the highest concentration in a large-scale system at the same photometric redshift ($2.1 \leq z_{phot} \leq 2.4$) that spans several arcminutes on the sky (see Fig. 11), with multiple secondary peaks well above $5\sigma$. In particular, a second peak (with a significance of over $8 \sigma$) lies at the edge of the {\em Planck} beam, separated from the main peak by 2.8$^{\prime}$; at a redshift of 2.27, this translates to a physical distance of 1.4 Mpc. Other, less prominent structures are scattered throughout the same area,  suggesting that we may be witnessing a young cluster in its assembly phase from the large-scale structure.

\begin{figure}
\includegraphics[width=8.5cm]{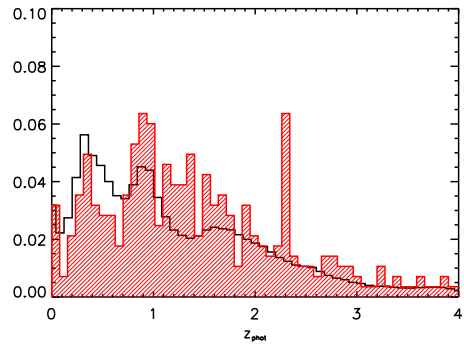}
\caption{Photometric redshift distribution in the Bo\"{o}tes field (in normalised counts). The black histogram shows the distribution of all objects in the 9 square degrees of the NDWFS, the red shaded one is the distribution of objects along the l.o.s. of the {\em Planck} clump. While the spikes at $z<1$ reflect the overall redshift distribution in the NDWFS field, a peak of  at $z \sim 2.3$ is only found on the l.o.s. of the clump.}
\end{figure}

\begin{figure}
\includegraphics[width=9.5cm]{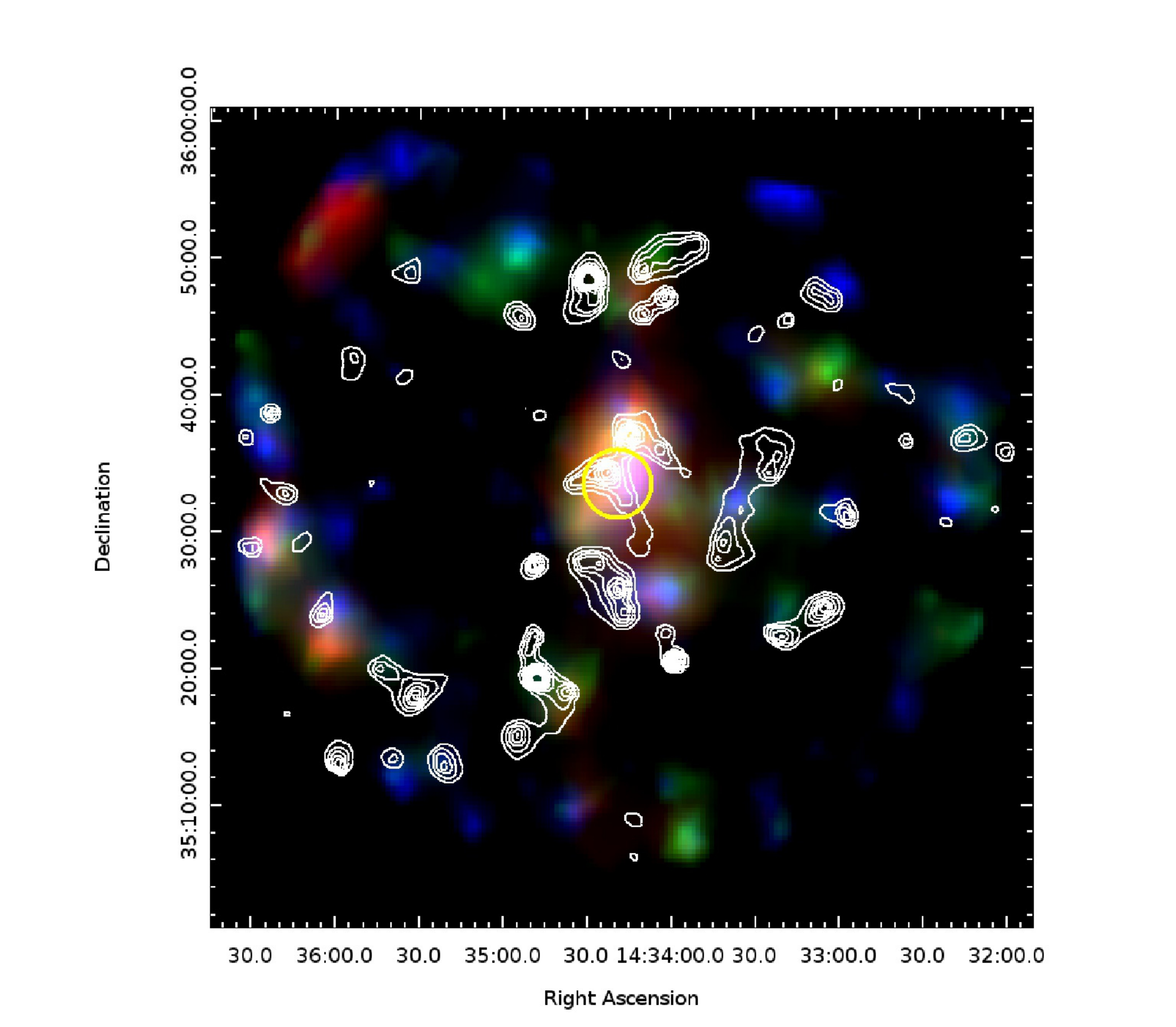}
\caption{The RGB density map of {\em Herschel} sources within 25$^{\prime}$ of the Bo\"{o}tes clump, as shown in Fig. 4, together with the distribution of objects at the photometric redshift of the clump. White contours mark overdensity significance levels from 3 to 10, in steps of $1\sigma$. The most significant spike is found inside the clump, as defined by the {\em Planck} beam (yellow circle), and has a significance of $10\sigma$. A second $\sim 8\sigma$ overdensity is found at a physical distance of 1.4 Mpc, north of the main overdensity.}
\end{figure}

\subsubsection{CDF-S}

Although less complete than the Bo\"{o}tes dataset, the photometry available for the CDF-S clump allows us to obtain accurate photometric redshifts with an intrinsic scatter $\Delta z/(1+z) = 0.077$ out to $z \sim 3$ (the lack of deep NIR data in the JHK bands systematically increases the scatter in the photometric redshifts relative to that found for Bo\"{o}tes). Where available, we prefer the SWIRE photometric redshifts which tests against collected archival spectroscopic redshifts show have somewhat smaller scatter. We find a spike of objects at a mean photometric redshift $1.04 \pm 0.11$, matched by a strong ($\sim 12\sigma$), compact spatial overdensity of objects at the position of the {\em Planck} beam. The structure appears to be isolated from other strong density peaks at the same redshift, but a web of lower density filaments populates the surrounding area. As in the case of Bo\"{o}tes, this is the only peak in redshift associated with a concentration of galaxies inside the {\em Planck} beam. Our Monte Carlo analysis indicates that the chance of this redshift spike being a false detection is $\sim$1\%.

\begin{figure}
\includegraphics[width=8.5cm]{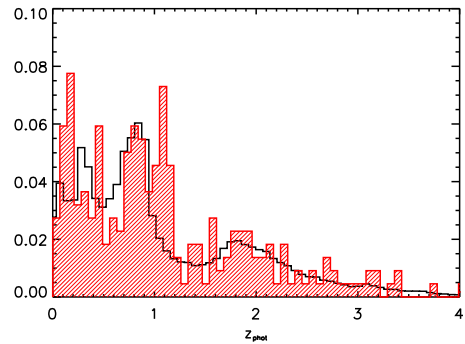}
\caption{Photometric redshift distribution in the CDFS field (in normalised counts). The black histogram shows the distribution of all objects in SWIRE survey field, the red shaded one is the distribution of objects along the l.o.s. of the {\em Planck} clump. While the spikes at $z<1$ reflect the overall redshift distribution in the NDWFS field, a peak at $z \sim 1.1$ is only found along the l.o.s. the clump.}
\end{figure}

\begin{figure}
\includegraphics[width=9cm]{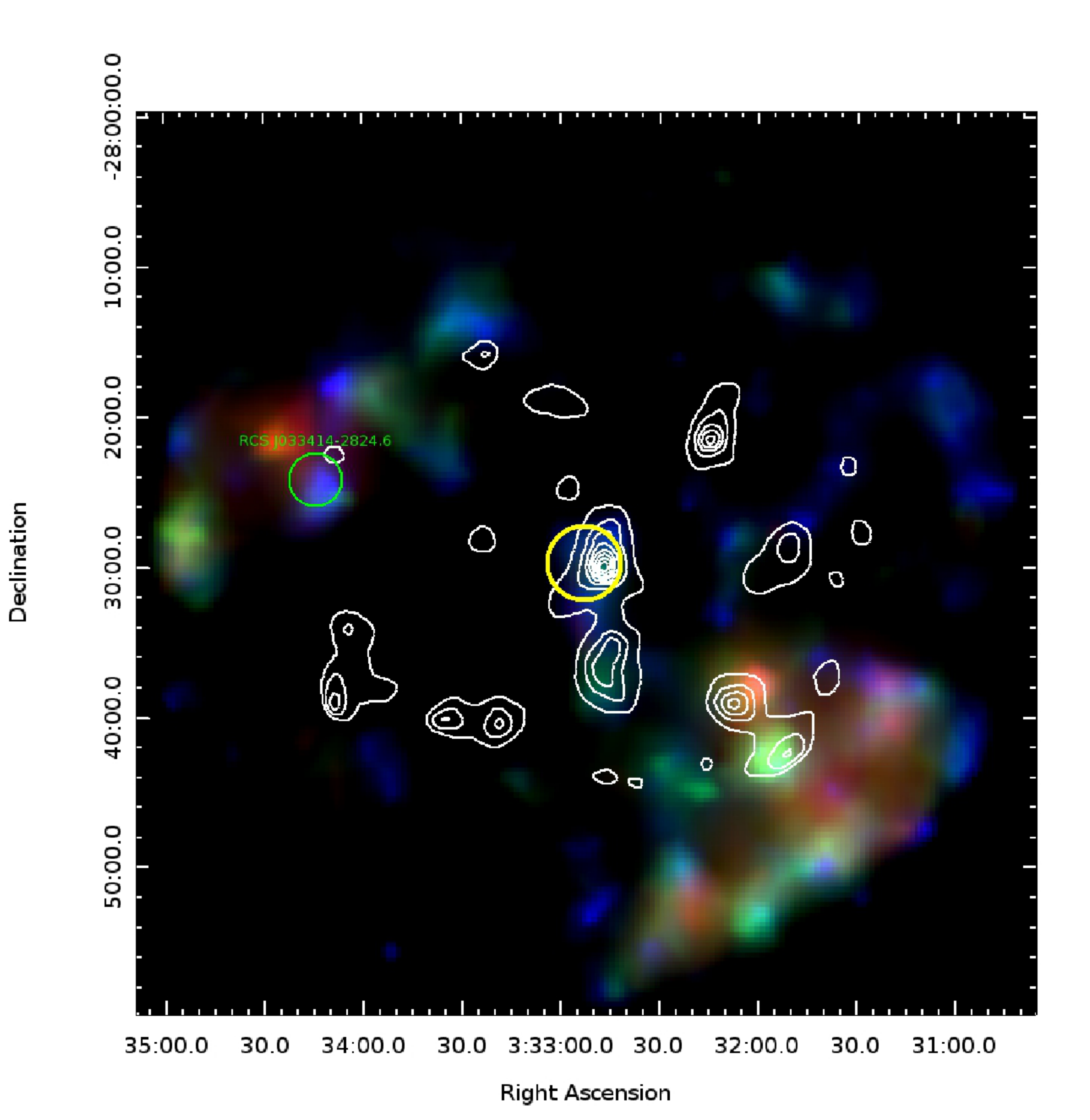}
\caption{The RGB density map of {\em Herschel} sources within 25$^{\prime}$ of the CDFS clump as shown  in Fig. 4, together with the distribution of objects at the photometric redshift of the clump. White contours mark overdensity significance levels from 3 to 12, in steps of $1\sigma$. The most significant overdensity is found inside the clump, as defined by the {\em Planck} beam (yellow circle), and has a significance of $13\sigma$. Also shown with a green circle is the position of the z=0.67 galaxy cluster RCSJ033414-2824.6 that is associated with an enhancement in the local density of {\em Herschel} sources and discussed in Section 6.4.}
\end{figure}

\section{Discussion}

The analysis above demonstrates our ability to uncover previously unknown clusters of galaxies using the combination of {\em Herschel} and {\em Planck} data. This provides us with a new tool for the examination of the history of both galaxy and galaxy cluster evolution. The current sample size in terms of clusters, though, is small, so we cannot yet draw definitive conclusions. Nevertheless, our results are capable of providing some interesting new constraints on galaxy and galaxy cluster models.

\subsection{The Clump Population}

We have uncovered four starburst clusters in our examination of 91.1 sq. deg. of HerMES data, which implies an area density of such objects of 0.044$\pm$0.022 per square degree or, equivalently, 23$_{-7}^{+22}$ square degrees per source. This would imply that there are roughly 2000 of these sources over the whole sky. The {\em Planck} all sky surveys can, in principle, detect these sources, though galactic foregrounds will impede such a search over a significant fraction of the sky. Nevertheless, followup observations of sources selected by {\em Planck} colours, as mentioned in Section 1 are likely to uncover a large number of these sources. Work by Herranz et al. (2013) in the H-ATLAS regions and by Montier et al. (in prep) using the {\em Planck} maps are already finding further such sources.

We compare our observations to the predictions of Negrello et al. (2005) for the number counts of dusty galaxy clusters detected by {\em Planck} in Fig. 14. The average 353 GHz (ie. 850$\mu$m) flux for our four clusters is 200mJy, and, converting to the units used by Negrello et al., they have a number density of 144$\pm$70 sources per steradian. This places the observed number counts of our sources at the lower end of the predicted counts in Figure 1 of Negrello et al. (2005). Specifically we can strongly exclude the $Q=1$ analytical model and the numerical model used by Negrello et al.. The $Q=1$ model corresponds to the case where the amplitude of the three-point angular correlation function $\xi$ does not evolve with redshift. This is something that is expected to be true for dark matter (Juszkiewicz, Bouchet \& Colombi, 1993; Colombi, Bouchet \& Hernquist, 1996) but not for luminous matter whose three-point correlation function should behave more as $1/b$ or $1/b^2$, where $b$ is the bias parameter (Fry \& Gaztanjaga, 1993; Szapudi et al. 2001). These possibilities provide the other two analytical models in Figure 1 of Negrello et al. (2005) which lie closer to our observational results. Larger sample sizes and a wider range of fluxes are needed to test these models more precisely but, as noted by Negrello et al. (2005), more sophisticated and physically realistic models are also required.

\begin{figure}
\includegraphics[width=8cm]{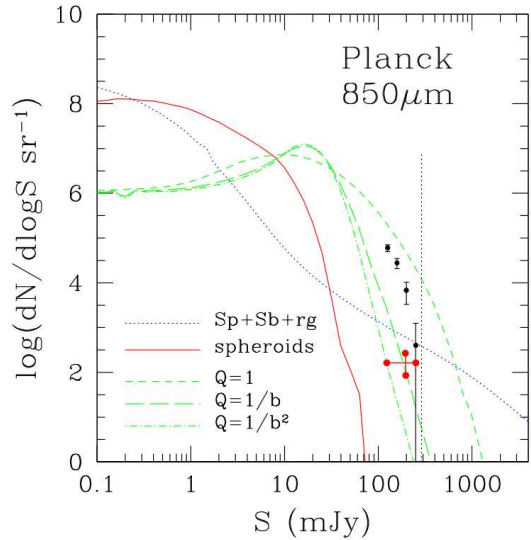}
\caption{Prediction for the counts of starbursting clusters from Negrello et al. (2005) with the observed counts of these sources from the present study overplotted as a red point. The curves represent different analytical predictions based on different assumptions for how the three-point angular correlation function evolves with redshift, while the black points are the result of their numerical simulations. See Section 6.1 for more details.}
\end{figure}

\subsection{Star-formation rates}

We derive a simple estimate of the integrated SFR for each clump by considering all {\it {\em Herschel}} sources inside the {\it {\em Planck}} beam as belonging to the clumps themselves. We consider only the sources with a $\geq 2 \sigma$ detection in all SPIRE bands (cf. Figure 5) and fit their SEDs with a modified blackbody, fixing the dust emission index $\beta = 2$ (different values of $\beta$ do not significantly affect the fit and return almost the same temperature and total FIR luminosity). We then derive the SFR from the integrated FIR luminosity using the relations of Bell (2003), which provide a mild correction (within 10\%) to the standard Kennicutt values. Errors on the SED fit are derived using 1000 Monte Carlo realisations around the best-fit value for each individual galaxy, and the corresponding confidence intervals are translated to SFR with the same relations of Bell. Table 7 summarises the total SFR and luminosities of each clump, together with their photometric redshift.

\begin{table}
\label{sfrclumps}
\begin{tabular}{cccc} \hline
Field&$z$&$L_{FIR}$ & SFR ($M_{\odot}/yr$)\\
\hline
Bo\"{o}tes & 2.27$\pm$0.12$^!$ & 73$\pm$11 & 11632$\pm$1800\\
EGS & 0.76$\pm$0.10$^*$ & 3.7$\pm$0.9 & 620$\pm$138\\
Lockman & 2.05$\pm$0.09$^*$ & 31$\pm$6 & 4924$\pm$946\\
CDF-S & 1.04$\pm$0.11$^!$ & 10$\pm$2 & 1631$\pm$356\\
\hline
\end{tabular}
\caption{Redshift estimates and integrated FIR luminosities and SFR for the four clumps. $^!$ indicates a redshift derived using a photometric redshift method, while $^*$ indicates values derived using the observed red sequences. The values of $L_{FIR}$ are expressed in units of 10$^{12} L_{\odot}$.}
\end{table}

Preliminary stacking of available IRAC-selected sources in the Bo\"{o}tes clump (Braglia et al., in prep.) suggests that these estimates of SFR could be underestimated by at least 30\% and possibly up to a factor of 2, down to the detection limit of the available IRAC photometry.

Comparison with the available literature shows a significant increase in the star-formation rate density (SFRD) in clusters with increasing redshift, up to at least $z \sim 2$. In particular, we add the SFR of other clusters for which MIR or FIR measurements are available: the IRAS measurements of Perseus by Meusinger et al. (2000); BLAST measurements of A3112 (Braglia et al. 2011); ISO measurements of A1689 (Fadda et al. 2000); and {\it Spitzer} measurements of A1758 (Haines et al. 2009), the Bullet cluster (Chung et al. 2010), Cl0024+16 and MS0451-03 (Geach et al. 2006). We also use the results of Stevens et al. (2010), based on SCUBA observations around high-redshift QSOs, to obtain a comparison sample in a redshift range similar to ours. We derive star-formation rate estimates for all galaxies in their catalogues based on the observed $F_{850}$ flux and using an Arp 220 spectral template, thus obtaining an estimate for the total FIR luminosity and SFR using the same relations as before.

We calculate the SFRD for each cluster by assuming that the clusters are spherical. We then derive an angular radius from the aperture within which the observations were made, and convert that to a proper distance at the redshift of the cluster. From this, we simply calculate the associated volume by assuming this distance to be the radius of the cluster. This is consistent with independent measurements of R$_{200}$ for the literature clusters. Table 8 shows, for each clump or cluster, the redshift, total SFR (in M$_{\odot}$ yr$^{-1}$), angular radius $\theta$ (in arcmin) and associated physical volume (in units of Mpc$^{3}$) used for this calculation. The derived SFRD plotted against redshift is shown in Figure 15.

\begin{table}
\label{sfrdvol}
\begin{tabular}{ccccc} \hline
Field&$z$&SFR&$\theta$ (arcmin) & V ($Mpc^{3}$)\\
\hline
Perseus & 0.017 & 22 & 150 & 126 \\
A3112 & 0.075 & 58 & 41 & 215 \\
A1689 & 0.18 & 280 & 13 & 49 \\
A1758 & 0.28 & 910 & 20 & 557 \\
Bullet & 0.30 & 267 & 13 & 157 \\
Cl0024+16 & 0.39 & 1000 & 6.3 & 34 \\
MS 0451-03 & 0.55 & 460 & 5.2 & 34 \\
\hline
RXJ1218 & 1.74 & 4055 & 1.4 & 1.5 \\
RXJ0941 & 1.82 & 7126 & 1.4 & 1.5 \\
RXJ0057 & 2.19 & 3898 & 1.4 & 1.4 \\
RXJ1249 & 2.21 & 6901 & 1.4 & 1.4 \\
RXJ1633 & 2.80 & 8546 & 1.4 & 1.2 \\
\hline
EGS & 0.76 & 620 & 2.1 & 2.9 \\
CDFS & 1.04 & 1631 & 2.1 & 3.8 \\
Lockman & 2.05 & 4924 & 2.1 & 4.2 \\
Bo\"{o}tes & 2.27 & 11632 & 2.1 & 4.2 \\
\hline
\end{tabular}
\caption{Summary of the sizes and star formation rates for both literature galaxy clusters (see text for details) and those associated with {\em Planck} clumps (last four entries).}
\end{table}

We find a steady increase in the SFRD in clusters from $z \sim 1$ to $z \sim 3$; this reflects the overall trend of the cosmic SFRD (e.g. Hopkins \& Beacom 2006; Bouwens et al. 2011; Magnelli et al. 2011), although we obtain values much higher than the field since clusters are overdense regions, where the galaxy density is orders of magnitude higher than in the field. Below $z \sim 1$, the SFRD in clusters drops rapidly towards negligible levels with a steeper slope than the high-z or field galaxy trend. This would be expected for older clusters where star-formation is mostly quenched. The clusters plotted here do not, of course, represent a homogeneous sample of objects of the same mass, since the clusters themselves are evolving over this wide range of redshifts. The low z clusters shown here, with star formation rates derived from the literature, have masses ranging from 6 $\times 10^{14}$M$_{\odot}$ to a few 10$^{15}$M$_{\odot}$. In contrast the Planck clumps and quasar fields are estimated to have masses around 1 $\times 10^{14}$M$_{\odot}$, though with large uncertainties since these mass estimates, using the $M_{200} / R_{200}$ relation from Carlberg et al. (1997), assume that these systems are virialized. Spectroscopic observations of our clumps will be necessary before more accurate mass estimates, and an assessment of their virialization, can be attempted.

It is worth noting that our lower-z clumps, EGS and CDFS, lie in a region previously devoid of data, and thus help fill the gap between the high-$z$ clusters, where the SFRD is consistently above 1000 M$_{\odot}$ yr$^{-1}$ Mpc$^{-3}$, and the low-$z$ sample of evolved clusters. We also note that clusters known to be undergoing a merger seem to have a higher SFRD even at lower redshift (in particular Cl0024), while clusters in a state of kinematic relaxation (e.g. MS0451) show a relatively early quenching of their SFR (cf. e.g. Mahajan, Raychaudhury \& Pimbblet 2012).

\begin{figure}
\label{mpclumps}
\includegraphics[width=8.5cm]{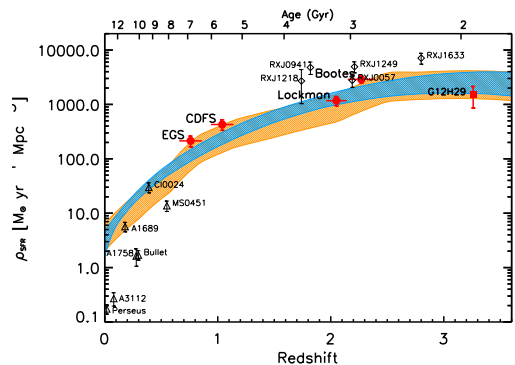}
\caption{Integrated SFRD vs. redshift for the four HerMES clumps, other clusters in literature with MIR or FIR measurements, and one candidate higher redshift clump (Clements et al., in prep). Red dots show the HerMES clumps (horizontal bars represent the error in the redshift determination); open triangles are for the $z<0.6$ clusters from literature, as explained in the text; open diamonds show the data from Stevens et al. (2010). Also shown is the field SFRDs from Hopkins \& Beacom (2006) (blue) and from Bouwens et al. (2011) (yellow) scaled to an SFRD of 100 at z=0.7 to match the cluster values. As can be seen the shape of the field SFRD as a function of redshift is broadly similar to that seen here for galaxy clusters down to about z$\sim$1.}
\end{figure}

\subsection{Colours and the red-sequence}

Figure 16 shows the mean colour and scatter in the $I - 3.6\mu$m observed colour for the Bo\"{o}tes and CDFS clumps, compared with a selection of clusters from the ISCS survey and with colour tracks for a single-starburst galaxy at different formation epochs. For the purpose of calculating the mean value of the RS and its scatter in the clumps, we use only the objects with high photo-z reliability (i.e. those for which at least 80\% of the photo-z probability density is contained within the 1-$\sigma$ interval). The values for the ISCS clusters are drawn from Eisenhard et al. (2008). It is immediately evident that, while the colour of $z \sim 1- 1.5$ clusters on average can provide lower limits to their formation age (all ISCS clusters are consistent with a formation epoch between $z = 2$ and $z = 3$), the leverage offered by $z \sim 2$ structures provides tighter constraints on the formation epoch. For instance, the Bo\"{o}tes clump, even accounting for both colour and photo-z error, fits a slightly higher formation redshift between 2.5 and 3. On the other hand, the colour of the red-sequence in the CDFS clump is significantly bluer than the typical values for the ISCS clusters, suggesting a relatively recent formation epoch of $z = 1.5$. Both clumps are thus consistent with having an age of about 1-1.5 Gyr.

The scatter that we observe in the RS of our clumps seems larger than in other clusters. Previous studies (e.g. Bell et al. 2004) have shown that the RS is already in place at $z \sim 1$, and that its scatter does not change much from $z \sim 1$ to $z = 0$. This seems to be confirmed at higher $z$ by recent findings (cf. Andreon et al. 2011b; Gobat et al. 2011; Santos et al. 2011; Pierini et al. 2012), where well-defined red-sequences are detected in clusters out to $z \sim 2$. However, it must be noted that all these detections are either based on the presence of a red-sequence, or on X-ray emission. In both cases, this implies a bias towards well-evolved clusters where at least partial virialization has been reached. 

Conversely, if our clumps actually represent young clusters in their early stages of formation, then we can expect their member galaxies to be on average dustier (which will make their colour redder), while at the same time several galaxies, even along the RS, will be undergoing massive bursts of star-formation (as confirmed from the SFR measurements discussed in Section 6.2). The combined effect of these two processes will be to increase the scatter of the RS in both directions, as observed in both the Bo\"{o}tes and CDFS clumps. Still younger clusters, whose member galaxies have yet to establish an old stellar population, might lack a significant red sequence. In these cases a photo-z approach, similar to our analysis of the Bo\"{o}tes and CDFS clumps, rather than a search for a red sequence, would be necessary.

All this confirms the expectation that our method is identifying young clusters in the early stages of formation, regardless of the presence of a red-sequence.

\begin{figure}
\label{rsclumps}
\includegraphics[width=8.5cm]{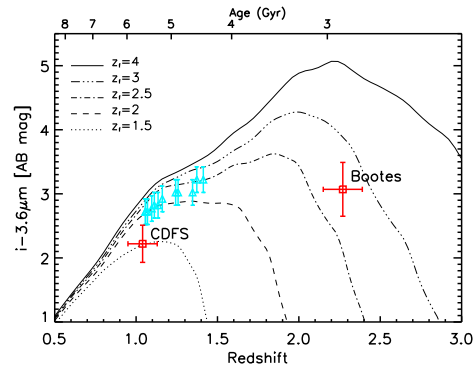}
\caption{The mean I-3.6$\mu$m colour of the red-sequence for the CDF-S and Bo\"{o}tes clumps (red squares). Different tracks for galaxies with different formation redshifts are shown (see legend). The cyan triangles mark the mean colour of the RS for a selection of clusters from the ISCS survey (Eisenhardt et al. 2008).}
\end{figure}

\subsection{Extending the Clump Search}

While there is clearly considerable potential in using {\em Planck}, and improved catalog releases such as the 2013 Planck Catalog of Compact Sources (PCCS, Planck Collaboration, 2013), to search the entire sky for the kind of dusty galaxy clusters discovered here, a common factor among all the models of such objects in Negrello et al. (2005) is that their number counts are quite steep, and that many more such objects are likely to be found at fainter fluxes. The eventual {\em Planck} sensitivity to these objects is unlikely to reach much below the sensitivities achieved here, thanks to the effects of confusion in the large {\em Planck} beams. However, surveys with {\em Herschel}, including HerMES, H-ATLAS and the HeLMS extension to HerMES (270 sq. deg coverage of part of the Stripe 82 region of SDSS (Oliver et al., 2012)), and others, mean that over 1000 sq. deg. of sky will eventually be covered by SPIRE to considerably greater depth than {\em Planck} can achieve in comparable bands, and with much higher angular resolution. These {\em Herschel} maps can be searched for fainter dusty galaxy clusters than can be found by {\em Planck}, or for structures that are not well matched in angular size to the {\em Planck} beam,  using a variety of techniques, ranging from simple smoothing, to wavelet transform methods. Indeed, our current analysis includes a serendipitous demonstration of this capability. Figure 13 shows our adaptively smoothed source density map of the CDF-S region. In addition to the {\em Planck} clump we also highlight an additional structure to its north east. This is prominent at 250$\mu$m where it appears as a significant overdensity. Its position matches that of a known z=0.67 galaxy cluster RCSJ033414-2824.6 (Gilbank et al., 2007). This cluster is not associated with a {\em Planck} ERCSC source. While a detailed analysis of this system is beyond the scope of the present paper, this detection clearly demonstrates the ability of {\em Herschel} to uncover lower luminosity clumps than can be detected by {\em Planck}. Pointed observations with {\em Herschel} and other instruments are also beginning to find similar objects. Rigby et al. (2013) have found evidence for {\em Herschel} source overdensities around z$\sim$2 to 4 radio galaxies, consistent with the presence of star forming galaxies clusters comparable to those discussed here. Additionally, Valtchanov et al. (2013) have found similar evidence for {\em Herschel} sources associated with the z=2.156 Spiderweb galaxy, while Noble et al. (2013) have used a combination of {\em Herschel} and SCUBA2 data to suggest the presence of a z$\sim$ 3 grouping of star forming galaxies behind the z=0.9 supercluster RCS231953+00. Far-IR/submm observations, whether using {\em Planck, Herschel} or other data sets, thus seem poised to provide major new insights into the history of star formation in dense environments at high redshift.  

\section{Conclusions}

We have investigated the nature of {\em Planck} ERCSC sources that lie in $\sim$90 sq. deg. of sky observed by {\em Herschel} as part of the HerMES survey. Of the 16 ERCSC sources that lie in this area we find that four are not associated with nearby discrete objects. Instead they are associated with local overdensities of {\em Herschel} sources, forming clumps of objects whose {\em Herschel} colours suggest they lie at z$>>$0. We investigated the nature of these sources by examining archival multifrequency data or through observations in the near-IR. This reveals evidence from both photometric redshift analysis and from the presence of red sequence galaxies that the {\em Planck} clumps and associated {\em Herschel} sources are clusters of galaxies at redshifts $\sim$1---2. The far-IR emission in these systems, which leads to their detection in the large, 5 arcmin., {\em Planck} beams as compact sources, and as an overdensity of separate sources by {\em Herschel}, results from several of the cluster members experiencing contemporaneous bursts of star formation. A starbursting phase such as this has been suggested by Granato et al. (2004) and others as an important stage in the formation and evolution of galaxy clusters and of the galaxies within them. This phase has hitherto been difficult to uncover. The combination of {\em Planck} and {\em Herschel} observations, as demonstrated here, and analysis of {\em Herschel} data on its own, are capable of detecting such sources and thus providing a new tool for testing models of galaxy and galaxy cluster formation and evolution. We use the ancillary data available for two of our clusters to determine the formation epoch of their constituent galaxies, confirming that we are identifying young clusters in the process of formation. We also compare the star formation rate uncovered in our clusters, to other clusters in the literature. Between redshifts of $\sim$1 and 2 we find that the cluster star formation rate density is roughly constant, but at lower redshift, star formation in clusters rapidly falls.

Our analysis is so far based on a relatively small number of {\em Planck} clumps shown to be galaxy clusters. There is considerable potential for this work to be expanded to large sample sizes using both the larger areas being covered by the H-ATLAS and HeLMS surveys with {\em Herschel}, and through followup observations of the sources in the all-sky {\em Planck} survey. This work will provide important new insights into the evolution of clusters and cluster galaxies.

{\bf Acknowledgements}

SPIRE has been developed by a consortium of institutes led
by Cardiff Univ. (UK) and including: Univ. Lethbridge (Canada);
NAOC (China); CEA, LAM (France); IFSI, Univ. Padua (Italy);
IAC (Spain); Stockholm Observatory (Sweden); Imperial College
London, RAL, UCL-MSSL, UKATC, Univ. Sussex (UK); and Caltech,
JPL, NHSC, Univ. Colorado (USA). This development has been
supported by national funding agencies: CSA (Canada); NAOC
(China); CEA, CNES, CNRS (France); ASI (Italy); MCINN (Spain);
SNSB (Sweden); STFC, UKSA (UK); and NASA (USA).
The development of {\em Planck} has been supported by: ESA; CNES and CNRS/INSU-IN2P3-INP (France); ASI, CNR, and INAF (Italy); NASA and DoE (USA); STFC and UKSA (UK); CSIC, MICINN and JA (Spain); Tekes, AoF and CSC (Finland); DLR and MPG (Germany); CSA (Canada); DTU Space (Denmark); SER/SSO (Switzerland); RCN (Norway); SFI (Ireland); FCT/MCTES (Portugal); and The development of {\em Planck} has been supported by: ESA; CNES and CNRS/INSU-IN2P3-INP (France); ASI, CNR, and INAF (Italy); NASA and DoE (USA); STFC and UKSA (UK); CSIC, MICINN and JA (Spain); Tekes, AoF and CSC (Finland); DLR and MPG (Germany); CSA (Canada); DTU Space (Denmark); SER/SSO (Switzerland); RCN (Norway); SFI (Ireland); FCT/MCTES (Portugal); and PRACE (EU). The data presented in this paper will be released through the {\em Herschel} Database in Marseille HeDaM ({hedam.oamp.fr/HerMES}). 
The authors would like to thank Mattia Negrello for the provision of Fig. 14. This work is funded in part by the UK STFC and UKSA. SJO, LW and AS acknowledge support from the Science and Technology Facilities Council  [grant number ST/I000976/1], GdZ acknowledges financial support by ASI/INAF agreement I/072/09/0. Lucia Marchetti, and Mattia Vaccari were supported by the Italian Space Agency (ASI ÒHerschel ScienceÓ Contract I/005/07/0). In contrast, the arXiv helpdesk were no help whatsoever.
\\~\\

\end{document}